\documentclass[oneside, a4paper, onecolumn, 11pt]{article}
\usepackage[left=2.5cm,top=2.5cm,bottom=2.5cm,right=2.5cm]{geometry}
\usepackage{amsmath}
\usepackage{graphicx}
\usepackage{fancyhdr}
\usepackage{calc}
\usepackage{amssymb}
\usepackage[super,sort&compress,comma]{natbib}
\usepackage{setspace}
\usepackage{amsfonts}
\usepackage{commath}
\usepackage[innercaption]{sidecap}
\usepackage[framemethod=TikZ]{mdframed}
\usepackage{hyperref}
\usepackage{parskip}
\usepackage{pifont}

\newcommand{\Rom}[1]{\expandafter\@slowromancap\romannumeral #1@}
\makeatother
{ \normalfont} 

\newcommand{\m}[3]{#1_{#2 #3}}

\newcommand{\R}{\mathcal R}

\newcommand{\M}{\mathcal{M}}
\expandafter\def\expandafter\normalsize\expandafter{%
	\normalsize
	\setlength\abovedisplayskip{0pt}
	\setlength\belowdisplayskip{5pt}
	\setlength\abovedisplayshortskip{0pt}
	\setlength\belowdisplayshortskip{5pt}
}

\definecolor{Gray}{gray}{0.75}

\newmdenv[backgroundcolor=Gray, leftmargin = 0pt, rightmargin = 0pt, linewidth = 0pt, roundcorner = 2 pt, innerleftmargin=5pt, innerrightmargin=5pt, innertopmargin=5pt, innerbottommargin=5pt]{Frame}

\begin{document}

\newcommand{\kk}{\langle k \rangle}
\newcommand{\kkk}{\langle k^2 \rangle}
\newcommand{\er}{Erd\H{o}s-R\'{e}nyi}
\newcommand{\red}{\color{red}\footnotesize}
\newcommand{\blue}[1]{{\color{blue} #1}}
\newcommand{\subfigimg}[3][,]{%

	\setbox1=\hbox{\includegraphics[#1]{#3}}
	\leavevmode\rlap{\usebox1}
	\rlap{\hspace*{30pt}\raisebox{\dimexpr\ht1-2\baselineskip}{#2}}
	\phantom{\usebox1}
}

\linespread{1.2}


\begin{center}
{\color{blue} \huge \textbf{Reviving a failed network through \\ \vspace{3mm} microscopic interventions}}

\vspace{2mm}
\footnotesize
Hillel Sanhedrai$^{1}$, Jianxi Gao$^{2,3}$, Amir Bashan$^{1}$, Moshe Schwartz$^{4}$, Shlomo Havlin$^{1}$ \& Baruch Barzel$^{5,6,*}$
\end{center}

{\footnotesize
\begin{enumerate}
\item
\textit{Department of Physics, Bar-Ilan University, Ramat-Gan, Israel}
\item
\textit{Network Science and Technology Center, Rensselaer Polytechnic Institute, Troy, USA}
\item
\textit{Department of Computer Science, Rensselaer Polytechnic Institute (RPI), Troy, USA}
\item
\textit{Department of Physics, Tel-Aviv University, Tel-Aviv, Israel}
\item
\textit{Department of Mathematics, Bar-Ilan University, Ramat-Gan, Israel}
\item
\textit{Gonda Multidisciplinary Brain Research Center, Bar-Ilan University, Ramat-Gan, Israel}
\end{enumerate}

\begin{itemize}
\item[\textbf{*}]
\textbf{Correspondence}:\ \textit{baruchbarzel@gmail.com}
\end{itemize} 
}


\textbf{
From mass extinction to cell death, complex networked systems often exhibit abrupt dynamic transitions between desirable and undesirable states. Such transitions are often caused by topological perturbations, such as node or link removal, or decreasing link strengths. The problem is that reversing the topological damage, namely retrieving the lost nodes or links, or reinforcing the weakened interactions, does not guarantee the spontaneous recovery to the desired functional state. Indeed, many of the relevant systems exhibit a hysteresis phenomenon, remaining in the dysfunctional state, despite reconstructing their damaged topology. To address this challenge, we develop a two-step recovery scheme:\ first - topological reconstruction to the point where the system can be revived, then dynamic interventions, to reignite the system's lost functionality. Applying this method to a range of nonlinear network dynamics, we identify the recoverable phase of a complex system, a state in which the system can be reignited by microscopic interventions, for instance, controlling just a single node. Mapping the boundaries of this dynamical phase, we obtain guidelines for our two-step recovery.	
}

\pagenumbering{arabic}

Complex systems, biological, social or technological, often experience perturbations and disturbances, from overload failures in power systems \cite{VanMieghem2010,Zhao2016,Dobson2007} to species extinction in ecological networks. \cite{shih2016ecological,jiang2019harnessing,May1977} The impact of such perturbations is often subtle, the system exhibits a minor response, but continues to sustain its global functionality. \cite{cohen2001breakdown,schreier2017exploratory} However, in extreme cases, a large enough perturbation may lead to a major collapse, with the system abruptly transitioning from a functional to a dysfunctional dynamic state \cite{Motter2002,Crucitti2004,Achlioptas2009,Boccaletti2016,Gao2016} (Fig.\ \ref{Illustration}a-d). 
When such collapse occurs, the na\"{i}ve instinct is to reverse the damage, retrieve the failed nodes and reconstruct the lost links. This, however, is seldom efficient, as (i) we rarely have access to all system components, \cite{Cornelius2013} limiting our ability to reconstruct the perturbed network; (ii) even if we \textit{could} reverse the damage, due to hysteresis, in many cases, the system will not spontaneously regain its lost functionality. 

To address this challenge, we derive here a two-step recovery process:\ 
\textbf{Step I.\ Restructuring} (Fig.\ \ref{Illustration}e).\ Retrieving the weighted topology to a point where the system \textit{can} potentially regain its functionality. 
\textbf{Step II.\ Reigniting} (Fig.\ \ref{Illustration}f).\ Introducing dynamic interventions to steer the system back to its functional state. Considering the fact that in most practical scenarios we cannot control the majority of the system components, we design our reigniting around micro-interventions, \textit{i.e}.\ controlling just a small number of components. To achieve this we uncover the recoverable phase, a dynamic state in which the system can be driven towards functionality by controlling just a single node. 

\vspace{5mm}
\textbf{\color{blue} \Large The challenge of irreversible collapse}

\newcommand{\x}{\mathbf{x}}
\newcommand{\xd}{\mathbf{x_{\rm D}}}
\newcommand{\xu}{\mathbf{x_{\rm U}}}
\newcommand{\Basin}{\mathfrak{B}}

\noindent
Consider a complex system of $N$ components (nodes), interacting via the adjacency matrix $A$, a sparse, potentially directed random network with an arbitrary degree distribution $P(k_{\rm in},k_{\rm out})$ (see Supplementary Section 1). 
Each node is assigned an \textit{activity} $x_i(t)$, whose meaning depends on context - \textit{e.g.}, a species abundance in a microbial network or a gene's expression level in a biological setting. We then track the evolution of $x_i(t)$ following \cite{Barzel2013a,Harush2017,Hens2019}

\begin{equation}
\dod{x_i}{t} = M_0 \big( x_i(t) \big) + 
\sum\limits_{j = 1}^{N} \m Aij \m Wij M_1 \big( x_i(t) \big) M_2 \big( x_j(t) \big),
\label{Dynamics}
\end{equation}

in which the interaction dynamics is characterized by three potentially nonlinear functions $M_0(x),M_1(x)$ and $M_2(x)$. The first function, $M_0(x_i)$ captures node $i$'s self-dynamics, describing mechanisms such as protein degradation \cite{Barzel2011} (cellular), individual recovery \cite{PastorSatorras2015} (epidemic) or birth/death processes \cite{Gardner2000} (population dynamics). The product $M_1(x_i)M_2(x_j)$ describes the $i,j$ interaction mechanism, \textit{e.g.}, genetic activation, \cite{Karlebach2008} infection \cite{PastorSatorras2015} or symbiosis. \cite{Holling1959} The strength of the $i,j$ interaction is governed by $\m Wij$, a random weight extracted from the density function $P(w)$, whose average we denote by $\omega = \int_0^{\infty} w P(w) \dif w$. 

In the context of recoverability, we seek to revive the activity of all nodes by activating a selected set of nodes, hence we focus on \textit{cooperative} interactions, in which nodes positively contribute to each others activity. This is expressed in Eq.\ (\ref{Dynamics}) through $\m Wij M_1(x_i) M_2(x_j) \ge 0$ (Supplementary Section 1). Later, in our discussion of the microbiome recoverability, we relax this condition and examine the impact of mixed-sign interactions.

Setting the derivative on the l.h.s.\ of (\ref{Dynamics}) to zero, we obtain the system's fixed-points, $\x_{\alpha} = (x_{\alpha,1},\dots,x_{\alpha, N})^\top$, which, if dynamically stable, represent different \textit{states}, desirable or undesirable, in which the system can naturally reside. Transitions between these states often result from perturbations to $A$ or $W$, such as node/link deletion or reduction in link weight. When this occurs, it is difficult to reverse the unwanted transition. This is because the system often avoids spontaneous recovery, even if we retrieve the lost nodes, links or weights. To illustrate this difficulty we refer to a concrete example below.

\noindent
\textbf{\color{blue} Example 1.\ Cellular dynamics} (Fig.\ \ref{Regulatory}).\
As our first example we consider the Michaelis-Menten model for gene-regulation, capturing activation interactions between genes (Fig.\ \ref{Regulatory}a). Here $M_0(x_i) = -B x_i^a$, $M_1(x_i) = 1$ and $M_2(x_j) = x_j^h/(1 + x_j^h)$, a \textit{switch-like} function that saturates to $M_2(x_j) \rightarrow 1$ for large $x_j$, representing the process of activation (see discussion in Supplementary Section 3.1). 

For sufficiently connected $A$ or large average weight $\omega$ the system exhibits an \textit{active} state $\x_1$, in which all $x_{1,i} > 0$ - capturing a \textit{living} cell. However, perturbations to $A,W$, such as link/node removal or weight loss can cause a sharp transition to the \textit{inactive} state $\x_0 = (0,\dots,0)^\top$, \textit{i.e}.\ cell death. To track this transition systematically we measured the average activity $\bar{x}_{\alpha} = (1/N) \sum_{i = 1}^N x_{\alpha,i}$, which follows $\bar{x}_{\alpha} > 0$ for $\alpha = 1$ and $\bar{x}_{\alpha} = 0$ for $\alpha = 0$. As we subject the system to increasing levels of  stress - here, reducing all weights $W$ by a factor $0 \le q \le 1$, we observe a sudden transition at $q = q_c$ from $\x_1$ (Fig.\ \ref{Regulatory}b, green) to $\x_0$ (red). Next, we attempt to revive the node activities by retrieving the lost weights, finding that the system fails to recover. The reason is that while $\x_1$ is only stable under $q \le q_c$, $\x_0$ is \textit{always} stable - both below and above this threshold. This leads to a hysteresis phenomenon, where the system remains inactive despite the reversal of the perturbation.   

Example 1, above, while representing a specific scenario, illustrates the family of challenges that we tackle here:\ system's with irreversible transitions, driven by perturbations to their weighted topology. To revive such systems, we must not just restructure their lost topology, but also dynamically reignite them by exerting external control over the activities $x_i(t)$.

\vspace{5mm}
\textbf{\color{blue} \Large Recoverability}

The most natural way to reignite the system is to drive all activities $x_i$ towards an initial condition from which the system naturally recovers to the desired $\x_1$. Namely, we must steer the system into $\x_1$'s basin of attraction  

\begin{equation}
\Basin_1 = \Big\{ \x(t = 0) \,\, \Big| \,\, \x(t \rightarrow \infty) = \x_1 \Big\}
\label{Basin}
\end{equation}

which comprises all initial conditions $\x(t = 0)$ from which Eq.\ (\ref{Dynamics}) converges to $\x(t \rightarrow \infty) = \x_1$ (Fig.\ \ref{Illustration}h). The problem is that such level of control over the dynamics of all nodes is seldom attainable, hence we seek to recover the system's functionality by driving just a microscopic fraction $f \rightarrow 0$ of \textit{forced} nodes. 

To achieve this, we consider the limit $f \sim 1/N$, in which case our reigniting is attempted through, typically, a single, randomly selected source node $s$, whose activity we force to follow $x_s(t) = \phi(t)$. In many practical applications our ability to exert such control, even on a single node, is restrictive, limiting the potential forms of $\phi(t)$. Hence, below we reignite (\ref{Dynamics}) using an extremely simple input, $\phi(t) = {\rm Const}$. Other practically accessible forms of $\phi(t)$ are further considered in Supplementary Section 4. As simple as it is, the constant forcing itself is also constrained, as our \textit{forcing capacity} is, often, bounded by $\phi(t) \le \Delta$. Hence, below we seek the conditions where such restricted interventions - controlling just one node and with a forcing bounded by $\Delta$ - can push the remaining nodes into the desired $\Basin_1$.

During our intervention, the remaining $N - 1$ nodes continue to follow the \textit{natural} system dynamics, \textit{i.e}.\ Eq.\ (\ref{Dynamics}), as they respond to the $s$-forcing. In technical terms, the failed state of the system, $\x_0$, captures Eq.\ (\ref{Dynamics})'s initial condition, and the forced node imposes a strict boundary condition at $s$. In a recoverable system, after some time, the activities will enter $\Basin_1$, at which point we can cease our external control and allow the system to naturally transition to $\x_1$, following its internal dynamics. If, however, the system is non-recoverable, such single-node reigniting is insufficient, the system remains at $\Basin_0$, and once we lift our forcing, it relaxes back to $\x_0$, a failed reigniting.

\begin{figure}
\centering
\includegraphics[width=0.75\textwidth]{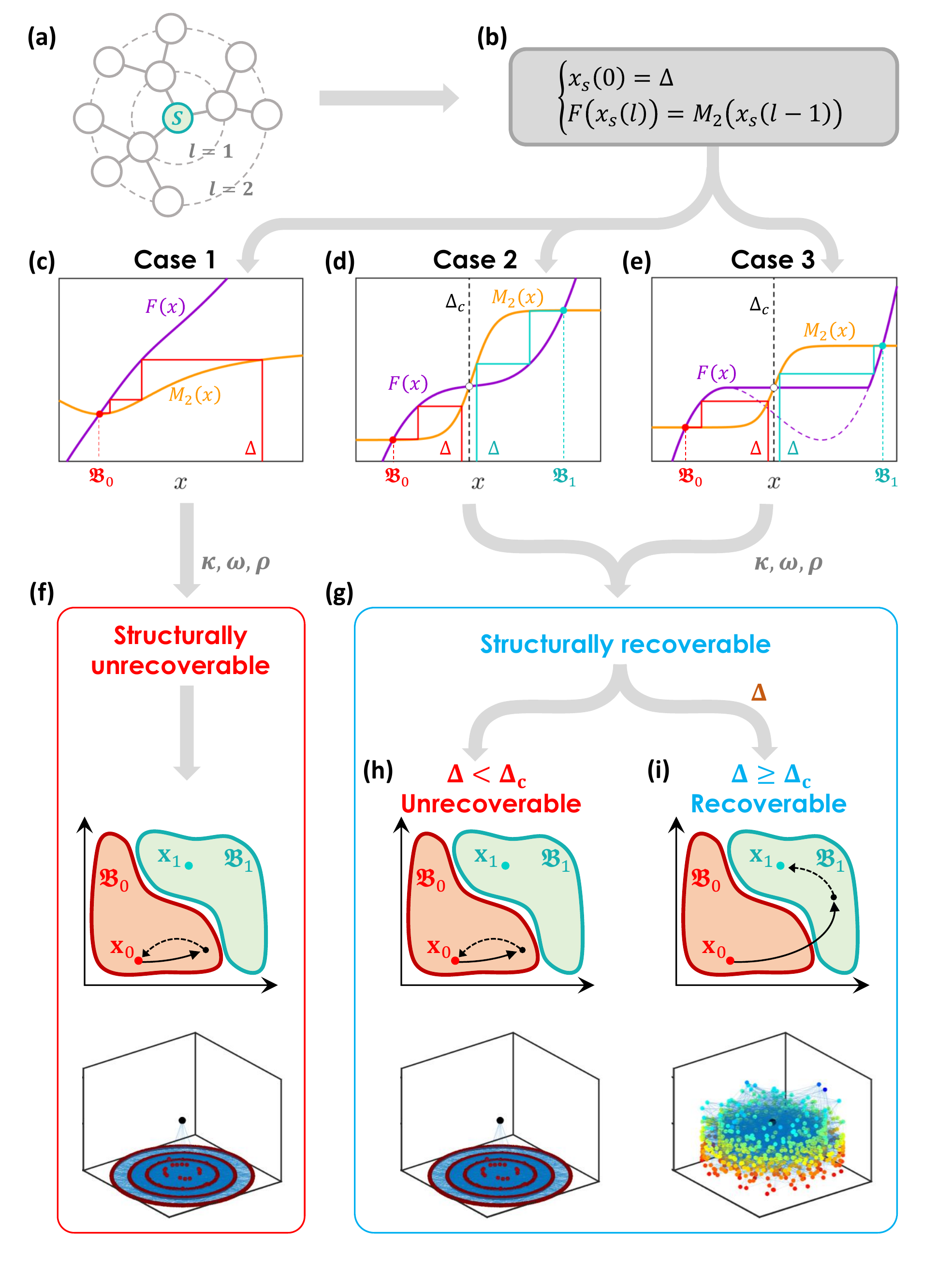}
\vspace{-5mm}
\caption{\footnotesize \textbf{Can a system be reignited by a single node}?\
(a) During reigniting we select a single \textit{source} node $s$ and, for a limited time, artificially force it to sustain a permanent activity $\Delta$. To track the system's response, we divide the network into \textit{shells} $K_s(l)$ comprising all nodes at distance $l$ from $s$. 
(b) The average activity $x_s(l)$ of all nodes at $K_s(l)$ follows the recurrence relation of Eq.\ (\ref{Recursion2}), starting from our forcing at $s$ (top equation), and tracking its propagation as it penetrates the network shells (bottom equation). Successful reigniting requires $x_s(l \rightarrow \infty) \in \Basin_1$, \textit{i.e}.\ that the distant shells were driven towards the desired basin. 
(c) - (e) To track the convergence of the recurrence relation we plot $F(x)$ (purple) and $M_2(x)$ (yellow). The forcing $\Delta$ determines our initial starting point, and the recurrence follows the red or green trajectories. The final state $x_s(l \rightarrow \infty)$ is reached when the two functions, $F(x)$ and $M_2(x)$, intersect. We observe three potential scenarios (Supplementary Section 2.5):\
\textbf{Case 1}.\ $F(x)$ and $M_2(x)$ have a single intersection in $\Basin_0$ (red). Under these conditions, regardless of $\Delta$ the recurrence converges to $\x_s(l \rightarrow \infty) \in \Basin_0$, and hence the system is structurally unrecoverable.
\textbf{Case 2}.\ $F(x)$ and $M_2(x)$ exhibit two intersections, corresponding to the system's two stable fixed points $\x_0$ (red) and $x_1$ (green); the intermediate intersection (white dot) is unstable. Here, for $\Delta < \Delta_c$ the system converges to $\Basin_0$, \textit{i.e}.\ unrecoverable (red trajectory), while for $\Delta \ge \Delta_c$ it approaches $\Basin_1$ - hence it is recoverable (green trajectory).     
\textbf{Case 3}.\ In case $F(x)$ is non-monotonic the critical forcing $\Delta_c$ is determined by $F(x)$'s local maximum point, see Supplementary Section 2.5 for a detailed analysis of Case 3. 
(f) Structural recoverability is determined by the network topology $A,W$ through $\kappa,\omega$ and $\rho$. A structurally unrecoverable (Case 1) system remains confined to $\Basin_0$ and hence cannot be revived, even under arbitrarily large $\Delta$. 
(g) A structurally recoverable system (Cases 2,3) is unrecoverable if $\Delta < \Delta_c$ (left), and recoverable otherwise (right). Hence, the recoverable-phase is driven by three structural parameters ($\kappa,\omega,\rho$, grey) and a single dynamic parameter ($\Delta$, orange).
}
\label{FigRecurssion}	
\end{figure}

\begin{figure}
\centering
\vspace{-15mm}
\includegraphics[width=0.88\textwidth]{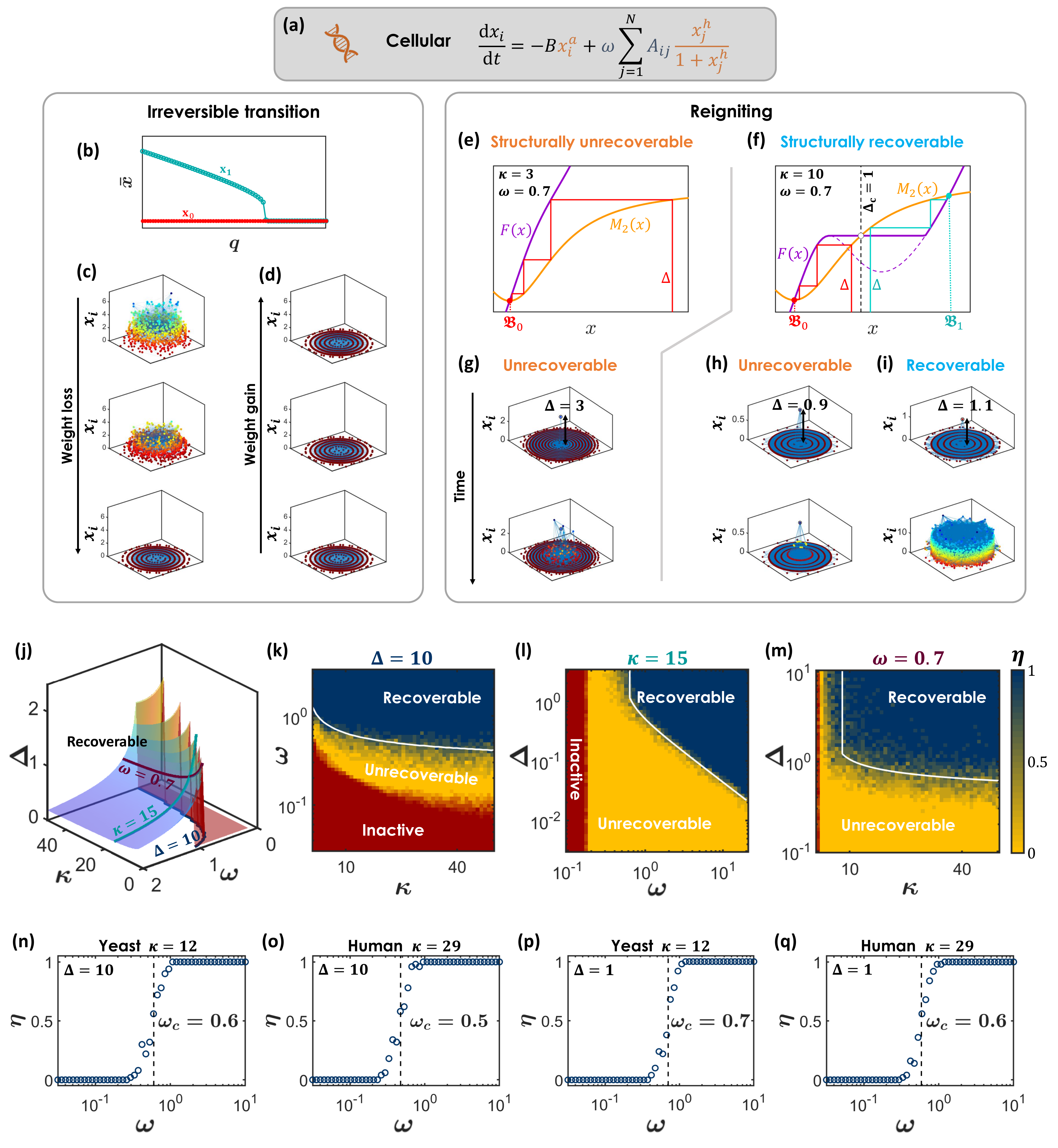}
\vspace{-5mm}
\caption{\footnotesize \textbf{Recoverability of cellular dynamics}.\
(a) The Michaelis-Menten model for sub-cellular dynamics.
(b) The average activity $\bar{x}$ vs.\ the weight reduction $q$ ($q = 0$ - no reduction). $\x_1$ is only stable for $q \le q_c$ (green); $\x_0$ is always stable (red). 
(c) We track $x_i$ for three specific values of $q$. For small $q$ (top) all $x_i > 0$, an \textit{active} state $\x_1$. As $q$ increases $x_i$ become smaller (center), until the system collapses into the inactive $\x_0$ (bottom). 
(d) Retrieving the lost weights does not revive the system since $\x_0$ is always stable. 
To revive the failed system we apply single-node reigniting:\ 
(e) $F(x)$ (purple) and $M_2(x)$ (yellow) as obtained from the left/right hand sides of Eq.\ (\ref{FxM2}) with $\kappa = 3, \omega = 0.7, \rho = 1$. We observe Case 1 of Fig.\ \ref{FigRecurssion}c, with a single intersection at $\Basin_0$ (red). This describes a structurally unrecoverable system.
(f) Under a denser $A$ ($\kappa = 10$), $F(x)$ changes form and the system now follows Case 3 - structurally recoverable with $\Delta_c = 1$ (vertical dashed-line).
(g) Forcing $s$ (black node) to a permanent $\Delta$ activity, no matter how large, fails to reignite the structurally unrecoverable system of panel e.
(h) - (i) Reigniting fails for $\Delta = 0.9$, but succeeds for $\Delta = 1.1$, confirming our predicted $\Delta_c = 1$ in panel f.
(j) The $\kappa,\omega,\Delta$ phase-space. The red area in $\kappa,\omega$ captures the structurally unrecoverable regime. We show three specific cross-sections, fixing $\Delta$ (blue), $\kappa$ (green) and $\omega$ (red), to be expanded in panels k-m.
(k) The $\kappa,\omega$ phase-diagram, as obtained from $5 \times 10^4$ scale-free networks (Supplementary Section 5.3). Each data-point represents the fraction $\eta$ of successful reigniting instances across $20$ independent attempts. We observe the predicted phases:\ inactive (red), where only $\x_0$ is stable, unrecoverable ($\eta = 0$, yellow), and recoverable ($\eta = 1$, blue), where single-node reigniting is possible. Our theoretical prediction is also shown (white solid line). 
(l) - (m) The $\omega,\Delta$ and $\kappa,\Delta$ diagrams. For $\omega < 0.6$ (l) and for $\kappa < 8$ (m) the system is structurally unrecoverable, and hence $\Delta_c \to \infty$. The theoretical prediction is also shown (white solid lines). 
(n) - (o) $\eta$ vs.\ $\omega$ as obtained from the Yeast and Human protein interaction networks ($\kappa =  12$ and $29$; $\Delta = 10$). For $\omega < \omega_c$ (vertical dashed-lines) reigniting fails ($\eta = 0$), above it - it is successful ($\eta = 1$), as predicted. 
(p) - (q) Similar results under $\Delta = 1$. 
}	
\label{Regulatory}
\end{figure}

To analytically track the system's response to our forcing at $s$, we divide the rest of the network into \textit{shells} $K_s(l) = \{ j \,\, | \,\, \m Lsj = l \}$, comprising all nodes located at distance $l$ from $s$ (Fig. \ref{FigRecurssion}a). In this notation $K_s(0) = \{s\}$, $K_s(1)$ is the group of $s$'s nearest outgoing neighbors, $K_s(2)$ its next neighbors, and so on. Then, starting with $x_s(t) = \Delta$, we track the average activity of nodes in $K_s(l)$, via 

\begin{equation}
x_s(l,t) = \dfrac{1}{|K_s(l)|} \sum_{i \in K_s(l)} x_i(t),
\label{Xsl}
\end{equation} 

\noindent
where $|K_s(l)|$ represents the number of nodes in $K_s(l)$. The shells adjacent to the source (small $l$) will be forced to respond to $s$'s activation $\Delta$. The distant shells, however, at $l \to \infty$, may be unaffected, and therefore still within $\Basin_0$. Under these conditions, upon termination of our $\Delta$-forcing, all shells retreat back to $\x_0$, rendering the system unrecoverable. Successful reigniting, therefore, requires that

\begin{equation}
x_s(l \to \infty, t \to \infty) \in \Basin_1, 
\label{Xslinfty}
\end{equation}

capturing a state in which the forcing at $s$ was able to penetrate the network and affect the activity of even the most distant nodes at $K_s(l \rightarrow \infty)$. This represents a recoverable system that will naturally revert to $\x_1$ once the forcing $\Delta$ is deactivated. \cite{FootNoteIn}    

To obtain the final shell states $x_s(l) = x_s(l, t \to \infty)$, we use the fact that, despite its potentially broad degree/weight distribution, our network is otherwise wired, and assigned link weights, at random (Supplementary Section 1). Therefore, (i) $A$ features insignificant degree-correlations, and hence the nodes in $K_s(l)$ are statistically similar to those in $K_s(l^\prime)$, (for $l,l^\prime \ge 1$); (ii) $A$ is locally tree-like, and therefore, asymptotically, there are almost no short-range loops surrounding the source $s$. Below we relax both approximations, when testing our method, numerically, against empirically constructed networks, which, indeed, feature both loops and measurable degree-correlations (Supplementary Section 5.5). However, to advance analytically, we use approximations (i) and (ii) to translate Eq.\ (\ref{Dynamics}) into a direct set of equations for $x_s(l)$. We arrive the recurrence relation (Supplementary Section 2)

\begin{equation}
\begin{cases}
x_s(0) = \Delta,
\\[5pt]
F \big( x_s(l) \big) = M_2 \big( x_s(l - 1) \big),
\end{cases}
\label{Recursion2}
\end{equation}

\noindent
where

\begin{equation}
F(x) = \dfrac{1}{\omega} R(x) - 
\rho \kappa M_2 
\left( 
R^{-1} 
\left(
\omega M_2 
(x) + \omega \kappa M_2 (\bar{\x}_0)
\right) 
\right)
- (1 - \rho) \kappa  M_2 \big( \bar{\x}_0 \big),
\label{Fx}
\end{equation}

\noindent
$R(x) = -M_0(x)/M_1(x)$ and $R^{-1}(x)$ is its inverse function. In (\ref{Fx}) the parameter $\bar{\x}_0$ represents the mean activity of nodes in $K_s(l > 1)$ under the failed state $\x_0$ (Supplementary Section 2.4). The remaining parameters are extracted directly from the weighted topology $A,W$:\ $\omega$ is is the average weight, 

\begin{equation}
\kappa = \dfrac{1}{N} \sum_{i = 1}^N \frac{1}{|K_i(1)|} \sum_{j = 1}^N \m Aij^{\top} k_{j,\rm in} - 1,
\label{kappa}
\end{equation}

represents the average neighbor's residual in-degree, \cite{Newman2010,Gao2016} and $\rho = P(\m Aij = 1|\m Aij^{\top} = 1)$ describes the network's \textit{reciprocity}, \textit{i.e}.\ the probability to observe a link $i \to j$, given the existence of $j \to i$ (for undirected networks $\rho = 1$). 

Equations (\ref{Recursion2}) and (\ref{Fx}) represent our key result. They approximate the recoverability of (\ref{Dynamics}), a multidimensional nonlinear dynamic equation, through a manageable first order recurrence relation. This recurrence takes the system's weighted topology ($\kappa,\rho,\omega$) and its nonlinear dynamics ($M_0,M_1,M_2$) as input, and, together with our intervention constraints ($\Delta$), predicts the system's recoverability as output. Indeed, for any given forcing $\Delta$, the recurrence (\ref{Recursion2}) either converges to $x_s(l \rightarrow \infty) \in \Basin_1$, satisfying the recoverability condition (\ref{Xslinfty}), or to $x_s(l \rightarrow \infty) \in \Basin_0$, indicating a failed recovery. 

To obtain $x_s(l \rightarrow \infty)$ we extract the stationary states of (\ref{Recursion2}), requiring

\begin{equation}
F(x) = M_2(x), 
\label{FixPoint}
\end{equation}

which in turn provides $x_s(l) = x_s(l - 1)$. \cite{May1976simple,strogatz2018nonlinear} Depending on $\kappa,\omega$ and $\rho$, we observe two characteristic behaviors (Fig.\ \ref{FigRecurssion}, Supplementary section 2.5):\ 
\textbf{Structurally unrecoverable} (Fig.\ \ref{FigRecurssion}c,f).\
In case $F(x)$ and $M_2(x)$ have a single intersection $x \in \Basin_0$, then the series in (\ref{Recursion2}) inevitably converges to that point. This captures structural unrecoverability, in which regardless of $\Delta$, single-node reigniting is prohibited.
\textbf{Structurally recoverable} (Fig.\ \ref{FigRecurssion}d,e,g).\
If, on the other hand, $F(x)$ and $M_2(x)$ have several intersections, then the convergence of Eq.\ (\ref{Recursion2}) depends on the boundary condition $x_s(0) = \Delta$, whose magnitude is determined by our forcing capacity. For $\Delta < \Delta_c$, our forcing is too small, and the system approaches $\Basin_0$, a failed reigniting. For $\Delta \ge \Delta_c$ it will reach $\Basin_1$, capturing a successful reigniting. 
\textbf{The recoverable-phase} (Fig.\ \ref{FigRecurssion}i).\
If a system is structurally recoverable and $\Delta \ge \Delta_c$ we say that it is in the recoverable-phase, a state in which one can revive the system by forcing just one node.

Taken together, for a given dynamics $M_0(x),M_1(x),M_2(x)$, our formalism predicts a four-dimensional phase-diagram in the $\kappa,\rho,\omega,\Delta$ phase-space, helping us identify the boundaries of recoverability. Next we investigate this phase space on a range of relevant systems, from cellular dynamics (Example 1), to neuronal and microbial systems. In our first examples below, we consider, for simplicity, undirected networks ($\rho = 1$), reducing our phase-space to three relevant dimensions, $\kappa,\omega$ and $\Delta$. Our final example, reviving a dysfunctional microbiome, examines recoverability under directed interactions ($\rho < 1$).   
 

\vspace{5mm}
\textbf{\color{blue} \Large Applications}

\textbf{\color{blue} Example 1.\ Cellular dynamics} (Fig.\ \ref{Regulatory}; Supplementary Section 3.1).\ 
As our first application we return to Example 1, regulatory dynamics, where $M_0(x) = -x^a, M_1(x) = 1$ and $M_2(x) = x^h/(1 + x^h)$, and therefore $R(x) = x^a$ and $R^{-1}(x) = x^{1/a}$. Equation (\ref{FixPoint}) under $\rho = 1$ becomes 

\begin{equation}
\dfrac{1}{\omega} x^a -
\dfrac{\kappa x^{\frac{h^2}{a}}}
{x^{\frac{h^2}{a}} + \omega^{-\frac{h}{a}} {\big( 1 + x^{h} \big)^{\frac{h}{a}}}}
= 
\dfrac{x^h}{1 + x^h},
\label{FxM2}
\end{equation}

whose roots ($x$) determine the potential fixed-points of the reignited system. Clearly, $x = 0$ is a solution, capturing the fact that the failed state $\x_0 = (0,\dots,0)^{\top} \in \Basin_0$ is always stable. Hence, the question is, under what conditions do we observe a second solution $x > 0$, representing a potential convergence to $\Basin_1$. To answer this, in Fig.\ \ref{Regulatory}e,f we plot $M_2(x)$ vs.\ $x$ (yellow) and observe its intersections with $F(x)$ (purple) as we vary the values of $\kappa,\omega$. This allows us to observe, graphically, the potential convergence of the system to $\Basin_0$ or $\Basin_1$. 

First we consider $\kappa = 3, \omega = 0.7$ (Fig.\ \ref{Regulatory}e). We find that (\ref{FxM2}) exhibits only one solution, represented by the single intersection at $x = 0$ (red dot). This guarantees that (\ref{Recursion2}) converges to $x_s(l \rightarrow \infty) = 0$, independently of $\Delta$. Consequently, the system is structurally unrecoverable. Indeed, Fig.\ \ref{Regulatory}g indicates that, despite the forcing $\Delta$ at $s$, the system fails to reactivate. 

Increasing the network density to $\kappa = 10$, however, changes the picture, as now (\ref{FxM2}) features three intersection points (Fig.\ \ref{Regulatory}f):\ an intermediate unstable point (white) and two stable points at $x = 0$ (red) and at $x > 0$ (green), representing convergence to $\Basin_0$ and $\Basin_1$, respectively. Hence now the system is structurally recoverable,  with critical forcing $\Delta_c = 1$ (vertical dashed line), above which it enters the recoverable-phase. This prediction is corroborated in Fig.\ \ref{Regulatory}h,i:\ under $\Delta = 0.9$ the system remains in $\Basin_0$, but for $\Delta = 1.1$, just above $\Delta_c$, it successfully reignites, precisely as predicted.

This uncovers the existence of a previously overlooked dynamic phase of the Michaelis-Menten model. Indeed, the regulatory system of Fig.\ \ref{Regulatory}a has been previously \cite{Gao2016} shown to follow two phases, \textit{inactive}, where only $\x_0$ is stable vs.\ \textit{bi-stable} where both $\x_0$ and $\x_1$ are stable (Fig.\ \ref{Regulatory}b-d). Yet we now unveil a third phase:\ \textit{recoverable}, a subspace of the bi-stable phase, in which the system can be reignited from $\x_0$ to $\x_1$ by controlling a single node. 

To examine this phase-space systematically, in Fig.\ \ref{Regulatory}j we present the recoverability phase-diagram. For small $\kappa,\omega$ we observe the structurally unrecoverable regime, in which recoverability is unattainable even with arbitrarily large $\Delta$ (red patch in the $\kappa,\omega$ plane). The remaining area in $\kappa,\omega$ represents the structurally recoverable regime, which is split between the unrecoverable phase, for $\Delta < \Delta_c$ (below surface), and the recoverable phase when $\Delta \ge \Delta_c$ (above surface). 

Setting $\Delta$ constant, in Fig.\ \ref{Regulatory}k we construct the $\omega,\kappa$ phase-diagram, by numerically simulating regulatory dynamics on an ensemble of $5 \times 10^4$ scale-free networks, covering $2,500$ distinct combinations of $\omega$ and $\kappa$ (Supplementary Section 5.3). Each data-point captures the fraction of successful recoveries $\eta$ among $20$ independent reigniting trails, from zero successes ($\eta = 0$, yellow) to $100\%$ successful recovery attempts ($\eta = 1$, blue). The white solid line represents our theoretical prediction, based on analyzing the intersections of (\ref{FxM2}). We find that the boundaries of recoverability (yellow/blue) can be well-approximated by our analytical framework. We also present the $\omega,\Delta$ and $\kappa,\Delta$ phase diagrams, further confirming our predicted transitions (Fig.\ \ref{Regulatory}l,m).

To test our predictions in an empirical setting, we collected data on two real biological networks, capturing protein interactions in Human \cite{Rual2005} ($\kappa = 29$) and Yeast \cite{Yu2008} ($\kappa = 12$) cells. Varying $\omega$ and $\Delta$ we measured the reigniting success rate $\eta$, observing in both networks the sharp transition into the recoverable phase (Fig.\ \ref{Regulatory}n-q), which falls precisely on the theoretically predicted phase boundary ($\omega_c$, vertical dashed lines). 

\begin{figure}
\centering
\vspace{-15mm}
\includegraphics[width=0.67\textwidth]{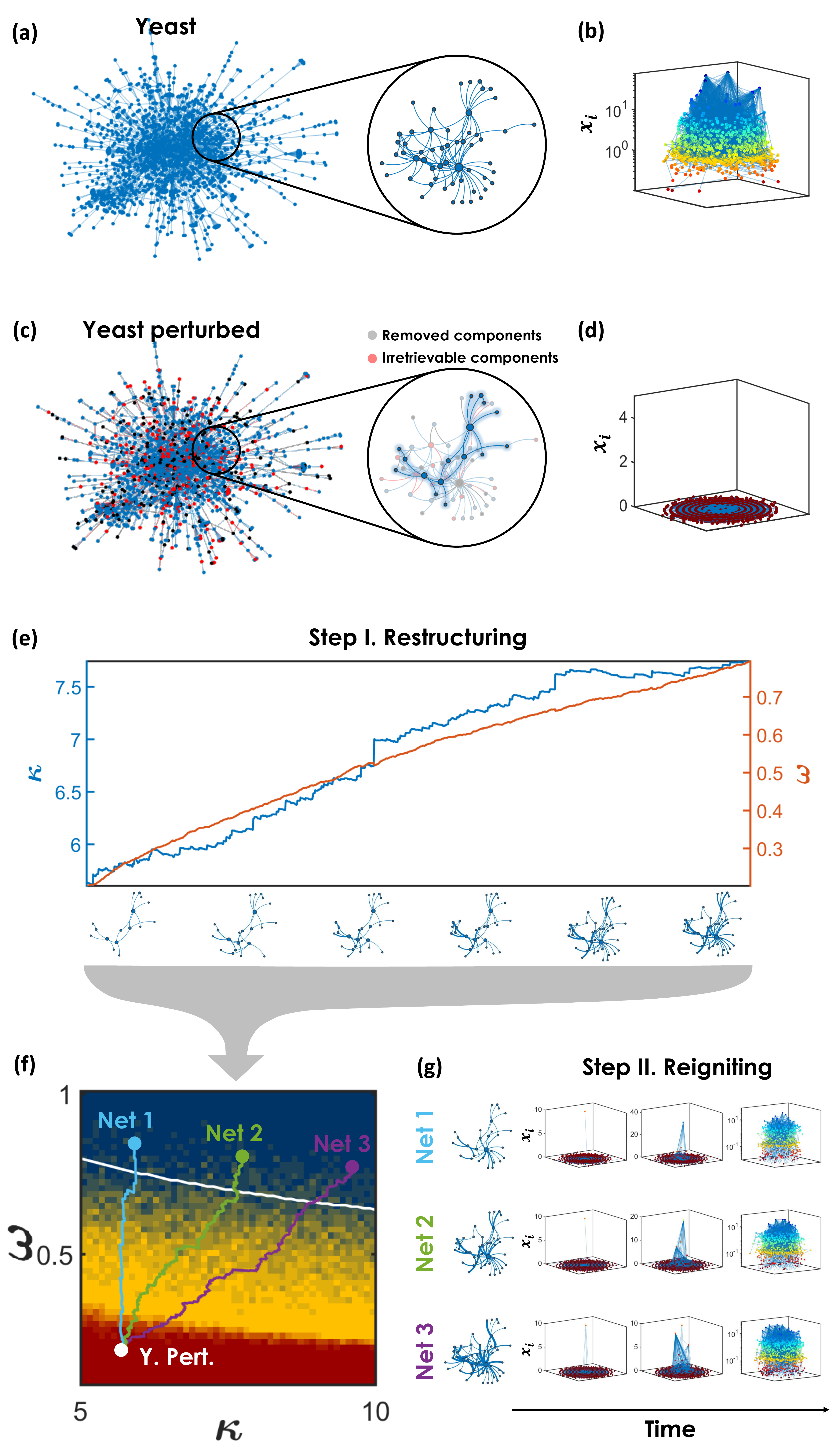}
\vspace{-2mm}
\caption{\footnotesize \textbf{Two-step recovery for reviving a failed cellular system}.\
(a) The unperturbed Yeast protein interaction network. For visibility we focus on the circled sub-network.
(b) As expected, the unperturbed network is in the active state $\x_1$, hence all $x_i > 0$.
(c) - (d) Following extensive perturbation in which $30\%$ of nodes/links (grey and red) were deleted the state of the network collapses to the inactive $\x_0$, \textit{i.e}.\ all $x_i = 0$. The challenge is that some of the deleted components (nodes/links) are inaccessible, and hence cannot be retrieved (red). This captures restructuring constraints that are, indeed, inevitable in realistic scenarios. Circle at center - we focus on the same sub-network shown in panel a, the deleted nodes/links appear in grey and red, the remaining unperturbed components are highlighted, .
(e) \textbf{Step I}.\ Within the given constraints we restructure the network by reintroducing nodes/links or strengthening link weights. We map these interventions into their impact on the two relevant control parameters $\kappa$ (blue) and $\omega$ (orange). For illustration, we show the highlighted sub-network of panel c as it restructures, acquiring nodes, links and increased weights (sub-networks along the $x$-axis).
(f) Restructuring paths in the $\kappa,\omega$ phase-diagram. A successful path leads the network from the collapsed phase (red) into the recoverable-phase (blue). Using our predicted phase-diagram we design several alternative paths, affording us flexibility to, \textit{e.g.}, focus on increasing weights $\omega$ (Net 1, light-blue path) or also on enhancing network density $\kappa$ (Net 3, purple path), all depending on the nature of our constraints. 
(g) \textbf{Step II}.\ Once the network is in the recoverable-phase, we can revive it via single-node reigniting, demonstrated here on each of our restructured networks, Net 1-3.  
}
\label{Restructuring}	
\end{figure}

\vspace{5mm}
\textbf{\color{blue} Restructuring guidelines} (Fig.\ \ref{Restructuring}).\
In case our system is not in the recoverable phase, we must design appropriate restructuring interventions to push $\omega$ or $\kappa$ towards recoverability. In a cellular environment this can be achieved through biochemical interventions that help catalyze or inhibit specific reactions. \cite{robinson2015enzymes} The phase-diagrams of Fig.\ \ref{Regulatory}j-m, mapping the boundaries of the recoverable-phase, can help us design such interventions. 

To illustrate this, in Fig.\ \ref{Restructuring}a-d we simulate a cellular network (Yeast) that has been driven towards inactivity due to a series of node/link deletion (grey nodes/links). Some of the removed components are inaccessible (red), and hence cannot be retrieve during restructuring. With these constraints in mind, we incorporate our two-step strategy:\
\textbf{Step I.\ Restructuring}.\ 
We design a sequence of accessible interventions on $A,W$ to bring the network closer to the recoverable phase. An example of such a sequence is shown in Fig.\ \ref{Restructuring}e (along the $x$-axis). On the $y$-axis we present the accumulating impact of these restructuring interventions on $\kappa$ (blue) and $\omega$ (orange). To revive the system, we seek paths of such accessible interventions that help deliver the network into the bounds of the recoverable-phase (Fig.\ \ref{Restructuring}f). Our goal, we emphasize, is not to simply retrieve the lost components, but to achieve recoverability. This designates, not a single point, but rather an entire sub-space in $\kappa,\omega$ (Fig.\ \ref{Restructuring}f, blue area), affording us some level of restructuring flexibility. Indeed, despite the network's irretrievable components, we were able to design three distinct restructuring paths, leading to different destinations - Net 1,2 or 3 - all within the recoverable sub-space (blue area).
\textbf{Step II.\ Reigniting}.\ 
Once in the recoverable phase we can revive the system via single-node reigniting as shown in Fig.\ \ref{Restructuring}g for Net 1,2 and 3. 

This example illustrates how the phase-diagrams of recoverability provide guidelines for restructuring. For example, in Fig.\ \ref{Restructuring}f path $1$ builds mainly on controlling the interaction strengths ($\omega$), but assumes little freedom to add nodes or links ($\kappa$). In contrast path $3$ involves a significant component of adding nodes/links to $A$, affecting not just $\omega$ but also $\kappa$. The optimal restructuring path is, therefore, determined by the nature of our constraints, \textit{e.g.}, the relative difficulty in adding weights vs.\ adding nodes/links. While the potential space of structural interventions in Step I is incomprehensibly vast, our phase-diagrams reduce this space into just two relevant control parameters - $\kappa$, characterizing $A$'s density, and $\omega$, capturing its link weights (and $\rho$ in case $A$ is directed). This reduction allows us to asses the contribution of all potential interventions by quantifying their effect on these two (or three) parameters - providing optimal pathways towards recoverability (see further discussion in Supplementary Section 3.1.3).

\begin{figure}
\centering
\vspace{-15mm}
\includegraphics[width=1\textwidth]{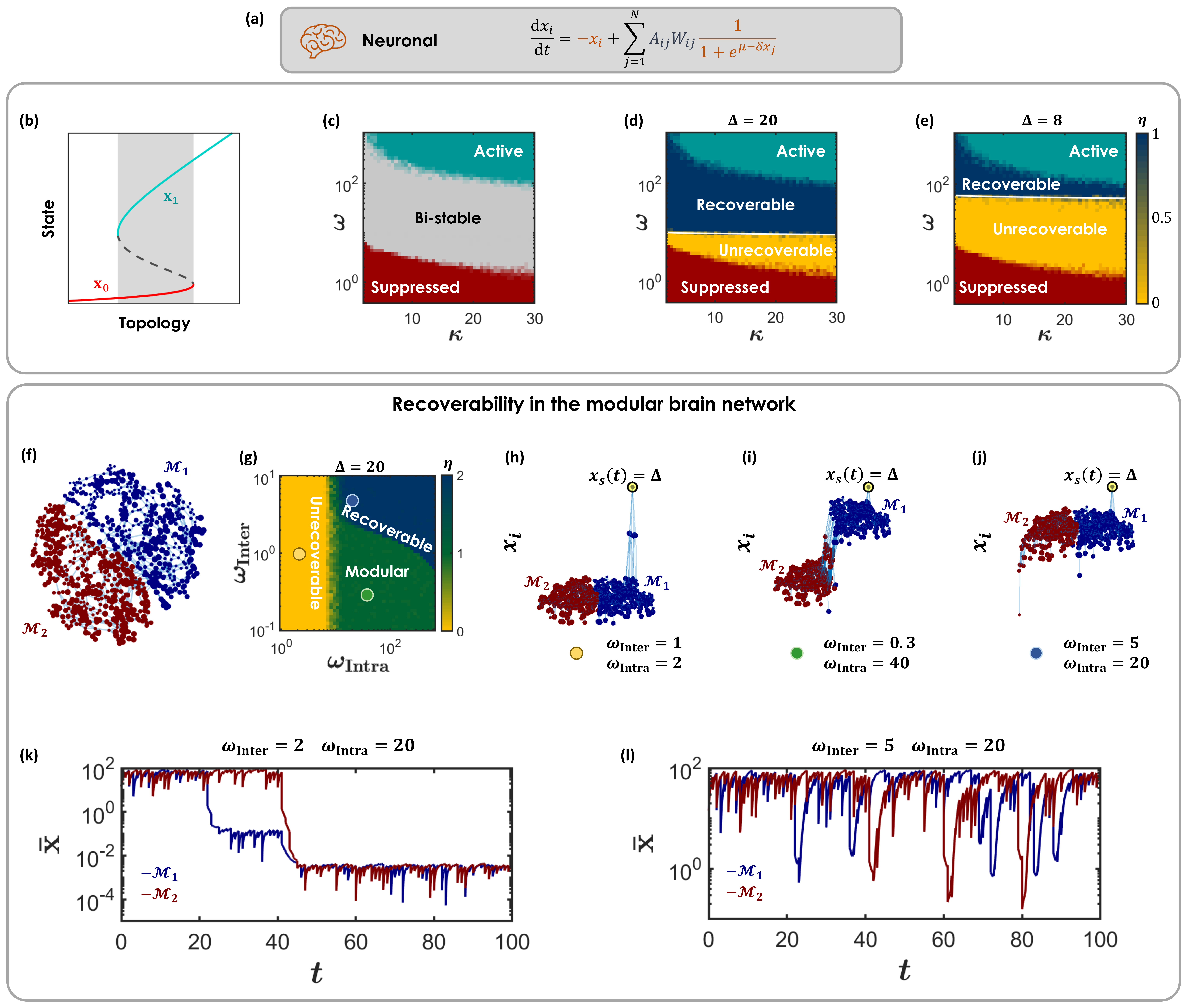}
\vspace{-7mm}
\caption{\footnotesize \textbf{Recoverability of neuronal dynamics}.\
(a) The Wilson-Cowan neuronal dynamics. \citep{FootNote}
(b) The system exhibits three states:\ suppressed $\x_0$ (red) when $\kappa,\omega$ are small; active $\x_1$ (green) under large $\kappa,\omega$; bi-stable (center, grey shaded) in which both $\x_0$ and $\x_1$ are stable.
(c) The three states, suppressed (red), active (green) and bi-stable (grey), as obtained from $5 \times 10^4$ scale-free networks capturing distinct combinations of $\kappa,\omega$. 
(d) - (e) The $\kappa,\omega$ recoverability phase-diagram under $\Delta = 20$ and $\Delta = 8$. Each data-point represents $20$ independent realizations. As predicted, the bi-stable state is split into two distinct dynamic phases - unrecoverable (yellow) vs.\ recoverable (blue). Simulation results (yellow-blue transition) are in perfect agreement with our theoretical predictions (white solid lines). 
(f) The empirically constructed brain network, and its two hemispheres $\M_1$ (blue) and $\M_2$ (red).
(g) We measured the reigniting success rate $\eta$ of the modular brain network along the $\omega_{\rm Intra},\omega_{\rm Inter}$ phase-diagram (see Methods). We observe three phases:\ unrecoverable, in which no module can be revived ($\eta = 0$, yellow); recoverable, in which both modules reactivate ($\eta = 2$, blue); modular, in which reigniting is constrained to a single module ($\eta = 1$, green).
(h) Setting $\omega_{\rm Inter} = 1, \omega_{\rm Intra} = 2$, in the unrecoverable phase (yellow circle), we find, indeed, that reigniting fails to revive both modules.
(i) In the modular phase (green circle), $\M_1$ recovers, but fails to reactivate $\M_2$.
(j) In the recoverable phase (blue circle) reigniting successfully crosses over from $\M_1$ to $\M_2$.    
(k) The average activity $\bar{\x}$ of $\M_1$ (blue) and $\M_2$ (red) vs.\ $t$, as obtained from numerically simulating neuronal dynamics on the brain network. Adding noise, we observe sporadic fluctuations, some causing a transition to the suppressed state. The first collapse occurs to $\M_1$ at $t \sim 20$. Soon after, $\M_2$ also collapses at $t \sim 40$, and the entire network irreversibly fails.
(l) Increasing the inter-module weights to $\omega_{\rm Inter} = 5$, we enter the recoverable phase. Now when one module fails, the other module reignites it. This captures a \textit{fail-safe} system, whose modular structure provides internally embedded self-recoverability.
}	
\label{FigNeuronal}
\end{figure}

  
\textbf{\color{blue} Example 2.\ Neuronal dynamics} (Fig.\ \ref{FigNeuronal}; Supplementary Section 3.2).\
As our second example we consider the Wilson-Cowan neuronal dynamics, \cite{wilson1972excitatory,wilson1973mathematical} in which (\ref{Dynamics}) follows the form shown in Fig.\ \ref{FigNeuronal}a. \cite{laurence2019spectral,FootNote} The system naturally exhibits two dynamic states (Fig.\ref{FigNeuronal}b):\ \textit{suppressed} ($\x_0$, red) in which all activities are constricted; \textit{active} ($\x_1$, green) where $x_i$ are relatively large (green). In between these two extremes lies a bi-stable-phase, in which both $\x_0$ and $\x_1$ are potentially stable (grey shaded). This predicts a hysteresis phenomenon, in which a system driven to the left of the grey area will avoid spontaneous recovery. 

To observe the predicted phases, in Fig.\ \ref{FigNeuronal}c we numerically analyze our ensemble of $5 \times 10^4$ scale-free networks. We find, indeed, the active (green) and suppressed (red) phases, separated by the strip of bi-stability (grey). Our formalism, however, predicts an additional dynamic phase - \textit{recoverable}. This phase, shown in Fig.\ \ref{FigNeuronal}d,e (blue area) identifies a sub-space within the bi-stable regime, under which the system can be driven to $\x_1$ via single-node reigniting. The theoretically predicted phase boundaries are also shown (white solid lines), precisely capturing the numerically observed transitions (yellow/blue). 

In the Methods Section we further demonstrate how the brain's modular structure impacts the recoverability phase-space. We also characterize conditions under which modularity provides a fail-safe mechanism, in which one module revives the other upon failure.

\begin{figure}
\centering
\vspace{-15mm}
\includegraphics[width=0.8\textwidth]{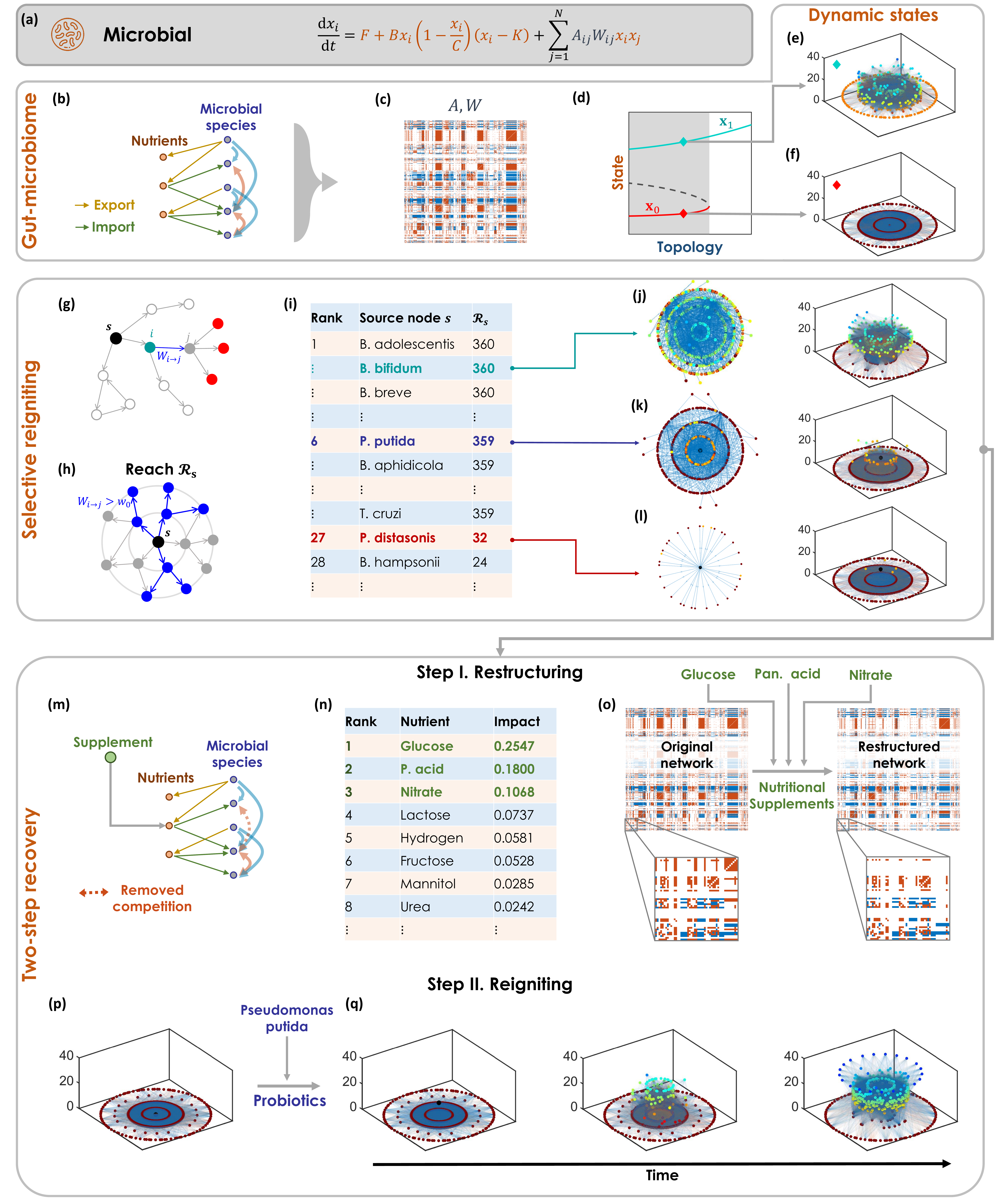}
\vspace{-3mm}
\caption{\footnotesize \textbf{Recoverability of microbial dynamics}.\
(a) Population dynamics.
(b) The export (orange) import (green) network, linking microbial species to the nutrients they produce/consume. Cooperative links (blue) are assigned when $i$'s exports are imported by $j$; adversarial links (red) - when $i$ and $j$ share mutual import/s. 
(c) The resulting gut-microbiome, a diverse weighted and directed network with mixed cooperative (blue) and adversarial (red) links.
(d) - (f) The system exhibits two states, the functional $\x_1$ (green diamond) or the dysfunctional $\x_0$ (red diamond), with a broad range of bi-stability (grey shaded)
(g) The propagation of a reigniting signal:\ $s \to i \to j$. In Methods we show that an effective $i \to j$ link must satisfy $W_{i \to j} > w_0$. 
(h) Counting only effective links, we obtain $s$'s reigniting capacity $\R_s$, as the number potentially reactivated nodes (blue). 
(i) We ranked all microbial species based on $\mathcal R_s$, identifying the network's most effective reigniters. The top $26$ nodes stand out with $\R_s$ orders of magnitude higher than the rest.  
(j) The visibly dense effective network surrounding \textit{B.\ bifidum} ($\R_s = 360$). As expected, \textit{B.\ bifidum} successfully reignites the microbiome (right).  
(k) - (l) We examined two lower ranked species, which failed to revive the network. Specifically, \textit{P.\ putida}, despite having high $\R_s$, causes only a local impact but fails to revive the network.
(m) To reignite with \textit{P.\ putida} we employ our two-step strategy. In Step I, we use nutritional supplements (green) to eliminate competitive links (red dashed-link) by ensuring nutrient availability 
(n) We ranked nutrients based on their contribution to the adversarial links. The top three ranking nutrients are highlighted (green).
(o) Supplementing these three nutrients via dietary interventions, we restructure the microbial network, reducing adversarial interactions (red). See insets for visible comparison. 
(p) In step 2 we administer probiotics to boost the population of a specific species. 
(q) After restructuring we find that \textit{P.\ putida}, an originally failed reigniter, can now successfully revive the microbiome.      
}
\label{FigMicrobial}	
\end{figure}

\textbf{\color{blue} Example 3.\ Microbial dynamics} (Fig.\ \ref{FigMicrobial}, Supplementary Section 3.3).\  
As our final example we consider the gut-microbiome, a microbial community, whose functional state has been shown to crucially impact human health. \cite{gould2018microbiome, garcia2018bacterial} Following perturbations, such as antibiotic treatment, the abundance of the different species may reach critical levels, potentially leading to a dysfunctional microbiome. \cite{willing2011shifting} We, therefore, examine below how our two-step revival strategy can help steer a failed microbiome back to functionality.

To construct the interaction network, we collected data \cite{lim2020large} describing the metabolic in/out flux of $N = 838$ microbial species, allowing us to map their complementary chemical interactions. A cooperative link $i \to j$ appears when species $i$ produces a resource consumed by $j$; \cite{Kehe2020positive} an adversarial link $i \leftrightarrow j$ is assigned if $i$ and $j$ both compete over the same resource. \cite{levy2013metabolic} The weight $\m Wij$ of each link captures the level of inter-species reliance:\ for example, if $i$ is the sole producer of $j$'s only consumed nutrient $m$, then $j$'s growth strongly depends on $i$. If, however, there are many alternative producers of $m$, or if $m$ is just one of $j$'s many consumed nutrients, then $\m Wij$ will be small. We arrive at a directed and highly diverse network with broadly distributed weights, and most crucially, with $\sim 75\%$ adversarial interactions, \textit{i.e}.\ $\m Wij < 0$ (Fig.\ \ref{FigMicrobial}b,c, Supplementary Section 3.3.3). This challenges our theoretical framework, which is primarily designed around cooperative interactions, and therefore helps us test its applicability limits. 

To track the microbial populations we used the dynamics shown in Fig.\ \ref{FigMicrobial}a. The self-dynamics combines logistic growth with the Allee effect, \cite{allee1949principles} and the inter-species cooperative/adversarial interactions follow the Lotka-Volterra dynamics. \cite{Holling1959} The parameter $F$ captures the externally introduced microbial influx. Of the original pool of $N$ species, we find that $32\%$ cannot be supported by the network and undergo extinction. The remaining $568$ species, comprising the actual microbiome composition, exhibit two potential fixed-points (Fig.\ \ref{FigMicrobial}d-f), a functional $\x_1$ and a dysfunctional $\x_0$. Our goal is to apply our two-step recoverability strategy to drive a dysfunctional microbiome at $\x_0$ back towards $\x_1$.

\textbf{\color{blue} Selective reigniting}.\ 
In the Methods Section we show how to evaluate each node's reigniting capacity $\mathcal{R}_s$, allowing us to rank all candidate reigniter species in the microbiome, based on their potential to revive the system (Fig.\ \ref{FigMicrobial}g,h). For a random topology, where all shells $K_s(l)$ are statistically similar, we expect minor differences in $\mathcal R_s$ between the nodes. Here, however, we find that $\R_s$ is highly diverse, with the top $26$ species having $\mathcal{R}_s > 300$, and the remaining species with $\R_s$ one or two orders of magnitude below (Fig.\ \ref{FigMicrobial}i). Such diversity, a consequence of the unique non-random structure of the microbiome, indicates that, in this system, the top $26$ nodes represent preferred candidates for reigniting. To examine this, we attempted reigniting the microbiome with all nodes sequentially, finding that, indeed, success was by far more likely among the top ranked reigniters (Fig.\ \ref{FigMicrobial}j-l, see also Methods).

One exception we identify among these top reigniters is \textit{P.\ putida}, which despite having $\mathcal{R}_s = 359$, ranking $6$ in the reigniters list, still falls short of reviving the system (Fig.\ \ref{FigMicrobial}k). This is due to the fact that \textit{P.\ putida}'s reigniting capacity is hindered by its many surrounding adversarial interactions. Indeed, reigniting, by its nature, builds on the positive activation that the source species $s$ exerts on its neighbors at $K_s(l)$. Such positive impact is undermined by negative links. Next, we use \textit{P.\ putida}'s restricted reigniting capacity as an opportunity to examine our two-step strategy in a realistic setting. 

\textbf{\color{blue} Two-step recovery of a dysfunctional microbiome} (Fig.\ \ref{FigMicrobial}m-q).\
Consider a microbial network at state $\x_0$, which we wish to recover. Among the many practical constraints on our potential interventions, one crucial constraint is that we lack control over the majority of microbial species, and hence we must achieve recovery with a handful of accessible reigniting nodes. Specifically, let us assume that our top accessible candidate for reigniting is precisely \textit{P.\ putida}, which, as Fig.\ \ref{FigMicrobial}k indicates, by itself, cannot revive the system. Hence we employ our two-step strategy.   
\textbf{Step I.\ Restructuring}.\
To enhance \textit{P.\ putida}'s reigniting capacity we wish to inhibit the adversarial interactions, which stand in the way of the system's reactivation. We consider two potential interventions:\ (i) Removing selected nodes that have many adversarial interactions (competition hubs) by means of targeted narrow spectrum antibiotic treatment. (ii) Deleting or weakening adversarial links through nutritional or biological interventions. \cite{costello2012application,hsu2019dynamic} For example, if $i$ and $j$ compete over metabolite $m$, we prescribe dietary supplements to ensure the availability of $m$, thus eliminating the $i,j$ competitive interaction (Fig.\ \ref{FigMicrobial}m). Since antibiotic intervention on an already dysfunctional microbiome may cause additional risks, here, we implement restructuring via option (ii). First, we rank all nutrients based on their relative contribution to the adversarial weights in $W$ (Fig.\ \ref{FigMicrobial}n, Supplementary Section 3.3.3). We then supplement the top three nutrients in the list (green) to restructure the network. Eliminating the competition over these, now freely available nutrients, we arrive at our restructured network, which, thanks to our interventions, has now a more suitable balance of cooperative vs.\ adversarial interactions (Fig.\ \ref{FigMicrobial}o). 
\textbf{Step II.\ Reigniting}.\
In the microbiome, forcing can be implemented by administering probiotics, a common therapeutic practice that helps sustain, artificially, a desired abundance of a selected species. \cite{el2017emerging} The rate of the probiotic intake determines the average activation force, set to be above the reigniting threshold $\Delta_c$. Having restructured the network, we now attempt, once again, to reignite it with \textit{P.\ putida}. We find that the originally failed reigniter, is now, capable of reviving the inactive microbiome (Fig.\ \ref{FigMicrobial}p,q).  
                
\textbf{\color{blue} Applicability limits}.\
Our theoretical analysis helps construct the precise phase-diagrams in the $\kappa,\omega,\Delta$ space, predicting the bounds of the recoverable-phase. These analytically tractable observations rely on a specific set of assumptions, mainly, that $A$ features little degree-correlations and has a scarcity of short-range loops, and that the dynamics is of the form (\ref{Dynamics}) and has primarily cooperative interactions. A precise description of these modeling assumptions is provided in Supplementary Section 1. 

Our applications, however, provide insights that extend well beyond these analytical limits. For example, our empirical networks, cellular, neuronal and microbial, all exhibit non-negligible deviations from the above assumptions, and yet, our analysis correctly predicted their recoverability (Supplementary Section 5.5). The gut-microbiome analysis helped us derive implementable guidelines for reigniting, including also the notion of selective reigniting, despite the network's strong degree-correlations and significant share of adversarial links (Fig.\ \ref{FigMicrobial}). In Supplementary Section 3.4 we further consider diffusive interactions, in which the governing equation is generalized beyond the form of Eq.\ (\ref{Dynamics}). We also analyze alternative forms of reigniting, using periodic activity boosts, which in certain applications may be more accessible than the time-independent forcing ($\Delta$) considered here (Supplementary Section 4).  

\vspace{3mm}
\textbf{\color{blue} \Large Discussion and outlook}

While the structure of complex networks has been deeply investigated over the years, our understanding of their dynamics is still emerging. The challenge is often focused on \textit{prediction}, aiming to foresee a network's dynamic behavior. Here, we go a step further, and focus on \textit{influence}, showing how to steer a system towards a desired behavior. 

Our solution seals a crucial gap in our pursue of nonlinear system controllability. \cite{liu2016control} Existing approaches often rely on specific system symmetries, \cite{isidori1995nonlinear,Hermann1977IEEE,whalen2015observability} which do not cover complex systems of the form (\ref{Dynamics}). Absent such symmetries, complex system control is frequently studied by means of linearization, \cite{coron2007control,sontag1998mathematical} using the linear approximation to help capture the system's local behavior in the vicinity of each of its fixed-point. Such local analysis, however, is insufficient in the context of recoverability, as here we seek to drive the system outside of its current basin and hence beyond the linear regime. To overcome this restriction of locality, cross-basin control was recently developed using time-varying inputs, that adapt until the system is pushed - step by step - into the desired basin of attraction. \cite{Cornelius2013} While highly effective, such interventions require a detailed control over the nodes' dynamics, first, monitoring the system's response, then updating, in real-time, the form of our intervention. Such level of observation/control is not always guaranteed. 

To break this gridlock, we seek non-local control, \textit{i.e}.\ across basins, but with simple dynamic interventions, that do not require highly detailed input signals. Our recoverablity phase-diagram addresses this by identifying unique conditions where such control is attainable. On the one hand, pushing the system across basins, but on the other hand, using an extremely crude and simple control input - a time-limited constant activation $\Delta$ that is applied to just one or few nodes. No fine-tuning or real-time updating of the input signal is required. Indeed, all that is needed is a strong enough jolt to the system, after which it naturally relaxes to its desired target state.  

The microscopic behavior of complex networks is driven by countless parameters, from the fine-structure of $A$ and $W$ to the specific rates of each node's dynamic processes. Our analysis, however, shows, that their large-scale functionality can be traced to a manageable set of \textit{relevant parameters}, \textit{i.e}.\ $\kappa,\omega,\rho$ and $\Delta$. Such dimension reduction is the fundamental premise of statistical-physics, allowing to analyze systems with endless degrees of freedom using a limited set of statistical entries. We believe, that such an approach to network dynamics, can help us understand, predict, and ultimately influence the behavior of these complex multi-dimensional systems.  

\vspace{10mm}

{\color{blue} \textbf{Acknowledgments}}.\
HS wishes to thank the support of the President fellowship of Bar-Ilan University, Israel, and the Mordecai and Monique Katz Graduate Fellowship Program. This research was supported by the Israel Science Foundation (grant no.\ 499/19 and 189/19), the US National Science Foundation-CRISP award (grant no.\ 1735505), the Bar-Ilan University Data Science Institute grant for research on network dynamics, the Binational Israel-China Science Foundation (grant no.\ 3132/19), the US-Israel NSF-BSF program (grant no.\ 2019740), the EU H2020 project RISE (grant no.\ 821115), the EU H2020 project DIT4TRAM (grant no.\ 953783) and the Defense Threat Reduction Agency (DTRA grant no.\ HDTRA-1-19-1-0016).

{\color{blue} \textbf{Author contributions}}.\
All authors designed the research. HS and BB conducted the mathematical analysis. HS performed the numerical simulations and analyzed the data. AB supervised the microbiome analysis. BB was the lead writer of the paper.

{\color{blue} \textbf{Competing interests}}.\
The authors declare no competing interests. 

{\color{blue} \textbf{Data availability}}.\
Empirical data required for constructing the real-world networks (Microbiome, Brain, Yeast PPI, Human PPI) will be uploaded to a freely accessible repository upon publication.

{\color{blue} \textbf{Code availability}}.\
All codes to reproduce, examine and improve our proposed analysis will be made freely available online upon publication.

\clearpage

\begin{figure}
\centering
\includegraphics[width=0.75\textwidth]{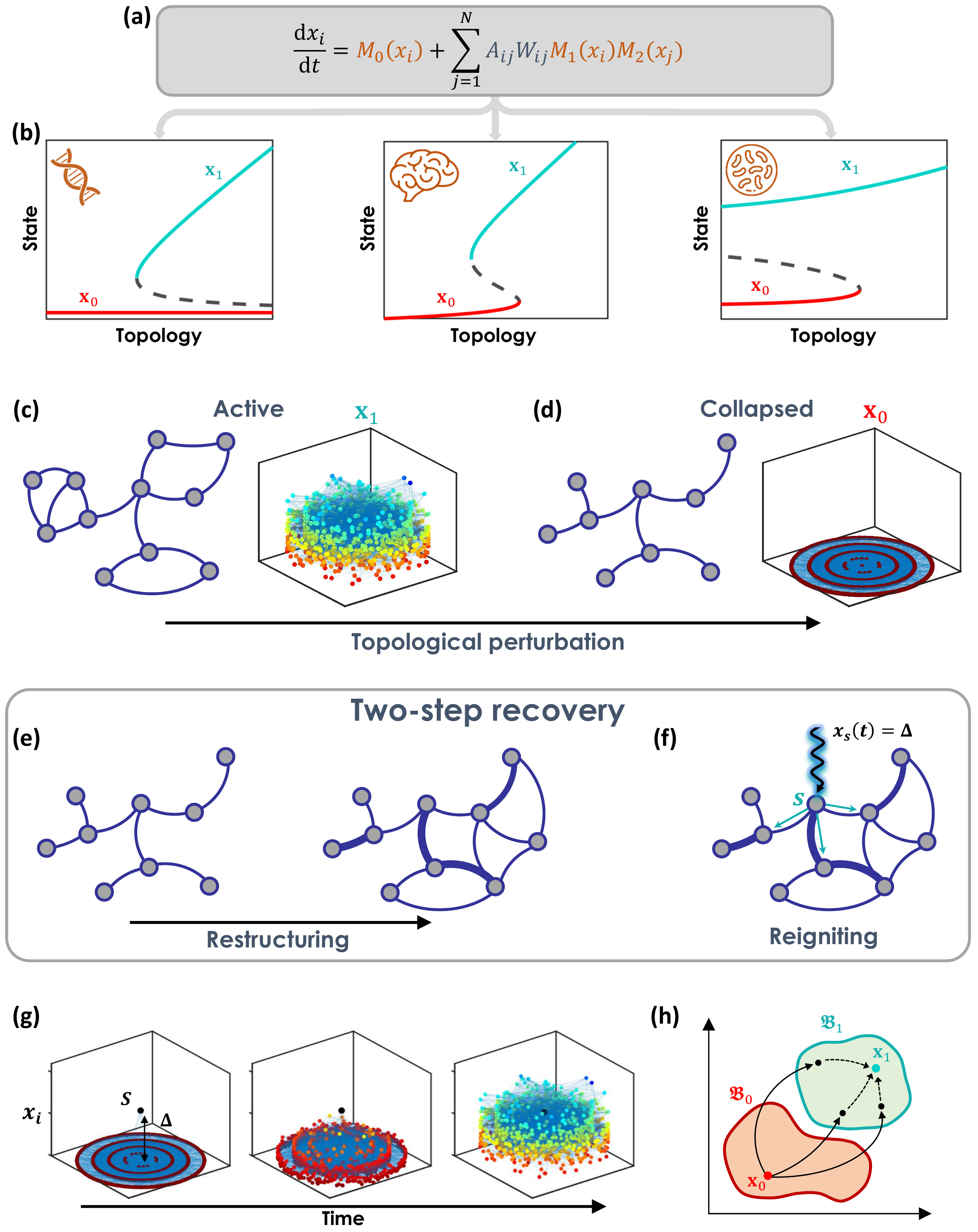}
\vspace{-3mm}
\caption{\footnotesize \textbf{Reviving a failed network}.\
(a) The components of our modeling framework:\ network structure/weights are captured by $A$ and $W$ (grey terms); interaction mechanisms are described by $M_0(x),M_1(x),M_2(x)$ (orange terms).
(b) Depending on the dynamics - \textit{e.g.}, cellular, neuronal or microbial - the system exhibits distinct fixed-points, active ($\x_1$, green) or failed ($\x_0$, red). Transitions between these states are driven by perturbations to $A,W$. 
(c) Unperturbed, the system resides in $\x_1$, where all $x_i > 0$. In this presentation, here and throughout, the network nodes are laid out on the $x,y$ plane, and their activities $x_i$ are captured by the $z$ coordinate. Hence, an active system has all nodes spread along the positive $z$-axis, while a failed network is laid-out around $z \to 0$. We also use color coding from red (small $x_i$) to blue (large $x_i$) as visual aid.  
(d) Perturbations to $A,W$, such as node/link removal or weight reduction, result in a collapse to the inactive $\x_0$. For this system, under $\x_0$, all activities vanish ($z = 0$). 
(e) Step I.\ Restructuring.\ To revive the failed system we first restructure $A,W$ to a point where it \textit{can} recover, namely a point where $\x_1$ is potentially stable. 
(f) Step II.\ Reigniting.\ After restructuring we revive the active state $\x_1$ by controlling a microscopic set of nodes, here the single node $s$. Externally \textit{forcing} $s$ to sustain constant activity $x_s(t) = \Delta$, we drive the network towards $\x_1$. 
(g) Following reigniting ($x_s(t) = \Delta$, black node at center), the forcing signal gradually spreads, until the system's activity $\x_1$ is restored.
(h) In this process we use the natural basin structure of our dynamics. To reignite $\x_1$ we steer the system from $\Basin_0$ (red) to any point within $\Basin_1$ (green). Once in $\Basin_1$, we cease our forcing, and the system naturally transitions to the desired $\x_1$ (dashed arrows).        
}
\label{Illustration}	
\end{figure}

\clearpage

\clearpage
 
\clearpage

\clearpage

\clearpage

\clearpage



\begin{thebibliography}{10}

\bibitem{VanMieghem2010}
{P. Van Mieghem}.
\newblock {\em {Graph Spectra for Complex Networks}}.
\newblock {Cambridge University Press}, {Cambridge, UK}, 2010.

\bibitem{Zhao2016}
{J. Zhao, D. Li, H. Sanhedrai, R. Cohen and S. Havlin}.
\newblock {Spatio-temporal propagation of cascading overload failures in
  spatially embedded networks}.
\newblock {\em {Nature Communications}}, 7:{10094 -- 99}, 2016.

\bibitem{Dobson2007}
{I. Dobson, B.A. Carreras, V.E. Lynch and D.E. Newman}.
\newblock {Complex systems analysis of series of blackouts: Cascading failure,
  critical points, and self-organization}.
\newblock {\em {Chaos}}, 17:026103, 2007.

\bibitem{shih2016ecological}
{H.-Y. Shih, T.-L. Hsieh and N. Goldenfeld}.
\newblock Ecological collapse and the emergence of travelling waves at the
  onset of shear turbulence.
\newblock {\em Nature Physics}, 12(3):245--248, 2016.

\bibitem{jiang2019harnessing}
{J. Jiang, A. Hastings and Y.-C. Lai}.
\newblock Harnessing tipping points in complex ecological networks.
\newblock {\em Journal of the Royal Society Interface}, 16(158):20190345, 2019.

\bibitem{May1977}
{R.M. May}.
\newblock {Thresholds and breakpoints in ecosystems with a multiplicity of
  stable states}.
\newblock {\em Nature}, {269}:{471--477}, 1977.

\bibitem{cohen2001breakdown}
{R. Cohen, K. Erez, D. Ben-Avraham and S. Havlin}.
\newblock Breakdown of the internet under intentional attack.
\newblock {\em Physical Review Letters}, 86(16):3682, 2001.

\bibitem{schreier2017exploratory}
{H.I. Schreier, Y. Soen and N. Brenner}.
\newblock Exploratory adaptation in large random networks.
\newblock {\em Nature Communications}, 8(1):1--9, 2017.

\bibitem{Motter2002}
{A.E. Motter and Y.-C. Lai}.
\newblock {Cascade-based attacks on complex networks}.
\newblock {\em {Physical Review E}}, 66:{065102}, 2002.

\bibitem{Crucitti2004}
{P. Crucitti, V. Latora and M. Marchiori}.
\newblock {Model for cascading failures in complex networks}.
\newblock {\em Physical Review E}, 69:045104--7, 2004.

\bibitem{Achlioptas2009}
{D. Achlioptas, R.M. D'Souza and J. Spencer}.
\newblock {Explosive percolation in random networks}.
\newblock {\em {Science}}, 323:{1453--1455}, 2009.

\bibitem{Boccaletti2016}
{S. Boccaletti, J.A. Almendral, S. Guana, I. Leyvad, Z. Liua, I.S. Nadal, Z.
  Wang and Y. Zou}.
\newblock {Explosive transitions in complex networks’ structure and dynamics:
  Percolation and synchronization}.
\newblock {\em {Physics Reports}}, 660:{1--94}, 2016.

\bibitem{Gao2016}
{J. Gao, B. Barzel and A.-L. Barab\'{a}si}.
\newblock {Universal resilience patterns in complex networks}.
\newblock {\em Nature}, 530:{307–--312}, 2016.

\bibitem{Cornelius2013}
{S.P. Cornelius, W.L. Kath and A.E. Motter}.
\newblock {Realistic control of network dynamics}.
\newblock {\em { Nature Communications}}, 4:{1942--1950}, 2013.

\bibitem{Barzel2013a}
{B. Barzel and A.-L. Barab\'{a}si}.
\newblock {Network link prediction by global silencing of indirect
  correlations}.
\newblock {\em {Nature Biotechnology}}, {31}:{720 -- 725}, 2013.

\bibitem{Harush2017}
{U. Harush and B. Barzel}.
\newblock {Dynamic patterns of information flow in complex networks}.
\newblock {\em {Nature Communications}}, {8}:{2181}, 2017.

\bibitem{Hens2019}
{C. Hens, U. Harush, R. Cohen, S. Haber and B. Barzel }.
\newblock {Spatiotemporal propagation of signals in complex networks}.
\newblock {\em {Nature Physics}}, 15:403, 2019.

\bibitem{Barzel2011}
{B. Barzel and O. Biham}.
\newblock {Binomial moment equations for stochastic reaction systems}.
\newblock {\em Physical Review Letters}, 106:150602--5, 2011.

\bibitem{PastorSatorras2015}
{R. Pastor-Satorras, C. Castellano, P. Van Mieghem and A. Vespignani}.
\newblock {Epidemic processes in complex networks}.
\newblock {\em Rev. Mod. Phys.}, 87:{925--958}, 2015.

\bibitem{Gardner2000}
{T.S. Gardner, C.R. Cantor and J.J. Collins}.
\newblock Construction of a genetic toggle switch in escherichia coli.
\newblock {\em Nature}, 403:339, 2000.

\bibitem{Karlebach2008}
{G. Karlebach and R. Shamir}.
\newblock {Modelling and analysis of gene regulatory networks}.
\newblock {\em {Nature Reviews}}, 9:770--780, 2008.

\bibitem{Holling1959}
{C.S. Holling}.
\newblock {Some characteristics of simple types of predation and parasitism}.
\newblock {\em {The Canadian Entomologist}}, 91:{385--398}, 1959.

\bibitem{FootNoteIn}
{Note that in Eq.\ (\ref{Xslinfty}) we use the $\in$ sign loosely, as strictly
  speaking, $\Basin_1$ includes initial conditions of the form $\x$, not scalar
  averages such as $x_s(l,t)$. Hence, Eq.\ (\ref{Xslinfty}) should be taken to
  mean:\ the shell average $x_s(l \rightarrow \infty, t \rightarrow \infty)$ is
  \textit{congruent} with a state $\x \in \Basin_1$.}

\bibitem{Newman2010}
{M.E.J. Newman}.
\newblock {\em {Networks - an introduction}}.
\newblock {Oxford University Press}, {New York}, 2010.

\bibitem{May1976simple}
{R.M. May}.
\newblock {Simple mathematical models with very complicated dynamics}.
\newblock {\em {Nature}}, {261}:{459--467}, 1976.

\bibitem{strogatz2018nonlinear}
{S.H. Strogatz}.
\newblock {\em Nonlinear dynamics and chaos with student solutions manual: With
  applications to physics, biology, chemistry, and engineering}.
\newblock CRC press, 2018.

\bibitem{Rual2005}
{J.F. Rual \textit{et al}.}
\newblock {Towards a proteome-scale map of the human protein-–protein
  interaction network}.
\newblock {\em Nature}, 437:{1173--1178}, 2005.

\bibitem{Yu2008}
{H. Yu {\it et al.}}
\newblock {High-quality binary protein interaction map of the yeast interactome
  network}.
\newblock {\em {Science}}, 322:104--110, 2008.

\bibitem{robinson2015enzymes}
{P.K. Robinson}.
\newblock Enzymes: principles and biotechnological applications.
\newblock {\em Essays in biochemistry}, 59:1, 2015.

\bibitem{wilson1972excitatory}
{H.R. Wilson and J.D. Cowan}.
\newblock Excitatory and inhibitory interactions in localized populations of
  model neurons.
\newblock {\em Biophysical Journal}, 12(1):1--24, 1972.

\bibitem{wilson1973mathematical}
{H.R. Wilson and J.D. Cowan}.
\newblock A mathematical theory of the functional dynamics of cortical and
  thalamic nervous tissue.
\newblock {\em Kybernetik}, 13(2):55--80, 1973.

\bibitem{laurence2019spectral}
{E. Laurence, N. Doyon, L.J. Dub{\'e} and P. Desrosiers}.
\newblock Spectral dimension reduction of complex dynamical networks.
\newblock {\em Physical Review X}, 9(1):011042, 2019.

\bibitem{FootNote}
{We use a modified version of the model, which can be cast if the form of Eq.\
  (1). See discussion in Supplementary Section 3.2.}

\bibitem{gould2018microbiome}
{A.L. Gould, V. Zhang, L. Lamberti, E.W. Jones, B. Obadia, N. Korasidis, A.
  Gavryushkin, J.M. Carlson, N. Beerenwinkel and W.B. Ludington}.
\newblock Microbiome interactions shape host fitness.
\newblock {\em Proceedings of the National Academy of Sciences},
  115(51):E11951--E11960, 2018.

\bibitem{garcia2018bacterial}
{L. Garc{\'\i}a-Bayona and L.E. Comstock}.
\newblock Bacterial antagonism in host-associated microbial communities.
\newblock {\em Science}, 361(6408), 2018.

\bibitem{willing2011shifting}
{B.P. Willing, S.L. Russell and B.B. Finlay}.
\newblock Shifting the balance: antibiotic effects on host--microbiota
  mutualism.
\newblock {\em Nature Reviews Microbiology}, 9(4):233--243, 2011.

\bibitem{lim2020large}
{R. Lim, J.J.T. Cabatbat, T.L.P. Martin, H. Kim, S. Kim, J. Sung, C.-M. Ghim
  and P.-J. Kim}.
\newblock Large-scale metabolic interaction network of the mouse and human gut
  microbiota.
\newblock {\em Scientific Data}, 7(1):1--8, 2020.

\bibitem{Kehe2020positive}
{J. Kehe, A. Ortiz, A. Kulesa, J. Gore, P.C. Blainey and J. Friedman}.
\newblock Positive interactions are common among culturable bacteria.
\newblock {\em bioRxiv}, 2020.

\bibitem{levy2013metabolic}
{R. Levy and E. Borenstein}.
\newblock Metabolic modeling of species interaction in the human microbiome
  elucidates community-level assembly rules.
\newblock {\em Proceedings of the National Academy of Sciences},
  110(31):12804--12809, 2013.

\bibitem{allee1949principles}
{W. C. Allee, O. Park, A.E. Emerson, T. Park and K.P. Schmidt}.
\newblock Principles of animal ecology.
\newblock 1949.

\bibitem{costello2012application}
{E.K. Costello, K. Stagaman, L. Dethlefsen, B.J. Bohannan and D.A. Relman}.
\newblock The application of ecological theory toward an understanding of the
  human microbiome.
\newblock {\em Science}, 336(6086):1255--1262, 2012.

\bibitem{hsu2019dynamic}
{B.B. Hsu, T.E. Gibson, V. Yeliseyev, Q. Liu, L. Lyon et al.}
\newblock Dynamic modulation of the gut microbiota and metabolome by
  bacteriophages in a mouse model.
\newblock {\em Cell host \& microbe}, 25(6):803--814, 2019.

\bibitem{el2017emerging}
{R. El Hage, E. Hernandez-Sanabria and T. Van de Wiele}.
\newblock Emerging trends in “smart probiotics”: functional consideration
  for the development of novel health and industrial applications.
\newblock {\em Frontiers in microbiology}, 8:1889, 2017.

\bibitem{liu2016control}
{Y.Y Liu and A.-L. Barab{\'a}si}.
\newblock Control principles of complex systems.
\newblock {\em Reviews of Modern Physics}, 88(3):035006, 2016.

\bibitem{isidori1995nonlinear}
{A. Isidori, E.D. Sontag, and M. Thoma}.
\newblock {\em Nonlinear Control Systems}, volume~3.
\newblock Springer, 1995.

\bibitem{Hermann1977IEEE}
{R. Hermann and A. Krener}.
\newblock Nonlinear controllability and observability.
\newblock {\em IEEE Transactions on Automatic Control}, 22(5):728--740, 1977.

\bibitem{whalen2015observability}
{A.J. Whalen, S.N. Brennan, T.D. Sauer and S.J. Schiff}.
\newblock Observability and controllability of nonlinear networks: The role of
  symmetry.
\newblock {\em Physical Review X}, 5(1):011005, 2015.

\bibitem{coron2007control}
J.-M. Coron.
\newblock {\em Control and nonlinearity}.
\newblock Number 136. American Mathematical Soc., 2007.

\bibitem{sontag1998mathematical}
E.D. Sontag.
\newblock Mathematical control theory, volume 6 of texts in applied
  mathematics, 1998.

\bibitem{bullmore2009complex}
E.~Bullmore and O.~Sporns.
\newblock Complex brain networks: graph theoretical analysis of structural and
  functional systems.
\newblock {\em Nature reviews neuroscience}, 10(3):186--198, 2009.

\end{thebibliography}


\begin{thebibliography}{10}

\bibitem{Barzel2013}
{B. Barzel and A.-L. Barab\'{a}si}.
\newblock {Universality in network dynamics}.
\newblock {\em {Nature Physics}}, {9}:{673 -- 681}, 2013.

\bibitem{Gao2016}
{J. Gao, B. Barzel and A.-L. Barab\'{a}si}.
\newblock {Universal resilience patterns in complex networks}.
\newblock {\em Nature}, 530:{307–--312}, 2016.

\bibitem{Tu2017pre}
{C. Tu, J. Grilli, F. Schuessler and S. Suweis}.
\newblock Collapse of resilience patterns in generalized lotka-volterra
  dynamics and beyond.
\newblock {\em Physical Review E}, 95:062307, Jun 2017.

\bibitem{Jiang2018pnas}
{J. Jiang, Z.-G. Huang, T.P. Seager, W. Lin, C. Grebogi, A. Hastings and Y.-C.
  Lai}.
\newblock Predicting tipping points in mutualistic networks through dimension
  reduction.
\newblock {\em Proceedings of the National Academy of Sciences},
  115(4):E639--E647, 2018.

\bibitem{laurence2019spectral}
{E. Laurence, N. Doyon, L.J. Dub{\'e} and P. Desrosiers}.
\newblock Spectral dimension reduction of complex dynamical networks.
\newblock {\em Physical Review X}, 9(1):011042, 2019.

\bibitem{Thibeault2020prr}
{V. Thibeault, G. St-Onge, L.J. Dub\'e and P. Desrosiers}.
\newblock Threefold way to the dimension reduction of dynamics on networks: An
  application to synchronization.
\newblock {\em Physical Review Research}, 2:043215, Nov 2020.

\bibitem{Tu2021iScience}
{C. Tu, P. D'Odorico and S. Suweis}.
\newblock Dimensionality reduction of complex dynamical systems.
\newblock {\em iScience}, 24(1):101912, 2021.

\bibitem{May1976simple}
{R.M. May}.
\newblock {Simple mathematical models with very complicated dynamics}.
\newblock {\em {Nature}}, {261}:{459--467}, 1976.

\bibitem{strogatz2018nonlinear}
{S.H. Strogatz}.
\newblock {\em Nonlinear dynamics and chaos with student solutions manual: With
  applications to physics, biology, chemistry, and engineering}.
\newblock CRC press, 2018.

\bibitem{Alon2006}
{U. Alon}.
\newblock {\em {An Introduction to Systems Biology: Design Principles of
  Biological Circuits}}.
\newblock {Chapman \& Hall}, {London, U.K.}, 2006.

\bibitem{Karlebach2008}
{G. Karlebach and R. Shamir}.
\newblock {Modelling and analysis of gene regulatory networks}.
\newblock {\em {Nature Reviews}}, 9:770--780, 2008.

\bibitem{Barzel2011}
{B. Barzel and O. Biham}.
\newblock {Binomial moment equations for stochastic reaction systems}.
\newblock {\em Physical Review Letters}, 106:150602--5, 2011.

\bibitem{wilson1972excitatory}
{H.R. Wilson and J.D. Cowan}.
\newblock Excitatory and inhibitory interactions in localized populations of
  model neurons.
\newblock {\em Biophysical Journal}, 12(1):1--24, 1972.

\bibitem{wilson1973mathematical}
{H.R. Wilson and J.D. Cowan}.
\newblock A mathematical theory of the functional dynamics of cortical and
  thalamic nervous tissue.
\newblock {\em Kybernetik}, 13(2):55--80, 1973.

\bibitem{allee1949principles}
{W. C. Allee, O. Park, A.E. Emerson, T. Park and K.P. Schmidt}.
\newblock Principles of animal ecology.
\newblock 1949.

\bibitem{Holling1959}
{C.S. Holling}.
\newblock {Some characteristics of simple types of predation and parasitism}.
\newblock {\em {The Canadian Entomologist}}, 91:{385--398}, 1959.

\bibitem{lim2020large}
{R. Lim, J.J.T. Cabatbat, T.L.P. Martin, H. Kim, S. Kim, J. Sung, C.-M. Ghim
  and P.-J. Kim}.
\newblock Large-scale metabolic interaction network of the mouse and human gut
  microbiota.
\newblock {\em Scientific Data}, 7(1):1--8, 2020.

\bibitem{willing2011shifting}
{B.P. Willing, S.L. Russell and B.B. Finlay}.
\newblock Shifting the balance: antibiotic effects on host--microbiota
  mutualism.
\newblock {\em Nature Reviews Microbiology}, 9(4):233--243, 2011.

\bibitem{blaser2016antibiotic}
{M.J. Blaser}.
\newblock Antibiotic use and its consequences for the normal microbiome.
\newblock {\em Science}, 352(6285):544--545, 2016.

\bibitem{lichtman2016host}
{J.S. Lichtman, J.A. Ferreyra, K.M. Ng, S.A. Smits, J.L. Sonnenburg and J.E.
  Elias}.
\newblock Host-microbiota interactions in the pathogenesis of
  antibiotic-associated diseases.
\newblock {\em Cell reports}, 14(5):1049--1061, 2016.

\bibitem{hord2009food}
{N.G. Hord, Y. Tang and N.S. Bryan}.
\newblock Food sources of nitrates and nitrites: the physiologic context for
  potential health benefits.
\newblock {\em The American journal of clinical nutrition}, 90(1):1--10, 2009.

\bibitem{Kessler1998pre}
{D.A. Kessler, Z. Ner and L.M. Sander}.
\newblock Front propagation: Precursors, cutoffs, and structural stability.
\newblock {\em Physical Review E}, 58:107--114, Jul 1998.

\bibitem{Yu2008}
{H. Yu {\it et al.}}
\newblock {High-quality binary protein interaction map of the yeast interactome
  network}.
\newblock {\em {Science}}, 322:104--110, 2008.

\bibitem{Rual2005}
{J.F. Rual \textit{et al}.}
\newblock {Towards a proteome-scale map of the human protein-–protein
  interaction network}.
\newblock {\em Nature}, 437:{1173--1178}, 2005.

\bibitem{bullmore2009complex}
E.~Bullmore and O.~Sporns.
\newblock Complex brain networks: graph theoretical analysis of structural and
  functional systems.
\newblock {\em Nature reviews neuroscience}, 10(3):186--198, 2009.

\bibitem{Newman2010}
{M.E.J. Newman}.
\newblock {\em {Networks - an introduction}}.
\newblock {Oxford University Press}, {New York}, 2010.

\end{thebibliography}

\clearpage

\begin{thebibliography}{10}

\bibitem{bullmore2009complex}
E.~Bullmore and O.~Sporns.
\newblock Complex brain networks: graph theoretical analysis of structural and
functional systems.
\newblock {\em Nature reviews neuroscience}, 10(3):186--198, 2009.

\end{thebibliography}

\clearpage

\textbf{\color{blue} \Large Methods}

\textbf{\color{blue} Recoverability of modular networks}.\
Applying our formalism to an empirically constructed brain network, \cite{bullmore2009complex} allows us to examine its predictive power beyond our analytical assumptions of a random $A$. Indeed, the brain, with its two hemispheres, provides a highly structured (non-random) modular network, partitioned into two clearly distinctive communities $\M_1$ and $\M_2$ (Fig.\ \ref{FigNeuronal}f). It, therefore, offers meaningful insights on recoverability across modules. The question is, can reigniting one module, say $\M_1$, spillover to also revive $\M_2$. Our analysis indicates that this depends on the average strength of the links \textit{within} the modules, $\omega_{\rm Intra}$, vs.\ that of the links \textit{between} the two modules, $\omega_{\rm Inter}$ (Fig.\ \ref{FigNeuronal}g). Clearly, if $\omega_{\rm Intra}$ is too small, then both $\M_1$ and $\M_2$ are, in and of themselves, unrecoverable, and reigniting will inevitably fail (Fig.\ \ref{FigNeuronal}h). If both $\omega_{\rm Intra}$ and $\omega_{\rm Inter}$ are sufficiently large, then the reactivated nodes at $\M_1$ will further reignite their neighbors at $\M_2$, allowing a complete recovery of both modules via single-node reigniting (Fig.\ \ref{FigNeuronal}j). In between these two extremes, we observe a third phase, in which reigniting is confined to $\M_1$, but fails to penetrate $\M_2$ (Fig.\ \ref{FigNeuronal}i). 

The result is a three state phase-space, as shown in Fig.\ \ref{FigNeuronal}g:\ recoverable (blue), in which reigniting at $\M_1$ can reactivate also $\M_2$, unrecoverable (yellow), in which both modules cannot be revived, and modular (green), where $\M_1$ recovers, but the reigniting signal fails to penetrate $\M_2$. To construct the phase-diagram of Fig.\ \ref{FigNeuronal}g we simulated neuronal dynamics on the brain network of Fig.\ \ref{FigNeuronal}f, scanning $2,500$ distinct combinations of $\omega_{\rm Intra}$ and $\omega_{\rm Inter}$. For each combination we attempted $20$ independent realizations of reigniting with randomly selected nodes. We then extracted $\eta$, as the average number of revived modules over the $20$ attempts. Hence, $\eta = 0$ means that no module was revived (yellow), and $\eta = 1$ indicates that (on average) reigniting was restricted to only the source node's module (green), but did not cross over to the other module. Finally, if $\eta = 2$ then both modules were reactivated (blue).

This observation highlights the potential benefits of network modularity for \textit{self-recovery}. Indeed, if one module fails, say $\M_2$, the other module, $\M_1$, if still active, can revive it. This is because $\M_1$'s active nodes can themselves help reignite the inactive $\M_2$. Hence, modularity offers a \textit{fail-safe} network architecture, in which, unless both modules simultaneously fail, one module can reactivate the other. This ensures a sustained activity in the face of sporadic failures. To observe this we simulated neuronal dynamics on our brain network, setting the inter-modular link weights to $\omega_{\rm Inter} = 2$. Starting at $\x_1$, we introduce external noise, that causes the activity of both modules to fluctuate, until a sufficiently large perturbation, occurring at random around $t \sim 20$, leads $\M_1$ to irreversibly fail (Fig.\ \ref{FigNeuronal}k, blue). Shortly after, a similar fate meets $\M_2$ (red) and the entire system collapses to $\x_0$. 

We now repeat the exact same experiment, this time with $\omega_{\rm Inter} = 5$, bringing the system into the fully recoverable-phase ($\eta = 2$, blue). Now, at every instance in which, say $\M_1$ fails, $\M_2$'s active nodes reignite it back into activity (Fig.\ \ref{FigNeuronal}l). Hence, modularity can afford the system a fail-safe dynamics, driven by its capacity for self-recoverability.

\textbf{\color{blue} Selective reigniting in the microbiome}.\ 
A unique aspect of the our empirically constructed microbiome network, is that it deviates significantly from a random topology. Indeed, in a random network, as degree-correlations are negligible, the statistical properties of the shells $K_s(l)$ become approximately independent of $s$ for $l > 1$. In simple terms, while nodes may have diverse degrees, \textit{i.e}.\ $K_s(1) \ne K_{s^\prime}(1)$, their second or third neighbors at $K_s(l > 1)$ are statistically similar. Under these conditions, once a system is in the recoverable phase, one can reignite it using any desired node, as, indeed, all nodes have roughly identical shells in their surrounding. If, however, the environments $K_s(l)$ vary significantly across the nodes, we expect that certain nodes become better reigniters than others. This gives rise to \textit{selective reigniting}, in which the system's recoverability is not just a function of the \textit{network}, but also depends on the specific source node $s$ and its unique \textit{reigniting capacity}.

Consider the reigniting signal as it propagates from the source $s$, and penetrates through the shells $K_s(1),K_s(2),\dots$. At a certain instance it revives a node $i \in K_s(l)$, then advances from $i$ to impact its neighbor $j \in K_s(l + 1)$ and so on (Fig.\ \ref{FigMicrobial}g). In Supplementary Section 3.3.3 we show that such propagation across shells, namely that a revived $i$ can, indeed, reactivate its more distant neighbor $j$, requires that the $i \to j$ link weight exceeds a threshold $w_0$. Hence, the links whose weight $W \ge w_0$, constitute \textit{effectual links}, that help propagate the reigniting signal. The remaining links with $W < w_0$ have but a marginal contribution to the reigniting. We, therefore, construct an effective network, comprising only the effectual links (Fig.\ \ref{FigMicrobial}h, blue links/nodes). This, more selective, network represents the relevant set of interactions that actively participate in the reigniting process. Constructing this network we obtain the effectual shells surrounding each node $s$, whose number of nodes captures $s$'s reigniting capacity $\mathcal R_s$. Nodes with large $\R_s$ are surrounded by many effectual links, and therefore they have a higher reigniting capacity.

In a random topology, where all shells are statistically similar, we expect minor differences in $\mathcal R_s$ between the nodes. In the microbiome, however, we find that $\R_s$ is highly diverse, with the top $26$ species having $\mathcal{R}_s > 300$, and the remaining species with $\R_s$ that is one or two orders of magnitude below (Fig.\ \ref{FigMicrobial}i). Such diversity, a consequence of the unique structure of the microbiome, indicates that, in this system, the top $26$ nodes represent preferred candidates for reigniting. To examine this in Supplementary Fig.\ 7 we show $\eta$ vs.\ the $\R_s$-ranking for all nodes in the microbiome. We clearly observe that the top ranked nodes have a much higher probability to successfully reignite the system.

\end{document}


\title{ \color{blue}  \vspace{40mm} \bf Reviving a failed network through \\ microscopic interventions \\ \vspace{10mm} Supplementary information}

\maketitle
\thispagestyle{empty}
\clearpage

\tableofcontents
\thispagestyle{empty}
\clearpage
\pagenumbering{arabic}

\section{Modeling framework}
\label{SecModelingFramework}

We consider a class of systems captured by 

\begin{equation}
\dod{x_i}{t} = M_0(x_i) + \sum_{j = 1}^N \m Aij \m Wij M_1(x_i) M_2(x_j),
\label{Dynamics}
\end{equation}

where $x_i(t)$ is node $i$'s dynamic \textit{activity} ($i = 1,\dots,N$) and the nonlinear functions $M_0(x)$, $M_1(x), M_2(x)$ describe the system's intrinsic dynamics, \textit{i.e}.\ its self-dynamics ($M_0$) and its interaction mechanisms ($M_1,M_2$). The matrices $A$ and $W$ provide the network's topology and weights, respectively, tracking which components interact ($\m Aij = 1$), and how strongly ($\m Wij$). In (\ref{Dynamics}) we focus on cooperative interactions, in which nodes positively impact each others activity, hence, we take the interaction term to be always positive, namely $\m Wij$ and the product $M_1(x) M_2(x)$ are all $\ge 0$. We also assume $M_2(x)$ is monotonous, excluding, \textit{e.g.}, oscillatory coupling functions. This ensures that the external activation of node $s$ positively impacts all other activities, and thus helps reactivate all other nodes, indeed, a crucial component of the reigniting strategy.   

\textbf{Weighted topology}.\
We take the network topology $A$, an $N \times N$ matrix with an average degree $\av k$, to be large ($N \rightarrow \infty$), sparse ($\av k \ll N$) and potentially directed. The network, we assume, is extracted from the configuration model ensemble, in which the nodes are characterized by an arbitrary degree sequence, but are otherwise randomly linked. Hence, $A$ can follow any in/out degree distribution, $P(k_{\rm in},k_{\rm out})$, but lacks higher order structure, such as degree-correlations, clustering or modularity. 

The weight matrix $W$ captures the fully positive rates of all existing interactions, as extracted from the probability density $P(w)$. From this density we obtain the probability $P(w)\dif w$ for a randomly selected weight to follow $\m Wij \in (w,w + \dif w)$. The weights are assigned at random from $P(w)$ to all links. We denote the expectation of $P(w)$ by 

\begin{equation}
\omega = E(w) = \int_{0}^{\infty} wP(w)\dif w \approx 
\dfrac{\mathbf{1}^{\top} (A \otimes W) \mathbf{1}}{\mathbf{1}^{\top} A \mathbf{1}},
\label{DefOmega} 
\end{equation}

where $\mathbf{1} = (1,\dots,1)^{\top}$ and $\otimes$ represents an element by element product (Hadamard). In the r.h.s.\ of (\ref{DefOmega}) we approximate the expectation $E(w)$ via the \textit{average} weight, summing over all existing link weights $\m Aij \m Wij$ (numerator) and dividing by the total number of links $\m Aij$ (denominator). 

\textbf{Directionality}.\
In (\ref{Dynamics}) a link $\m Aij$ captures a connection in which $j$ affects $i$'s activity $x_i$. In that sense $\m Aij = A_{i \leftarrow j}$ an \textit{incoming} link from $j$ to $i$. Hence $i$'s in-degree is defined as $k_{i,\rm in} = \sum_{j = 1}^N \m Aij$, counting all nodes that \textit{affect} $i$, and its out-degree as $k_{i,\rm out} = \sum_{j = 1}^N \m Aji = \sum_{j = 1}^N A^{\top}_{ij}$, namely all nodes \textit{affected by} $i$. Therefore, to track the pathways of signal propagation from one node to the other we use $A^\top$, as, indeed, $\m {A^\top}ij$ captures a link from $i$ \textit{to} $j$, \textit{i.e}.\ an interaction where $i$ impacts $j$. We can now use this transposed topology $A^{\top}$ to track the network paths, capturing a sequence of (outgoing) links $\m {A^\top}sj, \m {A^\top}jn, \dots, \m {A^\top}mi$, leading from the source $s$ to the target node $i$. Such paths track to potential propagation of a signal from $s$ to $i$. The minimal path length between $s$ and $i$ defines the distance $\m Lsi = L_{s \to i}$ between these two nodes. 

In a fully random directed (sparse) topology the majority of links are strictly one-directional. This is because the probability for creating a short loop of the form $\m Aij = \m {A^{\top}}ij = 1$ is tiny, scaling as $\sim 1/N$. Still, in real-world networks such reciprocity is rather common. Indeed, often if $i$ affects $j$, $j$ also affects $i$. We quantify this by the network's \textit{reciprocity coefficient}

\begin{equation}
\rho = P(\m Aij = 1 | \m {A^{\top}}ij = 1),
\label{Alpha}
\end{equation}

the conditional probability for a $j \to i$ link given the existence of one from $i \to j$. In a fully random topology, with no inclination for reciprocal links we have $\rho \to 0$, having almost no bi-directional links. Conversely in an undeirected network, where all links are reciprocal, we have $\rho = 1$. For example, in genetic interactions, most genes are either transcription factors (regulators) or regulated, hence genetic interactions are typically non-reciprocal, most links are one-directional and $\rho$ is small. In contrast, protein interactions, capturing chemical relationships between proteins, are fully reciprocal, $A$ is undirected and $\rho = 1$. Finally, in population dynamics, competition or symbiosis are, by their nature, bi-directional, while complementarity, an interaction in which species $i$ produces a product needed for species $j$, is directional. Hence in population dynamics we have $0 < \rho < 1$.  

\textbf{Shells}.\
Using $\m Lsi$ we define the \textit{shells} surrounding $s$ via

\begin{equation}
K_s(l) = \Big\{ i = (1,\dots,N) \Big| \m Lsi = l \Big\},
\label{Ksl}
\end{equation}

comprising all nodes at distance $l$ from $s$. Hence, we have $K_s(0) = \{s\}$, $K_s(1) = s$'s group of direct outgoing neighbors, $K_s(2)$ its second neighbors, etc. Nodes in $K_s(l)$ can be potentially \textit{affected} by $s$ through directed pathways of length $l$. In a similar fashion, we denote by 

\begin{equation}
Q_s(l) = \Big\{ i = (1,\dots,N) \Big| \m Lis = l \Big\},
\label{Qsl}
\end{equation}

the incoming $l$ shell around $s$. This group consists of all nodes that can \textit{affect} $s$ from a network distance $l$. Specifically $Q_s(1)$ is the groups of $s$'s incoming neighbors, allowing us, \textit{e.g}., to write Eq.\ (\ref{Dynamics}) as

\begin{equation}
\dod{x_i}{t} = M_0(x_i) + M_1(x_i) \sum_{j \in Q_i(1) } \m Wij M_2(x_j),
\label{Dynamics2}
\end{equation}

in which we replace the term $\sum_{j = 1}^N \m Aij \cdots$ by $\sum_{j \in Q_i(1)} \cdots$. We denote the size of group $X$ by $|X|$, writing, \textit{e.g}., $k_{s,\rm out} = |K_s(1)|$.

\subsection{Resulting approximations}
\label{ResultingApproximations}

Under these assumptions the weighted topology $A \otimes W$ features three characteristics, which we use below to derive our recoverability formalism:\

\begin{enumerate}
\item 
\textbf{Locally tree-like}.\ 
The random patterns of connectivity ensure the scarcity of short range loops (barring reciprocal links of the form $\m Aij = \m {A^{\top}}ij = 1$, which are technically a loop of length two). Hence, the shells $K_s(l)$ have (almost) no intra-shell links. Consequently, a typical node $i \in K_s(l)$ links to a single neighbor from $K_s(l - 1)$ and its remaining $k_{i,in} - 1$ neighbors are from $K_s(l + n), n = 1,2,\dots$. 

\item
\textbf{Shell statistics}.\
While the (weighted) degrees of all nodes may vary, at times - across orders of magnitude, their surrounding shells $K_s(l)$, comprising a random selection of neighboring nodes are statistically similar. Hence, for example, the average degree of nodes in $K_s(l)$, $l \ge 1$, is approximately independent of $s$ or of $l$. In simple terms:\ $s$ may have many more neighbors than $s^\prime$, \textit{i.e}.\ $|K_s(1)| \gg |K_{s^\prime}(1)|$, however $s$'s $|K_s(1)|$ neighbors and $s^\prime$'s $|K_{s^\prime}(1)|$ neighbors are extracted from the same statistical pool, and therefore have similar statistical properties.

\item
\textbf{Cooperativity}.\
With the interactions being of cooperative nature, the effect of reigniting is consistently positive, \textit{i.e}.\ $s$ \textit{activates} its neighbors at $K_s(1)$, which in turn, may also \textit{activate} their neighbors at $K_s(2)$ etc., capturing a sequence of recovering interactions. 

\end{enumerate} 

In Sec.\ \ref{SecRecoverability} we use all of the above features to derive the recurrence relation of Eq.\ (5) of the main text. Hence, our \textit{analytical framework} is exact in the limit where the listed  assumptions 1 - 3 hold, and becomes approximate in case the network $A$ or the weights $W$ do not adhere to the configuration model framework (characteristics 1,2), or if some interactions are of adversarial nature (characteristic 3). Still, our analysis shows that our derivation is quite robust against such discrepancies, and continues to offer both quantitative and qualitative insight. Specifically, we examine:\

\begin{itemize}
\item
Empirical networks, such as Yeast, Human, Brain and Microbiome, which are known to be \textit{imperfect}. All of these networks exhibit a non-negligible density of short loops (\textit{e.g}., clustering) and measurable degree-correlations. The Brain network also has a distinctive modular topology. Still, our results indicate that recoverability is equally relevant in these real-world network topologies (see Sec.\ \ref{SecDeviations}). 
 
\item
The Microbiome incorporates a mixture of cooperative and adversarial interactions, and yet, as we discuss in Sec.\ \ref{SecGutMicrobiome}, as long as the positive links dominate, recoverability is attainable, and continues to follow our theoretical guidelines. 

\end{itemize}  

\textbf{Factorizable interactions}.\
Equation (\ref{Dynamics}) assumes that the pairwise interactions can be factorized in the form $M_1(x_i)M_2(x_j)$. This excludes non-separable interaction dynamics, \textit{i.e}.\ $M(x_i,x_j)$. A classic example of such non-separability is encountered in synchronization dynamics, where the interaction follows $M(x_i,x_j) = \sin (x_j - x_i)$. This restriction to factorizable dynamics is crucial within our analytical framework, which, builds on extracting $M_1(x_i)$ outside the summation, as shown, \textit{e.g}., in Eq.\ (\ref{Dynamics2}). 

In Sec.\ \ref{SecFerromagnetism} we study the recoverability of ferromagnetic interactions, which violate this factorizability condition of Eq.\ (\ref{Dynamics}). While we cannot apply our analytical framework on such dynamics, we can examine their behavior numerically. We find, numerically, that also in these broader forms of dynamics, the system exhibits a recoverable phase. This indicates, once again, that the insights offered by out formalism range beyond the restrictions of our modeling framework.   

\textbf{States of the system}.\
The fixed-points $\x_{\alpha} = (x_{\alpha,1},\dots,x_{\alpha,N})^{\top}$ of the system can be obtained by setting the derivative on the l.h.s.\ of Eq.\ (\ref{Dynamics}) to zero. If dynamically stable, these fixed-points represent potential dynamic \textit{states} in which the system can reside. The specific activity of a node $x_{\alpha,i}$ in each of these states may generally depend on that node's weighted in/out degree $s_{i, \rm out} = \sum_{j \in K_i(1)} \m Wij^{\top}$, $s_{i, \rm in} = \sum_{j \in Q_i(1)} \m Wij$. Therefore, since $A,W$ may be extremely heterogeneous, the individual node activities can potentially be highly diverse. Fortunately, as explained above, in the configuration model, while individual nodes vary, the \textit{shell} statistics, are approximately uniform, namely the nodes in $K_s(l)$ are drawn from a similar statistical pool as those belonging to $K_{s^\prime}(l^\prime)$. We, therefore, in our derivation below, characterize the activity of the \textit{system} in each of its states $\x_{\alpha}$ using the \textit{shell-average}

\begin{equation}
\bar{\x}_{\alpha}(t) = \dfrac{1}{N} \sum_{s = 1}^N 
\dfrac{1}{|K_s(1)|} \sum_{i\in K_s(1)}x_{\alpha,i}(t).
\label{xbar}
\end{equation}   

This represents an average over the activities of all \textit{nearest neighbor nodes}, first summing over the direct neighbors of $s$, $K_s(1)$, then covering all $s$ from $1$ to $N$. In (\ref{xbar}) we use $|X|$ to denote the number of nodes in group $X$. To denote the fixed-point we simply omit the $t$ dependence on both sides on (\ref{xbar}), writing $\bar{\x}_\alpha$ instead of $\bar{\x}_\alpha(t)$. As a matter of convention, in our analysis we assume a two-state system:\ an undesirable state $\x_0$ and a desirable state $\x_1$. This can be readily generalized to treat more states, if relevant.

\textbf{Irreversible collapse}.\ 
The class of systems within (\ref{Dynamics}) that we are particularly interested in are ones that may suffer irreversible collapse. This rather common phenomenon occurs when the system exhibits multi-stability, for example, a bi-stable state where both the undesirable $\x_0$ and the desirable $\x_1$ are potentially stable. Such conditions arise quite frequently, as observed in the set of dynamical systems we consider here:\ Regulatory, Brain and Microbial. Since $\x_0$ is stable, the system does not recover on its own, and therefore mandates dynamic intervention or \textit{reigniting}, as we derive in Sec. \ref{SecRecoverability}. 

\textbf{Basins of attraction}.\ Under bi-stability, the state of the system, $\x_0$ or $\x_1$, depends on the initial condition $x_i(t = 0)$. This splits the $N$-dimensional state space of the system $\x$ into basins 

\begin{equation}
\Basin_{\alpha} = \Big\{ \x(t = 0) \,\, \Big| \,\, \x(t \rightarrow \infty) = \x_{\alpha} \Big\}  
\label{Basins}
\end{equation}

which comprise all initial conditions from which Eq.\ (\ref{Dynamics}) converges to the fixed-point $\x_{\alpha}$. Reigniting a system's failed dynamics, therefore, translates to steering it from the \textit{state} $\x_0$ to the \textit{basin} $\Basin_1$.

\clearpage

\section{Recoverability}
\label{SecRecoverability}

\subsection{Modeling single-node reigniting}

Consider a system of the form (\ref{Dynamics}), characterized by two stable states - an undesirable $\x_0$ and a desirable $\x_1$. Let us further assume that the system exhibits a bi-stable phase, in which both states are potentially stable, and that it is presently at the undesirable $\x_0$. We seek dynamic interventions, preferably minimal in nature, that will help us drive the system towards $\x_1$. To achieve this we assign a selected set of nodes $\mathcal{F}$ - the \textit{forced} nodes - whose dynamics we externally control. This, effectively, changes the system's dynamics (\ref{Dynamics}) into

\begin{equation}
\left\{
\begin{array}{cclr}
x_i &=& \phi_i(t) & i \in \mathcal{F}
\\[10pt]
\dod{x_i}{t} &=& M_0(x_i) + \displaystyle \sum_{j = 1}^N \m Aij \m Wij M_1(x_i) M_2(x_j) & 
i \notin \mathcal{F}
\end{array}
\right.,
\label{ForcedDynamics}
\end{equation}

\noindent
in which nodes in $\mathcal{F}$ are forced to follow the external control function $\phi_i(t)$, while the remaining $N - |\mathcal{F}|$ nodes continue to evolve via the system's natural interaction dynamics. 

In a realistic reigniting scenario we require $|\mathcal{F}| \ll N$ and $\phi_i(t)$ to be described by preferably simple functions, capturing the fact that often we have limited access or control over the dynamic behavior of the majority of the nodes. Taken to the limit, we choose

\begin{equation}
\mathcal{F} = \{ s \},
\label{F}
\end{equation}

\noindent
a single forced \textit{source} node, and

\begin{equation}
\phi_s(t) = \Delta,
\label{fst}
\end{equation}

\noindent
a time-independent forcing function. 

\begin{figure}[t]
\centering
\includegraphics[width=1\textwidth]{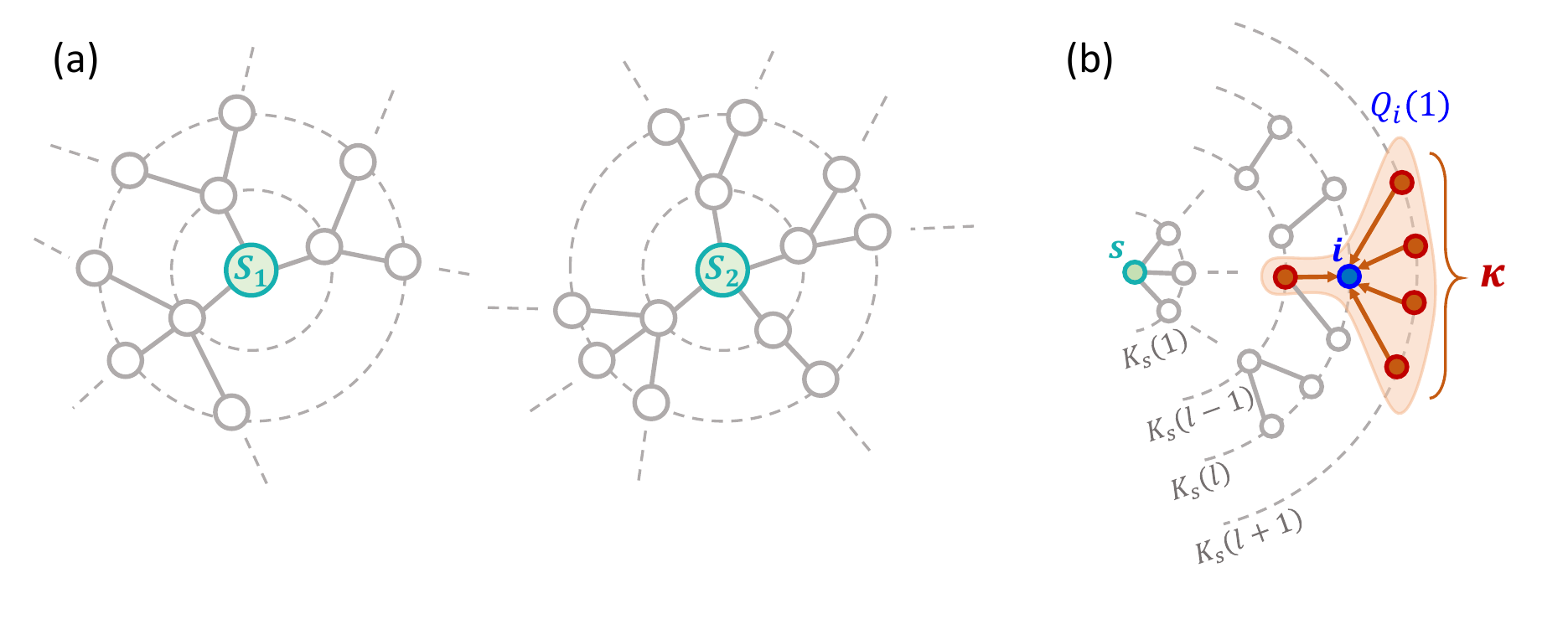}
\vspace{-10mm}
\caption{\footnotesize \color{blue} \textbf{How a single node impacts the network}.\ 
(a) Reigniting by two nodes $s_1, s_2$ (or any microscopic fraction of nodes) introduces two focal points of reactivation into the network. For a large network, the adjacent shells around $s_1$ and $s_2$ have little overlap, hence each only impacts its local neighborhood. Under these conditions $s_1$'s forcing does not meaningfully reinforce that of $s_2$, and each behaves as an independent single-node reigniting source. In case $s_1$ is able to excite the distant shells, to the extent that its impact interferes with that of $s_2$, this indicates that $s_1$'s activation had a non-local impact. This is the case where $s_1$ indeed successfully reignites the entire system. Therefore, we find that reigniting by a microscopic random set of nodes is, in effect, no different that reigniting with a \textit{single} node.  
(b) The impact of our reigniting at $s$ on a node $i \in K_s(l)$ is mediated by $i$'s direct incoming neighborhood $Q_i(1)$ (red). In a large random network, whose structure is locally tree-like, this neighborhood has typically a single node in $K_s(l - 1)$, closer to the source, and $\kappa$ nodes in $K_s(l + n)$, farther away from the source. This allows us to evaluate $\epsilon_n$ in (\ref{Epsilon}), and obtain the recurrence relation of Eq.\ (\ref{Recursion}). 
} 
\label{FigShells}
\end{figure}

Such single node reigniting is, in principle, no different than reigniting by few nodes, since, for a large network, the immediate neighborhoods of each two randomly selected nodes have negligible overlap. Therefore, the impact of one node's forced activity has little interference with that of the other. In other words, for a randomly selected microscopic set of nodes, the group $\mathcal{F}$ will be, most probably, spread throughout the network, comprising an isolated set of reigniting focal points, each impacting only its local neighborhood (Fig.\ \ref{FigShells}a). Under these conditions, the forcing of more than one node does not significantly contribute to the reigniting, and only begins to take effect if the recovered neighbors from one forced node overlap with those of another. Such overlap occurs only if each of the forced nodes has by itself reignited a large fraction of inactive nodes, which, in principle depicts an independent set of single-node reigniting instances. Therefore reigniting via $|\mathcal{F}| \ll N$ is, asymptotically identical to reigniting via $|\mathcal{F}| = 1$ - a single forced activity focal point that must penetrate the network and impact its distant neighbors towards the basin $\Basin_1$.


To evaluate the impact of $s$'s forcing we track the response of the \textit{shells} $K_s(l)$, as defined in Eq.\ (\ref{Ksl}). The average activity in $K_s(l)$ is captured by

\begin{equation}
x_s(l,t) = \dfrac{1}{|K_s(l)|} \sum_{i \in K_s(l)} x_i(t),   
\label{Xslt}
\end{equation}

\noindent
allowing us to evaluate the impact of the forcing $x_s(t) = \Delta$ at any distance $l$ from $s$. 

Being initially at the undesired state we begin with $x_s(l,t = 0) = \bar{\x}_0$ for all $l > 0$. Indeed, $\x_0$ in (\ref{xbar}) captures precisely the average shell activity in the $\alpha = 0$ state. For $l = 0$ we set $x_s(l = 0,t) = \Delta$, as per our forcing intervention. In a \textit{non-recoverable} system the $s$-forcing fails to reignite the system, its impact remains local, and hence the distant shells continue to be in the undesired state's basin of attraction, namely

\begin{equation}
x_s(l \rightarrow \infty, t \rightarrow \infty) \in \Basin_0.   
\label{NonRec}
\end{equation}

Conversely, if the system \textit{is} recoverable, $s$'s forcing penetrates the network to impact the entire system, thus leading to 

\begin{equation}
x_s(l \rightarrow \infty, t \rightarrow \infty) \in \Basin_1,   
\label{Rec}
\end{equation}

\noindent
in which the distant shells have been successfully reignited. Under these conditions we can terminate our forcing, to allow $s$'s local neighborhood, whose state is \textit{frozen} by our external intervention, to also transition to $\x_1$ via the system's natural, undisturbed, dynamics. Note that, strictly speaking, $\Basin_{\alpha}$ is a group of vectors of the form $\x$, while $x_s(l \rightarrow \infty, t \rightarrow \infty)$ is a scalar value, capturing the shell average. Therefore, the sign $\in$ in (\ref{NonRec}) and (\ref{Rec}) should be taken to mean that $x_s(l \rightarrow \infty, t \rightarrow \infty)$ represents the a shell average that is \textit{associated} with an initial condition $\x$ that is within $\Basin_{\alpha}$. 

To obtain a direct set of equations for $x_s(l,t)$ in (\ref{Xslt}), we write

\begin{equation}
\dod{x_s(l,t)}{t} = \dfrac{1}{|K_s(l)|} \sum_{i \in K_s(l)} \dod{x_i}{t},
\end{equation}

which using (\ref{Dynamics}) to express the r.h.s.\ derivative provides

\begin{equation}
\dod{x_s(l,t)}{t} =  \dfrac{1}{|K_s(l)|} \sum_{i \in K_s(l)} 
\Big( M_0(x_i) + \sum_{j = 1}^N \m Aij \m Wij M_1(x_i) M_2(x_j) \Big).   
\label{DDtXsl}
\end{equation}

\noindent
To approximate the summations over $M_0(x_i)$ and $M_1(x_i)$ we use a mean-field approach, writing 

\begin{equation}
\dfrac{1}{|K_s(l)|} \sum_{i \in K_s(l)} M_q \big(x_i(t) \big) \approx M_q \big(x_s(l,t) \big)  
\label{MF1}
\end{equation}

for $q = 0,1$, namely we take $\av{M_q(x)} \approx M_q(\av x)$. This approximation is exact if at least one of the following two conditions applies:\ (i) $M_0(x)$ and $M_1(x)$ are linear; (ii) $x_i(t)$ are uniform within the shell $K_s(l)$. Clearly, these conditions are not guaranteed, however, under many practical scenarios, they represent a sufficient approximation, designed to detect the macro-scale behavior of the system - as fully corroborated by our numerical examination. Indeed, while Eq.\ (\ref{Dynamics}) is, generally, nonlinear, its components, $M_0(x), M_1(x)$, in many of the useful models, are often sub-linear, linear or weakly super-linear, \textit{i.e}.\ involving powers that are not much higher than unity. This satisfies, approximately, condition (i). In other cases we may observe strong nonlinearities in $M_0(x)$ and $M_1(x)$, \textit{e.g.}, in our microbial dynamics, but in such cases, we often have bounded activities $x_i(t)$. This ensures a narrow distribution of $x_i(t)$, roughly satisfying condition (ii). We further elaborate on the relevance of these condition in the appropriate sections, where we analyze each of our specific dynamic systems (Sec.\ \ref{SecDynamicModels}). 

Using approximation (\ref{MF1}), we rewrite (\ref{DDtXsl}) as

\begin{equation}
\dod{x_s(l,t)}{t} = M_0 \big( x_s(l,t) \big) + M_1 \big( x_s(l,t) \big) 
\dfrac{1}{|K_s(l)|} \sum_{i \in K_s(l)} \sum_{j \in Q_i(1)} \m Wij M_2(x_j),   
\label{DDtXsl2}
\end{equation}

where we have also replaced the summation $\sum_{j = 1}^N \m Aij \cdots$ by $\sum_{j \in Q_i(1)} \cdots$, as shown in Eq.\ (\ref{Dynamics2}) above. According to our random network construction the weights $\m Wij$ are randomly assigned from $P(w)$. Therefore, summing over the weights as we do in (\ref{DDtXsl2}) can be replaced by extracting the average weight $\omega$ (\ref{DefOmega}) from the summation, providing

\begin{equation}
\dod{x_s(l,t)}{t} = M_0 \big(x_s(l,t) \big) + \omega M_1 \big(x_s(l,t) \big) 
\dfrac{1}{|K_s(l)|} \sum_{i \in K_s(l)} \sum_{j \in Q_i(1)} M_2(x_j),   
\label{DDtXsl3}
\end{equation}

Next, we consider the nodes in $Q_i(1)$, appearing in the second summation term of (\ref{DDtXsl3}). This sum includes all nodes $j$ that are nearest incoming-neighbors of $i \in K_s(l)$, namely neighbors of a node that is located at distance $l$ from $s$. Therefore, by definition, $j$ is in one of the shells $K_s(l - 1), K_s(l),K_s(l + 1),\dots$, allowing us to write 

\begin{equation}
j \in \bigcup_{n = -1}^{+\infty} K_s(l + n).
\label{WhereIsj}
\end{equation}

We can use (\ref{WhereIsj}) to express the sum over $j \in Q_i(1)$ appearing in Eq.\ (\ref{DDtXsl3}) as

\begin{equation}
\sum_{j \in Q_i(1)} M_2(x_j) = 
\sum_{n = -1}^{+\infty} \sum_{j \in Q_i(1) \cap K_s(l + n)} M_2(x_j), 
\end{equation}

splitting the group $Q_i(1)$ into a disjoint union $Q_i(1) = \bigcup_{n = -1}^{+\infty} Q_i(1) \cap K_s(l + n)$. Next we employ the mean-field approximation of (\ref{MF1}), this time on $M_2(x)$, to write

\begin{equation}
\sum_{j \in Q_i(1) \cap K_s(l + n)} M_2 \big( x_j(t) \big) \approx 
|Q_i(1) \cap K_s(l + n)| M_2  \big( x_s(l + n,t) \big),
\label{SumSplit}
\end{equation}  

in which we replace $x_j(t)$ from the l.h.s.\ with the shell average $x_s(l + n,t)$. Together, we can now use (\ref{SumSplit}) to express the sum over $j \in Q_i(1)$, which appears in Eq.\ (\ref{DDtXsl3}), as

\begin{equation}
\dfrac{1}{|K_s(l)|} \sum_{i \in K_s(l)} \sum_{j \in Q_i(1)} M_2(x_j) \approx 
\dfrac{1}{|K_s(l)|} \sum_{i \in K_s(l)}
\sum_{n = -1}^{+\infty} |Q_i(1)| \epsilon_{i,n}(s,l) M_2 \big( x_s(l + n,t) \big),
\label{Sum}
\end{equation}

where 

\begin{equation}
\epsilon_{i,n}(s,l) = 
\dfrac{|Q_i(1) \cap K_s(l + n)|}{|Q_i(1)|}
\label{Epsiloni}
\end{equation}

denotes the fraction of nodes within $i$'s neighborhood $Q_i(1)$ that belongs to the shells $K_s(l + n)$ \cite{Barzel2013}. 

The r.h.s.\ of Eq.\ (\ref{Sum}) represents an averaging of $\epsilon_{i,n}(s,l)$ and of $Q_i(1)$, as carried out over all nodes $i \in K_s(l)$. However, thanks to characteristic 2 of Sec.\ \ref{ResultingApproximations}, stating that the shell statistics are independent of $s$ or $l$ (for $l \ge 1$), we can replace the $K_s(l)$ average by the \textit{network} average. The meaning is that, for any node characteristic $X_i$, we can write

\begin{equation}
\dfrac{1}{|K_s(l)|} \sum_{i \in K_s(l)} X_i \approx 
\dfrac{1}{N} \sum_{s = 1}^N \dfrac{1}{|K_s(1)|} \sum_{i \in K_s(1)} X_i
\,\,\,\,\,\,\,\,\,\,\,\,\,\,\,\,\,\,\,\, (l \ge 1)
\label{ConfigModel}
\end{equation}

where the l.h.s.\ represents an average performed selectively over nodes in $K_s(l)$, a specific shell, and the r.h.s.\ uses the independence on $s$ and $l$ to (i) average over all potential source nodes $s$, and (ii) replace the specific $l$-shell with $l = 1$, as indeed, \textit{on average}, all shells are the same. Approximation (\ref{ConfigModel}) becomes exact under the conditions discussed in Sec.\ \ref{SecModelingFramework}. In the context of $\epsilon_{i,n}(s,l)$ and $Q_i(1)$, we can use (\ref{ConfigModel}) to further simplify the summation in (\ref{Sum}) into

\begin{equation}
\dfrac{1}{|K_s(l)|} \sum_{i \in K_s(l)} \sum_{j \in Q_i(1)} M_2(x_j) \approx 
\sum_{n = -1}^{+\infty} \epsilon_n (\kappa + 1) M_2 \big( x_s(l + n,t) \big),
\label{SumEpsilon}
\end{equation}

where

\begin{equation}
\epsilon_n = \dfrac{1}{N} \sum_{s = 1}^N \dfrac{1}{|K_s(1)|} \sum_{i \in K_s(1)} \epsilon_{i,n}(s,1)
\label{Epsilon}
\end{equation}

and 

\begin{equation}
\kappa = \dfrac{1}{N} \sum_{s = 1}^N \dfrac{1}{|K_s(1)|} \sum_{i \in K_s(1)} \big( |Q_i(1)| - 1 \big).
\label{Kappa}
\end{equation}
 
The two parameters $\epsilon_n$ and $\kappa$ represent \textit{global} network characteristics, independent of the selected source node $s$, or of the specific shell $K_s(l)$ - solely determined by the network topology $A$. The first, $\epsilon_n$, captures the network expansion, seeking, for a typical node at distance $l$ from any arbitrary source node $s$, how many of its direct neighbors are at greater ($n > 0$), equal ($n = 0$) or smaller ($n = -1$) distance from that source (Fig.\ \ref{FigShells}b). Note that in (\ref{Epsilon}) we focus specifically on $l = 1$, but we could have equally used $l = 2,3,\dots$ instead, under the assumption of the similar shell statistitics. In our discussion below, we, therefore, treat $\epsilon_n$ as extracted from an arbitrary shell $l > 0$, not necessarily $l = 1$. The second parameter, $\kappa$ is the average residual in-degree of the network, which quantifies, for a typical \textit{neighbor} node, \textit{i.e}\. one that was reached via an incoming link, the number of its remaining incoming links. 

To evaluate $\epsilon_n$ in (\ref{Epsilon}) we use characteristic 1 of Sec.\ \ref{ResultingApproximations} - the fact that our networks are locally tree like. In such networks the shells inflate exponentially as $|K_s(l)| \sim e^{\xi l}$, a consequence of the tree-like topology, and hence the randomly wired links have a vanishing probability to link to the inner shells, and preferably connect to nodes from $K_s(l + 1), K_s(l + 2), \dots$. Under these conditions, a typical node $i \in K_s(l)$ links primarily to the most distant shells at $l \to \infty$. Of course, it must have at least one neighbor in $K_s(l - 1)$, as indeed, absent such a neighbor $i \notin K_s(l)$, but then all its remaining $|Q_i(1)| - 1$ neighbors, will be situated at the most distance shells $K_s(l \to \infty)$. 

Our networks, however, are not fully random, due to their reciprocity $\rho$ in (\ref{Alpha}). The meaning of this is that of $i$'s $|Q_i(1)|$ nearest incoming neighbors, a fraction $\rho$ are also outgoing, \textit{i.e}.\ reciprocal. To understand how this deviation impacts the inter-shell link distribution, let us consider a node $j \in Q_i(1)$ who is within this $\rho$ fraction. The node $j$ is an incoming neighbor of $i$, and therefore $\m Aij = 1$, but it is also an outgoing neighbor, and hence $\m Aij^{\top} = 1$ as well. Now since $i \in K_s(l)$, there exists a path of $l$ outgoing links from $s$ to $i$, in the form $\m Asn^{\top},\m Amn^{\top},\dots,\m Aqi^{\top}$. We can now concatenate the link $\m Aij^{\top}$ to this sequence to obtain a path of length $l + 1$ leading from $s$ to $j$. Therefore we have $j \in K_s(l + 1)$. This implies that out of $i$'s $|Q_i(1)|$ nearest neighbors we have:\ 
a single neighbor in $K_s(l - 1)$, i.e.\ 

\begin{equation}
|Q_i(1) \cap K_s(l - 1)| = 1;
\end{equation} 

and the remaining $|Q_i(1)| - 1$ neighbors split into a $\rho$ fraction in $K_s(l + 1)$ and a $1 - \rho$ fraction in $K_s(l \to \infty)$, namely

\begin{eqnarray}
|Q_i(1) \cap K_s(l + 1)| &=& \rho \big( |Q_i(1)| - 1 \big)
\\[5pt]
|Q_i(1) \cap K_s(l \to \infty)| &=& (1 - \rho) \big( |Q_i(1)| - 1 \big).
\end{eqnarray}

Note that, on average, none of $i$'s neighbors is within the $K_s(l)$ shell, indeed - loops are scarce in a tree-like topology - and therefore $|Q_i(1) \cap K_s(l)| \to 0$. This predicts in (\ref{Epsilon}) that

\begin{eqnarray}
\epsilon_{-1} &=&  
\dfrac{1}{N} \sum_{s = 1}^N \dfrac{1}{|K_s(1)|} \sum_{i \in K_s(1)} \dfrac{1}{|Q_i(1)|}
\label{Epsilon1}
\\[5pt]
\epsilon_{0} &=& 0 
\\[5pt]
\label{Epsilon2}
\epsilon_{+1} &=&  
\dfrac{1}{N} \sum_{s = 1}^N \dfrac{1}{|K_s(1)|} \sum_{i \in K_s(1)} \rho \dfrac{|Q_i(1)| - 1}{|Q_i(1)|}
\\[5pt]
\label{Epsilon3}
\epsilon_{\infty} &=&  
\dfrac{1}{N} \sum_{s = 1}^N \dfrac{1}{|K_s(1)|} \sum_{i \in K_s(1)} 
(1 - \rho) \dfrac{|Q_i(1)| - 1}{|Q_i(1)|}.
\label{Epsilon4}
\end{eqnarray}
               
Using (\ref{Kappa}) we arrive at the network's average shell expansion  

\begin{eqnarray}
\epsilon_{-1} &\approx &  
\dfrac{1}{\kappa + 1}
\label{EpsilonMinus}
\\[5pt]
\epsilon_{0} &=& 0 
\\[5pt]
\label{EpsilonZero}
\epsilon_{+1} &\approx &  
\rho \dfrac{\kappa}{\kappa + 1}
\\[5pt]
\label{EpsilonPlus}
\epsilon_{\infty} &\approx &  
(1 - \rho) \dfrac{\kappa}{\kappa + 1}.
\label{EpsilonInf}
\end{eqnarray}

Equations (\ref{EpsilonMinus}) - (\ref{EpsilonInf}) describe the average fraction of links from $K_s(l)$ that lead to $K_s(l - 1),K_s(l),K_s(l + 1)$ and $K_s(l \to \infty)$, respectively. In the limit of $\rho = 1$, an undirected network, we have all links pointing either to $K_s(l - 1)$ or to $K_s(l + 1)$ (Fig.\ \ref{FigShells}b). In the opposite limit of $\rho = 0$, a purely directed network, we have all links, but the one directed to $K_s(l - 1)$, connecting to distant nodes at $l \to \infty$. For $0 < \rho < 1$, a directed network with finite reciprocity, we have a fraction of local links, connected to nodes from $K_s(l + 1)$, and a remaining group of distant neighbors at $K_s(l \to \infty)$.  

We now use (\ref{Kappa}) and (\ref{EpsilonMinus}) - (\ref{EpsilonInf}) to substitute (\ref{Sum}) into (\ref{DDtXsl3}), obtaining 

\begin{eqnarray}
\dod{x_s(l,t)}{t} &=& M_0 \big(x_s(l,t) \big) + 
\omega M_1 \big( x_s(l,t) \big) 
\Big( M_2 \big( x_s(l - 1,t) \big) + 
\rho \kappa M_2 \big( x_s(l + 1,t) \big) 
\nonumber \\[5pt]
&+&
(1 - \rho) \kappa  M_2 \big( x_s(l \to \infty,t) \big) \Big).   
\label{DDtXsl4}
\end{eqnarray}

\noindent
Beginning at an initial condition where the system is at the undesired $\x_0$, the shell-average, as defined in Eq.\ (\ref{xbar}), is initially $\bar{\x}_0$ at all shells $l \ge 1$. Of course, this state changes with $t$, however, for $x_s(l \to \infty,t)$, the most distant shells, we set the boundary condition to $x_s(l \to \infty,t) = \bar{\x}_0$. This provides us with 

\begin{eqnarray}
\dod{x_s(l,t)}{t} &=& M_0 \big(x_s(l,t) \big) + 
\omega M_1 \big( x_s(l,t) \big) 
\Big( M_2 \big( x_s(l - 1,t) \big) + 
\rho \kappa M_2 \big( x_s(l + 1,t) \big) 
\nonumber \\[5pt]
&+&
(1 - \rho) \kappa  M_2 \big( \bar{\x}_0 ) \big) \Big),   
\label{DDtXsl5}
\end{eqnarray}

tracking the dynamics of the average activity in the $l$-shell, as driven by the system's internal dynamic mechanisms $M_0(x),M_1(x)$ and $M_2(x)$. It uses the tree-like structure of the network shells to reduce the detailed network topology $\m Aij$ in (\ref{ForcedDynamics}) to a simplified form, focusing on the average node in $K_s(l)$, which, typically, interacts with a single node in $K_s(l - 1)$ and $\kappa$ residual nodes in $K_s(l + 1)$ or $K_s(l \to \infty)$. This equation is valid for all shells $K_s(l)$ at $l > 0$; for $K_s(0)$ we use the first equation of (\ref{ForcedDynamics}), providing $x_s(0,t) = \Delta$. Together we arrive at 

\begin{equation}
\left\{
\begin{array}{lcl}
x_s(0,t) &=& \Delta
\\[7pt]
\dod{x_s(l,t)}{t} &=& 
M_0 \big( x_s(l,t) \big) + \omega M_1 \big( x_s(l,t) \big)
\Big(
M_2 \big( x_s(l - 1,t) \big) + \rho \kappa M_2 \big( x_s(l + 1,t) \big) 
\\[6pt]
&+& 
(1 - \rho) \kappa  M_2 \big( \bar{\x}_0 \big)
\Big),
\end{array}
\right.
\label{Recursion}
\end{equation}

\noindent
capturing the time evolution of nodes at distance $l$ from the forced $s$; the initial condition is set to $x_s(l,t = 0) = \bar{\x}_0$ for all $l > 0$. Below in our simulation of cellular and brain dynamics (Sec.\ \ref{SecDynamicModels}) we consider Eq.\ (\ref{Recursion}) in the case of undirected networks. Under these conditions the equations can be simplified by setting $\rho = 1$, which provides

\begin{equation}
\left\{
\begin{array}{lcl}
x_s(0,t) &=& \Delta
\\[7pt]
\dod{x_s(l,t)}{t} &=& 
M_0 \big( x_s(l,t) \big) + \omega M_1 \big( x_s(l,t) \big)
\Big(
M_2 \big( x_s(l - 1,t) \big) + \kappa M_2 \big( x_s(l + 1,t) \big) \Big)
\end{array}
\right..
\label{RecursionUD}
\end{equation}

\subsection{Steady-state analysis}
\label{SecSSofRec}

To observe the system's response to our single-node reigniting we wish to obtain the fixed-point of our recurrence relation. Hence, we set the derivative on the l.h.s.\ of (\ref{Recursion}) to zero, yielding

\begin{equation}
\left\{
\begin{array}{ccl}
x_s(0) &=& \Delta
\\[7pt]
0 &=& M_0 \big( x_s(l) \big) + \omega M_1 \big( x_s(l) \big)
\Big(
M_2 \big( x_s(l - 1) \big) + \rho \kappa M_2 \big( x_s(l + 1) \big) 
\\[5pt]
&+& (1 - \rho) \kappa  M_2 \big( \bar{\x}_0 \big)
\Big)
\end{array}
\right..
\label{RecursionSS}
\end{equation}

Note that we have now omitted the term $t$ from $x_s(l,t)$, focusing on the steady-state 

\begin{equation}
x_s(l) = x_s(l, t \rightarrow \infty),
\label{Xsl}
\end{equation} 

\noindent
\textit{i.e}.\ the system's final, long term, activity patterns. To isolate $x_s(l)$ we rewrite the second equation in (\ref{RecursionSS}) as

\begin{equation}
R\big( x_s(l) \big) = \omega M_2 \big( x_s(l - 1) \big) + 
\omega \rho \kappa M_2 \big( x_s(l + 1) \big) + 
\omega (1 - \rho) \kappa  M_2 \big( \bar{\x}_0 \big),
\label{RXsl}
\end{equation}

where 

\begin{equation}
R(x) = - \dfrac{M_0(x)}{M_1(x)}.
\label{Rx} 
\end{equation}

This, by inversion, provides

\begin{equation}
x_s(l)  = R^{-1} \left(
\omega M_2 \big( x_s(l - 1) \big) + \omega \rho \kappa M_2 \big( x_s(l + 1) \big) + 
\omega (1 - \rho) \kappa  M_2 \big( \bar{\x}_0 \big) 
\right),
\label{XslRm1}
\end{equation}

a direct expression of $x_s(l)$ in function of its two neighboring terms, $x_s(l \pm 1)$.

We can now substitute (\ref{XslRm1}) into the second equation in (\ref{RecursionSS}) to transform it into a second order recurrence relation, obtaining

\begin{equation}
\left\{
\begin{array}{l}
x_s(0) = \Delta
\\[7pt]
x_s(l) = R^{-1} \left( 
\omega M_2 \big( x_s(l - 1) \big) + \omega \rho \kappa M_2 \big( x_s(l + 1) \big) + 
\omega (1 - \rho) \kappa  M_2 \big( \bar{\x}_0 \big) 
\right)
\end{array}
\right..
\label{RecursionSecondOrder}
\end{equation}

The challenge is that the recurrence (\ref{RecursionSecondOrder}) is ill-defined, as we only have one boundary condition, $x_s(0) = \Delta$, instead of the two anchoring points required to obtain a unique solution. Hence, in and of itself, Eq.\ (\ref{RecursionSecondOrder}) cannot predict the final shell states $x_s(l)$, and therefore it is insufficient to determine if our reigniting is successful or not. Next, we introduce an approximate approach that allows us to track the desired fixed-points of (\ref{RecursionSecondOrder}).

\subsection{Predicting the final shell states}
\label{SecFinalShell}

While in general (\ref{RecursionSecondOrder}) is under-determined, we can use our prior knowledge on the states of Eq.\ (\ref{Dynamics}) to constrain its potential solutions. Indeed, knowing that our system has potentially two stable fixed-points, $\x_0$ and $\x_1$, we assume that our forcing at $x_s(0)$ can lead, asymptotically, to only two outcomes:\ successful reigniting, in which

\begin{equation}
x_s(l \rightarrow \infty) \in \Basin_1,   
\label{Successful}
\end{equation}

\noindent
or unsuccessful reigniting, where

\begin{equation}
x_s(l \rightarrow \infty) \in \Basin_0.   
\label{UnSuccessful}
\end{equation}

\noindent
Therefore, we do not need to solve the recurrence relation fully, just to determine whether it assumes the asymptotic solution (\ref{Successful}) or (\ref{UnSuccessful}).

We begin by expressing $x_s(l + 1)$ in (\ref{RecursionSecondOrder}) as 

\begin{equation}
x_s(l + 1) = R^{-1} \left(
\omega M_2 \big( x_s(l) \big) + \omega \rho \kappa M_2 \big( x_s(l + 2) \big) + 
\omega (1 - \rho) \kappa  M_2 \big( \bar{\x}_0 \big)
\right),
\label{RXslPlusOne}
\end{equation}

obtained by substituting $l$ with $l + 1$ in the recurrence relation. This allows us to rewrite the recursive series as

\begin{equation}
\left\{
\begin{array}{lcl}
x_s(0) &=& \Delta
\\[7pt]
x_s(l) &=& R^{-1} 
\Bigg(
\omega M_2 \big( x_s(l - 1) \big) 
\\[4pt]		
&+& 
\omega \rho \kappa M_2 
\left( 
R^{-1} 
\left(
\omega M_2 
\big( x_s(l) \big) + \omega \rho \kappa M_2 \big( x_s(l + 2) \big)
+ \omega (1 - \rho) \kappa  M_2 \big( \bar{\x}_0 \big)
\right) 
\right)
\\[4pt]
&+& 
\omega (1 - \rho) \kappa  M_2 \big( \bar{\x}_0 \big)
\Bigg)
\end{array}
\right.,
\label{RecursionLPlusTwo}
\end{equation}

expressing the term $x_s(l + 1)$ in (\ref{RecursionSecondOrder}) via Eq.\ (\ref{RXslPlusOne}). This step provides $x_s(l)$ in terms of $x_s(l - 1)$, $x_s(l)$ and $x_s(l + 2)$. Consequently, the average activity at $K_s(l)$ is impacted by the state of the directly neighboring shell $K_s(l - 1)$, by the equidistant nodes at $K_s(l)$ itself, and by the indirectly interacting \textit{second} neighbors at $K_s(l + 2)$. \textit{Our main assumption is that of these three effects - the first two, which represent $K_s(l)$'s direct neighborhood, supersede that of the third $x_s(l + 2)$ term. Indeed, this term captures the state of the distant shell $K_s(l + 2)$ whose impact on $x_s(l)$ is marginal as compared to the other two terms}. We therefore approximate this term by 

\begin{equation}
x_s(l + 2) \approx \bar{\x}_0,
\label{Xslplus2}
\end{equation}

assuming this distant shell has not been significantly impacted by our reigniting, hence still at its initial undesired state $\x_0$, having on average activity $\bar{\x}_0$. This discrepancy, we expect, will have little impact on the accuracy of $x_s(l)$, as it only applies to $l$'s distant neighbors. Using (\ref{Xslplus2}) to rewrite (\ref{RecursionLPlusTwo}) we now have

\begin{equation}
\begin{array}{lcl}
x_s(l) &=& R^{-1} 
\Bigg(
\omega M_2 \big( x_s(l - 1) \big) + 
\omega \rho \kappa M_2 
\left( 
R^{-1} 
\left(
\omega M_2 
\big( x_s(l) \big) + \omega \kappa M_2 (\bar{\x}_0)
\right) 
\right)
\\[4pt]
&+& \omega (1 - \rho) \kappa  M_2 \big( \bar{\x}_0 \big)
\Bigg),
\end{array}
\label{XLPlusTwoApprox}
\end{equation}

from which we can extract $x_s(l)$ as

\begin{eqnarray}
&& \dfrac{1}{\omega} R \big( x_s(l) \big) - 
\rho \kappa M_2 
\left( 
R^{-1} 
\left(
\omega M_2 
\big( x_s(l) \big) + \omega \kappa M_2 (\bar{\x}_0)
\right) 
\right)
- (1 - \rho) \kappa  M_2 \big( \bar{\x}_0 \big)
\nonumber
\\[7pt]
&& = M_2 \big( x_s(l - 1) \big).
\label{XLPlusTwoApprox2}
\end{eqnarray}

Finally, our recurrence relation (\ref{RecursionLPlusTwo}) becomes 

\begin{Frame}
\begin{equation}
\left\{
\begin{array}{l}
x_s(0) = \Delta
\\[7pt]
F \big( x_s(l) \big) = M_2\big( x_s(l - 1) \big)
\end{array}
\right.,
\label{RecursionF}
\end{equation}
	
where

\begin{equation}
F(x) = \dfrac{1}{\omega} R(x) - 
\rho \kappa M_2 
\left( 
R^{-1} 
\left(
\omega M_2 
(x) + \omega \kappa M_2 (\bar{\x}_0)
\right) 
\right)
- (1 - \rho) \kappa  M_2 \big( \bar{\x}_0 \big).
\label{Fx}
\end{equation}
\end{Frame}

recovering Eqs.\ (5) and (6) of the main text. Under $\rho = 1$, an undirected network, $F(x)$ takes the form

\begin{equation}
F(x) = \dfrac{1}{\omega} R(x) - 
\kappa M_2 
\left( 
R^{-1} 
\left(
\omega M_2 
(x) + \omega \kappa M_2 (\bar{\x}_0)
\right) 
\right).
\label{FxUD}
\end{equation}

\subsection{Evaluating $\bar{\x}_\alpha$}
\label{EvaluatingX0}

The components of (\ref{RecursionF}) and (\ref{Fx}) can all be extracted directly from our given parameters:\ the functions $M_1(x),M_2(x),M_3(x)$ and hence also $R(x)$ are provided within Eq.\ (\ref{Dynamics}), $\kappa$ and $\omega$ are extracted from $A$ and $W$ via (\ref{Kappa}) and (\ref{DefOmega}), and $\Delta$ is determined by our forcing capacity. This leaves the shell-average $\bar{\x}_0$ in (\ref{Fx}), as the only \textit{unknown} in this recurrence. We, therefore, now discuss how to calculate $\bar{\x}_0$ and $\bar{\x}_1$ - the \textit{natural} states of the system, ans specifically how to extract $\bar{\x}_0$ in (\ref{Fx}). We offer three different options by which to evaluate these parameters:\

\subsubsection{Option 1.\ Setting $\bar{\x}_0$ to zero}

The most crude approach is to set $\bar{\x}_0 = 0$. Indeed, in the typical scenario we have the undesired state representing a system under suppressed activity in which all $x_i$ are small. Under these conditions $\bar{\x}_0 \ll \bar{\x}_1$, and can be approximated by a zero state. This approach represents a rough approximation for, \textit{e.g.}, Brain or Microbial dynamics, but becomes exact for Regulatory, where the undesired state is, indeed, $\x_0 = (0,\dots,0)^\top$. Such approximation allows us to solve the recurrence relation directly, predicting whether single-node reigniting will, indeed, revive the system. Yet, it offers no information on the recovered state $\bar{\x}_1$.

\subsubsection{Option 2.\ Mean-field approximation}
\label{SecMFApproximation}

We can approximate analytically the dynamic states of the system using a mean-field approach, which was originally presented in Ref.\ \cite{Gao2016}. This approach is exact under the modeling framework presented in Sec.\ \ref{SecModelingFramework}, and specifically builds on characteristic 2 in Sec.\ \ref{ResultingApproximations} - the statistical similarity of all shells $K_s(l)$. Our presented results indicate that it also holds quite well even when considering realistic networks, which show some discrepancies with regards Sec.\ \ref{SecModelingFramework}'s assumptions.

Starting from the definition of $\bar{\x}_\alpha$ in (\ref{xbar}) we can use (\ref{Dynamics2}) to write a direct equation for the shell-average as 

\begin{equation}
\dod{\bar{\x}_\alpha}{t} = \dfrac{1}{N} \sum_{s = 1}^N \dfrac{1}{|K_s(1)|} 
\sum_{i \in K_s(1)} \Big( M_0 \big( x_i(t) \big) + 
M_1 \big( x_i(t) \big) \sum_{j \in Q_i(1)} \m Wij M_2 \big( x_j(t) \big) \Big).
\label{DynamicsMF}
\end{equation}

Let us first treat the summation over $Q_i(1)$ on the r.h.s., by writing it as

\begin{equation}
\sum_{j \in Q_i(1)} \m Wij M_2 \big( x_j(t) \big) = 
\dfrac{1}{|Q_i(1)|} \sum_{j \in Q_i(1)} |Q_i(1)| \m Wij M_2 \big( x_j(t) \big).
\end{equation} 

Using the mean-field approximation of (\ref{MF1}) we obtain 

\begin{equation}
\sum_{j \in Q_i(1)} \m Wij M_2 \big( x_j(t) \big) \approx 
|Q_i(1)| \omega_i  M_2 \big( x_{i\leftarrow}(t) \big),
\label{MeanFieldi}
\end{equation}

where $\omega_i = |Q_i(1)|^{-1} \sum_{j \in Q_i(1)} \m Wij$ is the average weight of $i$'s incoming links, and $x_{i\leftarrow}(t)$ is the average activity of $i$'s incoming neighborhood $Q_i(1)$ (analogous to $x_s(l,t)$ in (\ref{Xslt}), defined for \textit{outgoing} neighborhoods). Substituting (\ref{MeanFieldi}) into (\ref{DynamicsMF}) we arrive at

\begin{equation}
\dod{\x_\alpha}{t} \approx 
\dfrac{1}{N} \sum_{s = 1}^N \dfrac{1}{|K_s(1)|} 
\sum_{i \in K_s(1)} \Big( M_0 \big( x_i(t) \big) + 
M_1 \big( x_i(t) \big) |Q_i(1)| \omega_i  M_2 \big( x_{i\leftarrow}(t) \big) \Big).
\label{DynamicsMF2}
\end{equation}

Averaging over all $i \in K_s(1)$, as we do in (\ref{DynamicsMF2}), allows to substitute $\omega_i$, the average weight surrounding $i$, by $\omega(s,1)$, the average weight surrounding the $K_s(1)$ shell. Further, as we average over all $s$, we can extract the global weight $\omega$, as defined in (\ref{DefOmega}), outside of the sum. Indeed, characteristic 2 in Sec.\ \ref{ResultingApproximations} indicates that the shell statistics are largely independent of $s$ and $l$, and therefore the average weights of links in the different shells are approximately equal, and are all narrowly distributed around the global average $\omega$. As a result, aggregating $\omega(s,1)$ over all $s$ guaranties to fall close to the global average in (\ref{DefOmega}). In a similar fashion we use (\ref{Kappa}) to write

\begin{equation}
\dfrac{1}{N} \sum_{s = 1}^N \dfrac{1}{|K_s(1)|} \sum_{i \in K_s(1)} |Q_i(1)| = \kappa + 1.
\end{equation}  

Finally, absent degree-correlations $i$'s incoming neighbors $Q_i(1)$ are, on average, no different than  the incoming neighbors of any other node, allowing us to express the average over all $x_{i\leftarrow}(t)$ as

\begin{equation}
\dfrac{1}{N} \sum_{s = 1}^N \dfrac{1}{|K_s(1)|} \sum_{i \in K_s(1)} x_{i\leftarrow}(t) 
= \bar{\x}_\alpha(t), 
\label{xi1txalpha}
\end{equation}

the global shell-average of (\ref{xbar}). This, once again, is a direct consequence of our configuration model network ensemble, and specifically of characteristic 2 in Sec.\ \ref{ResultingApproximations}. We arrive at

\begin{equation}
\dod{\x_\alpha}{t} \approx 
\dfrac{1}{N} \sum_{s = 1}^N \dfrac{1}{|K_s(1)|} 
\sum_{i \in K_s(1)} \Big( M_0 \big( x_i(t) \big) + (\kappa + 1) \omega
M_1 \big( x_i(t) \big) M_2 \big( \bar{\x}_\alpha(t) \big) \Big),
\label{DynamicsMF3}
\end{equation}

on which we apply, once again, the mean-field approximation of (\ref{MF1}) to obtain

\begin{equation}
\dod{\x_\alpha}{t} = \dfrac{1}{N} \sum_{s = 1}^N 
\Big( M_0 \big( x_s(1,t) \big) + (\kappa + 1) \omega M_1 \big( x_s(1,t) \big) 
M_2 \big( \bar{\x}_\alpha(t) \big) \Big).
\label{DynamicsMF4}
\end{equation}

To construct a specific equation for $\bar{x}_\alpha(t)$ we use the averaging over all $s$ on the r.h.s.\ of (\ref{DynamicsMF4}) to substitute the $s$-shell average, $x_s(1,t)$, by the \textit{network} shell-average $\bar{x}_\alpha(t)$. This provides us with

\begin{equation}
\dod{\x_\alpha}{t} = M_0 \big( \bar{\x}_\alpha(t) \big) + 
\beta M_1 \big(\bar{\x}_\alpha(t) \big) 
M_2 \big( \bar{\x}_\alpha(t) \big),
\label{DynamicsMF5}
\end{equation}

where 

\begin{equation}
\beta = (\kappa + 1) \omega
\label{Beta}
\end{equation}

is a parameter fully determined by the weighted network topology $A \otimes W$.

Equation (\ref{DynamicsMF5}) provides us with a single equation for the shell-average, which we can solve to directly obtain $\bar{\x}_\alpha(t)$. The equation is based on characteristic 2 of Sec.\ \ref{ResultingApproximations}, which predicts that the shell-averages around all nodes, \textit{e.g}., in $K_s(l)$ or in $Q_i(1)$, are all narrowly distributed and can, therefore be approximated by their ensemble average. This reduction is a direct consequence of the random nature of $A$ and $W$, in which $K_s(l)$ and $K_{s^\prime}(l^\prime)$ are statistically indistinguishable. While, generally, (\ref{DynamicsMF5}) represents an approximation, it is \textit{exact} in the limit of a large ($N \rightarrow \infty$) and perfectly random configuration model network, \textit{i.e}.\ the framework discussed in Sec.\ \ref{SecModelingFramework}. A detailed derivation of Eq.\ (\ref{DynamicsMF5}), including a discussion on its validity limits appears in Ref.\ \cite{Gao2016}, and further generalized in Refs.\ \cite{Tu2017pre,Jiang2018pnas,laurence2019spectral,Thibeault2020prr,Tu2021iScience}.

\textbf{Extracting $\bar{\x}_\alpha$}.\ 
Setting the derivative on the l.h.s.\ of (\ref{DynamicsMF5}) to zero, we obtain an algebraic equation for $\bar{\x}_\alpha$. This allows us to obtain the different fixed-points in function of the system's network parameters $\kappa$ and $\omega$. The stability of each potential fixed-point is then evaluated using linear stability analysis. This requires that 

\begin{equation}
\left.
\dod{}{x} \Big[ M_0(x) + \beta M_1(x)M_2(x) \Big]
\right|_{x = \bar{\x}_{\alpha}} < 0,
\label{Stability}
\end{equation}

\noindent
a negative derivative, ensuring that deviations from $\bar{\x}_{\alpha}$ decay exponentially. Satisfying (\ref{Stability}), for a given state $\bar{\x}_{\alpha}$, depends on the value of $\kappa$ and $\omega$ through $\beta$, and hence on $A$ and $W$. Equation (\ref{Stability}), therefore, exposes the range of $\kappa,\omega$ values (if any) under which each fixed-point $\bar{\x}_{\alpha}$ is stable. Hence it exposes the naturally occurring states of the system, providing us with the \textit{state-space} on which we apply our recoverability formalism. Specifically, it also helps us extract the undesired $\bar{\x}_0$, to be used in Eq.\ (\ref{Fx}).

\subsubsection{Option 3.\ Numerical analysis}

The most accurate evaluation of $\bar{\x}_0$ can be obtained from the actual observed state of the failed system. Indeed, before we apply our reigniting, we can use the detailed unperturbed state of the system $\x_0 = (x_{0,1},\dots,x_{0,N})^\top$, to accurately construct $\bar{\x}_0$ via (\ref{xbar}). Such direct calculation provides an \textit{exact} alternative to the mean-filed approximation offered in Option 2 above. It is especially useful if the network shows significant deviations from the configuration model. Yet, we emphasize, that our result clearly indicate that even \textit{imperfect} networks continue to be well-approximated by Option 2. 

\begin{Frame}
Throughout our analysis we used Option 2, the mean-field approximation, to evaluate $\bar{\x}_0$ analytically via Eqs.\ (\ref{DynamicsMF5}) and (\ref{Stability}), and solve the recurrence relation to predict recoverability. The actual results for $\bar{\x}_0$ and $\bar{\x}_1$, presented, \textit{e.g}., in Fig.\ 5c of the main text were obtained from numerical simulations of a large scale-free network ensemble.
\end{Frame}


\begin{figure}[t]
\centering
\includegraphics[width=0.7\textwidth]{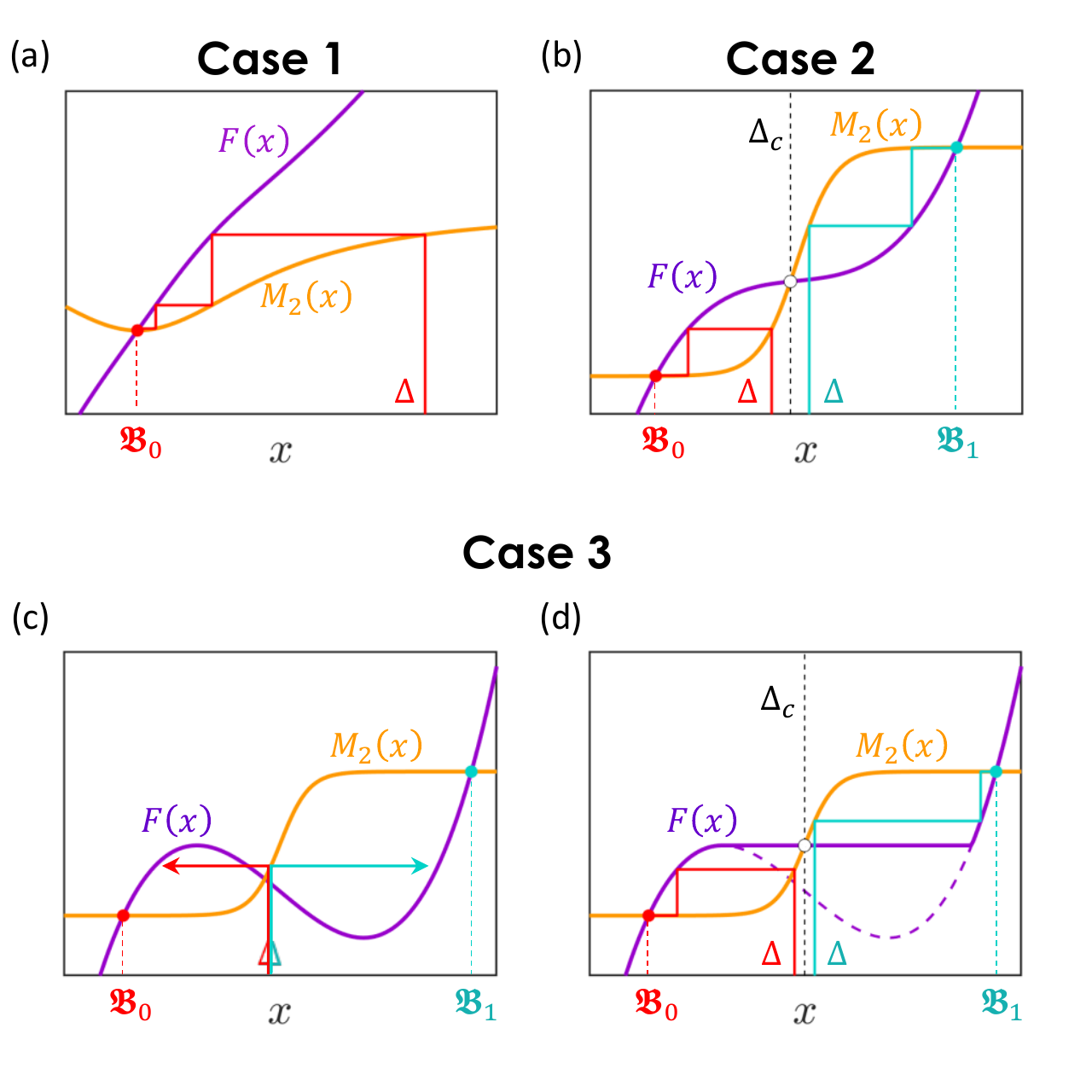}
\vspace{-7mm}
\caption{\color{blue} \footnotesize \textbf{Solving the recurrence relation via cobweb plots}.\ To solve (\ref{RecursionF}) we begin at $x_s(0) = \Delta$, then track the recurrence by following the value of $M_2(x)$ (vertical steps) and equating this with $F(x)$ (horizontal steps), as per the second equation in (\ref{RecursionF}). 
(a) \textbf{Case 1}.\ Regardless of the starting point $\Delta$, the system unconditionally converges to $\Basin_0$, and hence it is structurally unrecoverable (red path).
(b) \textbf{Case 2}.\ For $\Delta > \Delta_c$ the system reaches $\Basin_1$, \textit{i.e}.\ recoverable (green path).
(c) \textbf{Case 3}.\ In case $F(x)$ is non-monotonic we encounter dual solutions, in which both the red and green pathways coexist. The real system, \textit{i.e}.\ Eq.\ (\ref{Dynamics}), however, will only follow the red path of the failed reigniting, as its initial condition is of all nodes residing in $\Basin_0$. 
(d) To treat this ambiguity we reconstruct $F(x)$ by introducing a plateau over the non-monotonic range, leading to the monotonic $F(x)$ (purple solid line) instead of the original function (purple-dashed line).
} 
\label{FigCobweb}
\end{figure}


\subsection{Solving the recurrence relation}

Equation (\ref{RecursionF}) replaces the original second order recurrence of Eq.\ (\ref{RecursionSecondOrder}), which was under-determined, by an approximate first order recurrence, now solvable using standard tools for analyzing recurrence relations, as outlined in the main text. First, we extract the fixed points of the recursion via \cite{May1976simple}

\begin{equation}
F(x) = M_2(x).
\label{FxM2x}
\end{equation}

\noindent
Maintaining consistency with the original dynamics (\ref{Dynamics}), from which (\ref{RecursionF}) is derived, the fixed point obtained via (\ref{FxM2x}) must, by definition coincide with the intrinsic fixed-points $\bar{\x}_{\alpha}$ of the system. Indeed, our reigniting may either fail, leading to $\bar{\x}_0$, or succeed, arriving at the vicinity of $\bar{\x}_1$ (within $\Basin_1$). It cannot, however, lead to any other points, which are not within the potential states intrinsic to (\ref{Dynamics}). Hence the condition (\ref{FxM2x}) may have only two outcomes:\ the system is non-recoverable if (\ref{FxM2x})'s only stable fixed-point is $\bar{\x}_0$. However, if it also features a stable $\bar{\x}_1$, then we can use the structure of $F(x)$ and $M_2(x)$ to obtain the critical $\Delta$ above which (\ref{RecursionF}) converges to $\bar{\x}_1$, and hence successfully reignites. Hence, analyzing the intersection/s of $F(x)$ and $M_2(x)$ we can predict the system's recoverability, and if recoverable, the required critical forcing.

\textbf{Cobweb plots} (Fig.\ \ref{FigCobweb}).\ 
We solve the recurrence relation of (\ref{RecursionF}) using cobweb plots \cite{strogatz2018nonlinear}. Starting from an initial setting of $x_s(0) = \Delta$, our maximal forcing, we track the evolution of the recurrence. First obtaining $M_s(\Delta)$ (vertical path), then shifting horizontally to $F(x)$, extracting $x_s(1)$. Continuing the process we observe weather the recurrence converges to $\Basin_0$ or to $\Basin_1$. In case the function $F(x)$ is non-monotonic, this process may lead to convergence ambiguity, as illustrated in Case 3 of Fig.\ \ref{FigCobweb}c, with both red and green pathways enabled. Such ambiguity is, of course, a mathematical artifact, as the real system in (\ref{Dynamics}) will indeed follow only one of the potential tracks, not both. In reality, since we employ our reigniting from an initial condition which is in $\Basin_0$, the system, under any such instance of ambiguity, will converge back to $\Basin_0$, namely, it will select the red path and not recover. To remove this duality we use the construction of Fig.\ \ref{FigCobweb}d, in which we introduce a plateau along the non-monotonic range in $F(x)$. The \textit{corrected} $F(x)$ (solid purple line) is now monotonic and we can unambiguously analyze it via the proposed cobweb plots.  

\clearpage


\section{Dynamic models}
\label{SecDynamicModels}

To demonstrate our framework we examined the recoverability of four distinct dynamic systems, three within the form of Eq.\ (\ref{Dynamics}), and on that extends beyond this form. Below we detail the analytical/numerical treatment of each of these systems, starting from the \textit{free system}, in which we examine the states of the system absent our forcing $\Delta$, then treating the \textit{reignited system}, in which we introduce our single-node activation.

\subsection{Cellular dynamics}
\label{SecCellular}

We consider gene-regulatory dynamics, as captured by the Michaelis-Menten model \cite{Alon2006,Karlebach2008}, for which (\ref{Dynamics}) takes the form

\begin{equation}
\dod{x_i}{t} = -Bx_i^a + \sum_{j = 1}^N \m Aij \m Wij \dfrac{x_j^h}{1 + x_j^h}.
\label{MMexample}
\end{equation} 

\noindent
Under this framework $M_0(x_i) = -B x_i^a$, describing degradation ($a = 1$), dimerization ($a = 2$) or a more complex bio-chemical depletion process (fractional $a$), occurring at a rate $B$ \cite{Barzel2011}. For simplicity, in our simulations we set $B = 1$. The activation interaction is captured by a Hill function of the form $M_1(x_i) = 1$, $M_2(x_j) = x_j^h/(1 + x_j^h)$, a \textit{switch-like} function that saturates to $M_2(x_j) \rightarrow 1$ for large $x_j$, representing $j$'s positive, albeit bounded, contribution to $i$'s activity $x_i(t)$. 

\subsubsection{Free system}

\begin{figure}[h!]
\centering
\includegraphics[width=0.95\textwidth]{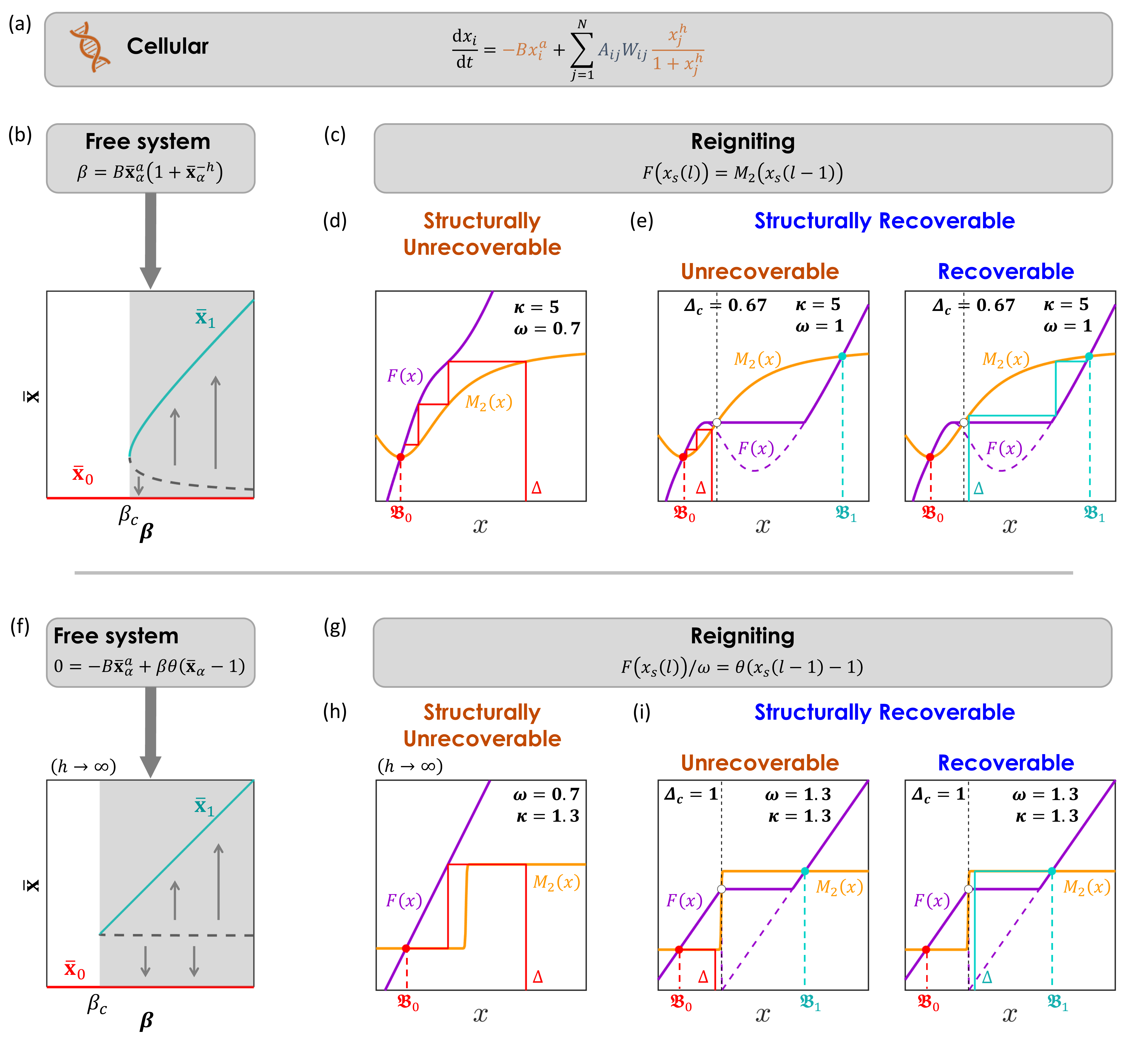}
\caption{\color{blue} \footnotesize \textbf{Cellular dynamics}.\ (a) The Michaelis-Menten model for gene-regulation. 
(b) The fixed-points, inactive (red), vs.\ active (green), as obtained from Eq.\ (\ref{MMFixedPoint}), top. When $\beta < 2$ the system undergoes an irreversible collapse.
(c) The system recoverability depends on the intersection points of $F(x)$ and $M_2(x)$. Hence we plot $F(x)$ as appears in (\ref{MMFx}) for different values of $\omega$ and $\kappa$ (purple), and $M_2(x) = x^h/(1 + x^h)$, for $h = 2$ (yellow). 
(d) For $\omega = 0.7$ the plot follows the pattern of Case 1 - structurally unrecoverable. 
(e) Increasing the average weight to $\omega = 1$ the system transitions to Case 2 - recoverable if $\Delta \ge 0.67$ (right), and unrecoverable otherwise (left). 
(f) - (i) 
The same analysis for the case $h \rightarrow \infty$. In this limit the activation function $M_2(x)$ behaves as a step function. We find that here the structure of $F(x)$, and hence the recoverability of the system, is independent of $\kappa$, affected solely by $\omega,\Delta$.
}
\label{FigMM}	
\end{figure}

First we seek the natural fixed-points of (\ref{MMexample}), by mapping it to the one dimensional Eq. \eqref{DynamicsMF5}, providing

\begin{equation}
\dod{\bar{\x}_\alpha}{t} = -B\bar{\x}_\alpha^a + \beta \frac{\bar{\x}_\alpha^h}{1+\bar{\x}_\alpha^h},
\label{CellularMF}
\end{equation}

whose fixed-points, obtained by setting ${\rm d}{\bar{\x}_\alpha}/{{\rm d}t} = 0$, follow

\begin{equation}
\beta = B \bar{\x}_\alpha^a \left( 1 + \dfrac{1}{\bar{\x}_\alpha^h} \right),
\label{MMFixedPoint}
\end{equation}

as shown in Fig.\ \ref{FigMM}b. Setting $B = 1, a = 1$ and $h = 2$ (see Ref.\ \cite{Gao2016} for a more general treatment), we extract $\bar{\x}$ from (\ref{MMFixedPoint}), predicting three fixed-points (Fig.\ \ref{FigMM}b):\ The inactive $\bar{\x}_0 = 0$ (undesirable, red) , the active $\bar{\x}_1 = \beta/2 + \sqrt{(\beta/2)^2 - 1}$ (desirable, green), and the intermediate $\bar{\x}_2 = \beta/2 - \sqrt{(\beta/2)^2 - 1}$ (grey dashed-line). 

Next we test the dynamic stability of each of these three solutions via Eq.\ (\ref{Stability}), providing the stability condition

\begin{equation}
- aB \bar{\x}_\alpha^{a - 1} + 
\beta \dfrac{h \bar{\x}_\alpha^{h - 1}}{\big (1 + \bar{\x}_\alpha^h \big)^2} < 0.
\label{MMStability}
\end{equation}

\noindent
Using our selected parameters ($B = a = 1, h = 2$) condition (\ref{MMStability}) reduces to 

\begin{equation}
-\left( 1 + \bar{\x}_\alpha^2 \right)^2 + 2\beta \bar{\x}_\alpha < 0,
\label{MMStability2}
\end{equation}

from which we can directly obtain the system's natural state-space. For the inactive $\bar{\x}_0 = 0$ we have the left hand side of (\ref{MMStability2}) being $-1$, independently of $\beta$, and hence this state is unconditionally stable. The stability of the other two states is obtained by substituting $\bar{\x}_1$ and $\bar{\x}_2$ into (\ref{MMStability2}), finding that $\bar{\x}_2$ is never stable, and hence it does not represent a potential state of the system, while the desirable $\bar{\x}_1$ is stable under the condition that $\beta > 2$ (Fig.\ \ref{FigMM}b). Therefore $\beta < 2$ forces a collapse on to $\x_0$, while $\beta > 2$ represents a bi-stable state, where the system can reside in both $\x_0$ and $\x_1$, depending on its initial conditions. 


\subsubsection{Reigniting}

To examine the behavior of our cellular dynamics (\ref{MMexample}) under reigniting we seek to construct the recurrence relation (\ref{RecursionF}), and specifically the function $F(x)$ in (\ref{Fx}). First we write

\begin{eqnarray}
R(x) &=& x^a
\\[5pt]
R^{-1}(x) &=& x^{1/a}	
\\[5pt]
M_2(x) &=& \dfrac{x^h}{(1 + x^h)},
\label{MMfunctions} 
\end{eqnarray}

where we used (\ref{Rx}) to obtain $R(x)$. Setting $a = 1$, we can now collect all the terms to construct $F(x)$, providing us with

\begin{gather}
F(x) = \dfrac{x}{\omega} - \kappa M_2 \left( \omega M_2(x) \right) = 
\dfrac{x}{\omega} - \kappa \dfrac{1}{1+\omega^{-h}(1+x^{-h})^{h}}
\label{MMFx}
\end{gather}

a function whose shape depends on the topological parameters $\kappa$ and $\omega$. Here we used an undirected interaction pattern, setting $\rho = 1$ in (\ref{Fx}). Also note that here $\bar{\x}_0 = 0$, and hence $M_2(\bar{\x}_0) = 0$ on the r.h.s.\ of (\ref{Fx}). Equation (\ref{MMFx}) maps to $F(x)$ in Eq.\ (9) of the main text. 

Varying $\kappa$ and $\omega$ we can observe the recoverability of our system as $F(x)$ transitions from Case 1 to Case 3 of Fig.\ \ref{FigCobweb}. Two specific examples are presented in Fig.\ \ref{FigMM}d and e. The first with $\kappa = 5$ and $\omega = 0.7$ falls under Case 1, and therefore it is structurally unrecoverable, and the second, in which the average weight is increased to $\omega = 1$ follows Case 3, \textit{i.e}.\ recoverable if $\Delta \ge \Delta_c$. 

To obtain the complete phase diagram and hence the boundaries of the recoverable phase, as we do in Fig.\ 3j-m of the main text, we systematically plot $F(x)$ in (\ref{MMFx}) for a range of $\kappa,\omega$ values, seeking for each $\kappa$ the critical $\omega$ in which $F(x)$ transitions to the form of Case 3. These critical transition points provide the theoretical phase boundaries (Fig.\ 3k-m, white solid lines). At the same time we tested numerically, for each $\kappa,\omega$ combination whether single-node reigniting indeed reactivates the system. To achieve this, for each $\kappa,\omega$ combination we conducted $20$ reigniting attempts with randomly selected source nodes, and plotted $\eta$, the fraction of successful recoveries ($\eta = 0$ yellow; $\eta = 1$ blue, see details in Sec.\ \ref{SecPhaseDiagrams}).

\subsubsection{The role of the Hill coefficient}

The Hill coefficient $h$ in (\ref{MMexample}) determines the saturation rate of the activation function $M_2(x)$. A small $h$ captures a mild activation, in which $M_2(x)$ increases gradually with $x$, while $h \rightarrow \infty$ describes an effective step-function of the form

\begin{equation}
M_2(x) = \lim_{h \rightarrow \infty} \left( \dfrac{x^h}{1 + x^h} \right) = \theta(x - 1),
\label{M2StepFunction}
\end{equation}

being $M_2(x) = 1$ (activation) if $x > 1$ and $M_2(x) = 0$ otherwise; $\theta(x)$ is the Heaviside step-function. Taking this limit in (\ref{MMFx}) we obtain

\begin{equation}
\lim_{h \rightarrow \infty} F(x) = x/w - \kappa\theta(\omega - 1)\theta(x - 1).
\label{MMFxStep}
\end{equation}

For $\omega < 1$ we have

\begin{equation}
F(x) = \dfrac{1}{\omega} x, 
\end{equation} 

a linear function whose slope is $1/\omega > 1$. This function has a single intersection with $M_2(x) =\theta(x - 1)$ at $x = 0$, therefore rendering the system unrecoverable (Fig.\ \ref{FigMM}h, left). In case $\omega \ge 1$ Eq.\ (\ref{MMFxStep}) becomes

\begin{equation}
F(x) = \dfrac{1}{\omega} x - \kappa \theta(x - 1),
\end{equation}

having two intersections with $M_2(x)$, describing a recoverable system for $\Delta \ge 1$ (Fig.\ \ref{FigMM}i, right). 

\begin{Frame}
This describes a limit in which $\kappa$ plays no role in recoverability, and reigniting is driven solely by $\omega$ and $\Delta$, as discussed in the main text under Restructuring guidelines. 
\end{Frame}


\subsection{Neuronal dynamics}

We consider a modified Wilson-Cowan model \cite{wilson1972excitatory,wilson1973mathematical} for excitation in neuronal networks, writing

\begin{equation}
\dod{x_i}{t} = -x_i + \sum_{j = 1}^N \m Aij \m Wij \dfrac{1}{1 + e^{\mu-\delta x_j}},
\label{Neuronal}
\end{equation}

which we examine under $\mu = 10$ and $\delta = 1$. In this version of the model, adapted to the form of Eq.\ (\ref{Dynamics}), the summation is extracted outside of the exponential function, namely we write $\sum_{j = 1}^N \m Aij \m Wij (1 + e^{\mu-\delta x_j})^{-1}$ instead of $(1 + e^{\mu-\delta \sum_{j = 1}^N \m Aij \m Wij x_j})^{-1}$. As a result, each node receives accumulating inputs from all its neighbors, and hence higher in-degree nodes gain a stronger overall activation signal from their surrounding $Q_i(1)$. This is in contrast to the standard version of the model, with the summation appears inside the exponent, and hence the effect of the multiple incoming signals quickly reaches saturation. In the present context, where we wish to observe the role of degree heterogeneity on reigniting, \textit{e.g}., through $\kappa$, the form of Eq.\ (\ref{Neuronal}) provides a more relevant testing ground. 


\begin{figure}[h!]
\centering
\includegraphics[width=0.95\textwidth]{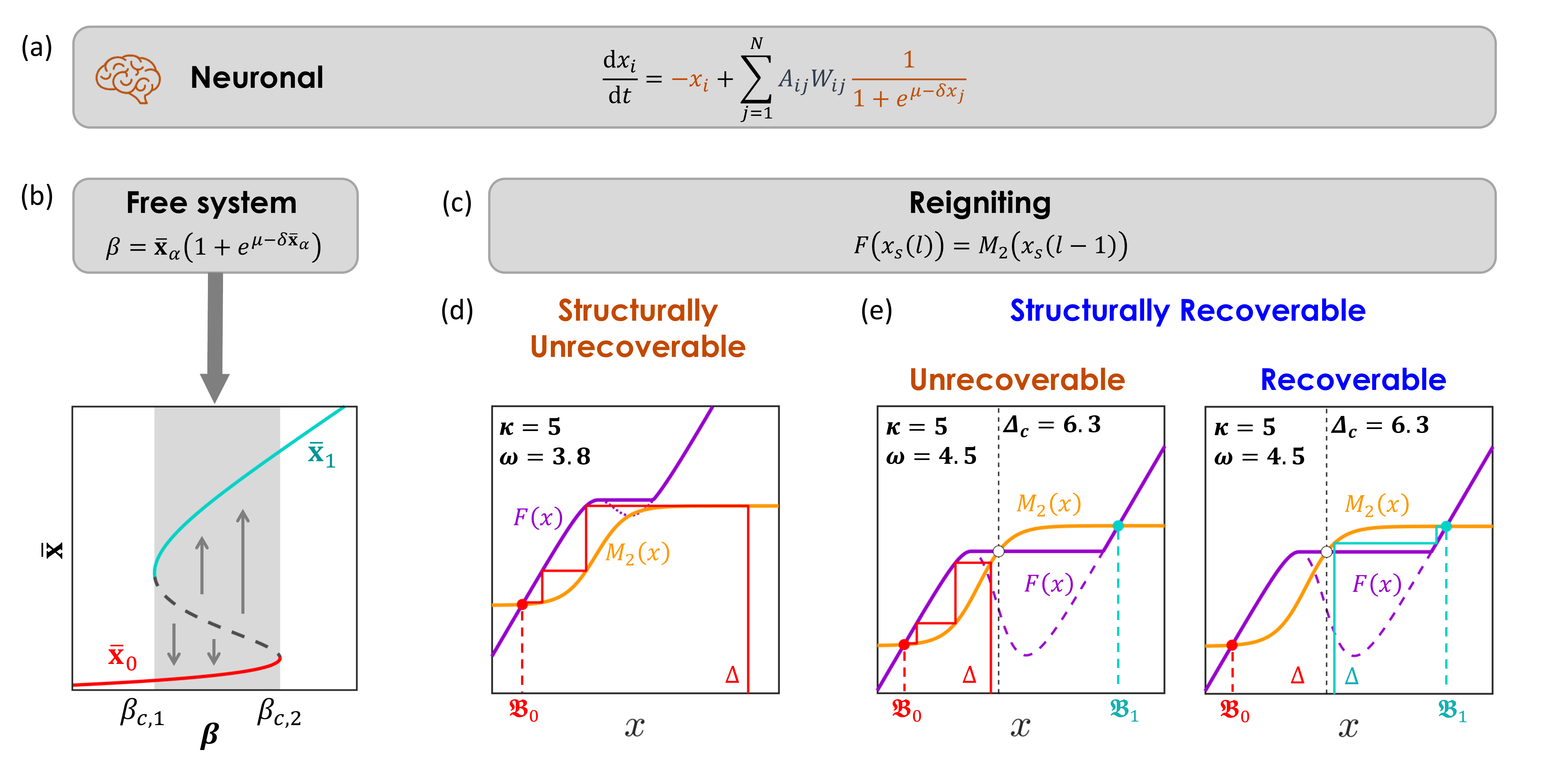}
\vspace{-5mm}
\caption{\color{blue} \textbf{Neuronal dynamics}.\ 
(a) The modified Wilson-Cowan model. 
(b) The shell average $\bar{\x}_\alpha$ vs.\ $\beta$ as obtained from Eq.\ (\ref{betaXBrain}) with $\mu = 10, \delta = 1$. The system exhibits two potential states $\bar{\x}_0$ (red) and $\bar{\x}_1$ (green), separated by a bi-stable phase (grey shaded).
(c)-(d) $F(x)$ (purple) from (\ref{BrainFx}) and $M_2(x)$ (yellow) from (\ref{BrainM2x}) for a structurally unrecoverable system ($\kappa = 5,\omega = 3.8$).
(e) Setting $\omega= 4.5$ the system becomes structurally recoverable. It is unrecoverable for $\Delta < 6.3$ (left) and recoverable otherwise (right).
}
\label{FigBrain}	
\end{figure}


\subsubsection{Free system}	

To obtain the fixed-points of the system we use the mapping of (\ref{DynamicsMF5}) to reduce (\ref{Neuronal}) into

\begin{equation}
\dod{\bar{\x}_\alpha}{t} = -\bar{\x}_\alpha + \beta \frac{1}{1+e^{\mu-\delta\bar{\x}_\alpha}},
\label{BrainMF}
\end{equation}

whose fixed-points follow

\begin{equation} \label{betaXBrain}
\beta = \bar{\x}_{\alpha} (1+e^{\mu -\delta \bar{\x}_{\alpha}}).
\end{equation}

Plotting $\bar{\x}_\alpha$ vs.\ $\beta$ (Fig. \ref{FigBrain}b) we obtain the dynamic phases of the system - the \textit{suppressed} state $\x_0$ (red), in which all activities are constricted, is obtained when the network is extremely sparse, \textit{i.e}.\ small $\beta$; the \textit{active} $\x_1$ (green), in which $x_i$ are relatively high, is observed when $\beta$ is large. In between these two extremes the system features a \textit{bi-stable} state, in which both $\x_0$ and $\x_1$ are potentially stable. These phases are separated by two critical points $\beta_{c,1} < \beta_{c,2}$, predicting a hysteresis phenomenon:\ if $\beta$ was driven below $\beta_{c,1}$ and the system has failed, it will not spontaneously recover unless we retrieve $\beta$ to be above $\beta_{c,2}$. Hence, we seek the sub-space \textit{within} the bi-stable regime, $ \beta_{c,1} < \beta < \beta_{c,2}$, in which the system can be reignited.

\subsubsection{Reigniting}

The relevant functions to construct $F(x)$ (\ref{Fx}) are 

\begin{eqnarray} 
R(x) &=& x,
\\[5pt]
R^{-1}(x) &=& x,
\\[5pt]
M_2(x) &=& \dfrac{1}{(1+e^{\mu-\delta x})},
\label{BrainM2x}
\end{eqnarray}	

from which we obtain, for an undirected network ($\rho = 1$)

\begin{equation}
F(x) = \dfrac{x}{\omega} - 
\kappa M_2 \left(  \omega M_2(x) + \omega\kappa M_2(\bar{\x}_0) \right).
\label{BrainFx}
\end{equation}

Once again we arrive at a function which depends on $\kappa$ and $\omega$, sometimes following Case 1, in which the system is structurally unrecoverable (Fig.\ \ref{FigBrain}d), and sometimes following the Case 3, in which it is potentially recoverable if $\Delta \ge \Delta_c$ (Fig.\ \ref{FigBrain}e).


\subsection{Microbial dynamics}

We capture cooperative population dynamics via

\begin{equation}
\dod{x_i}{t} = 
F + B x_i \left( 1 - \dfrac{x_i}{C} \right)(x_i - K) + \sum_{j = 1}^N \m Aij \m Wij x_i x_j.
\label{Ecology}
\end{equation}

\begin{figure}[h!]
\centering
\includegraphics[width=0.76\textwidth]{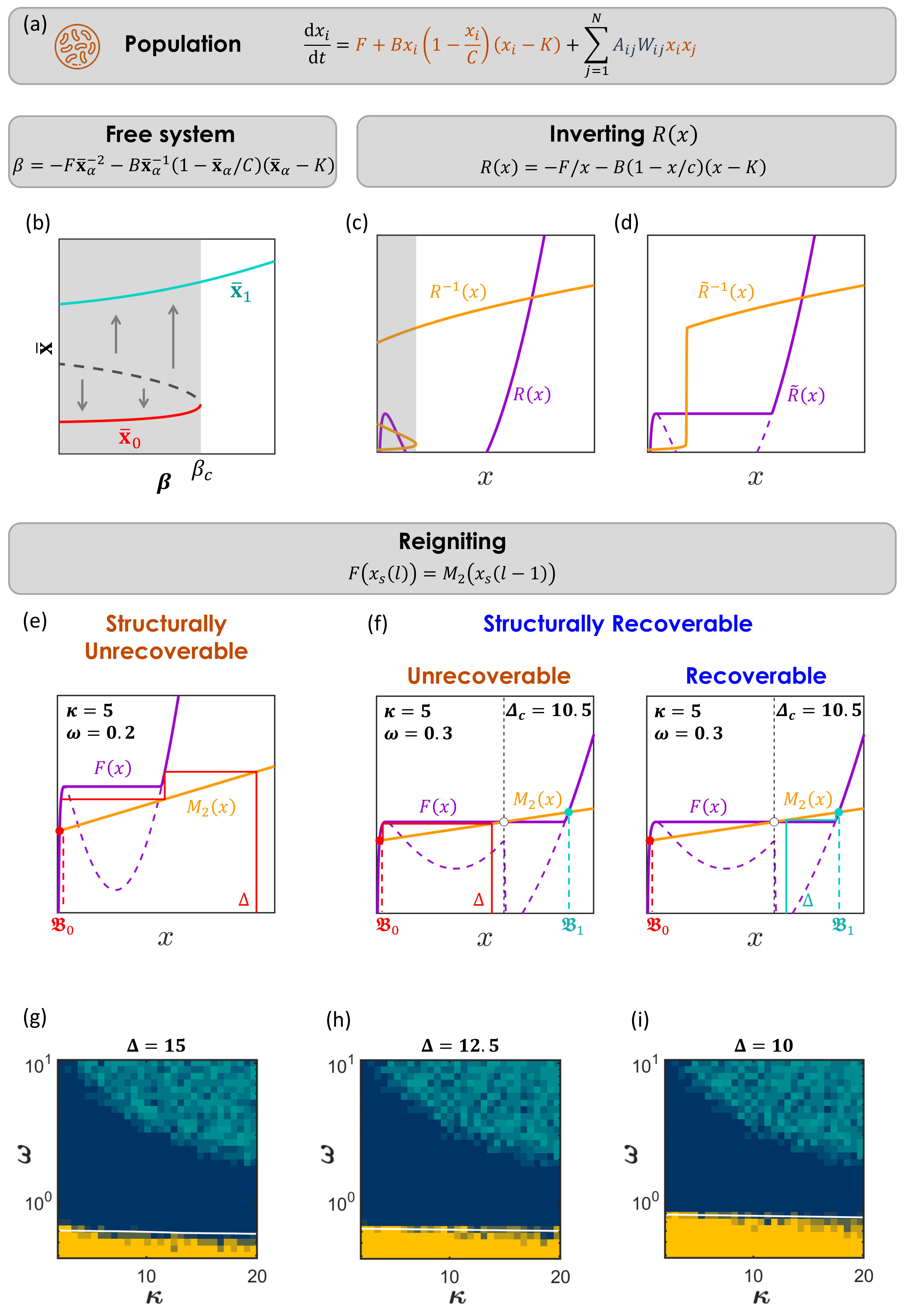}
\vspace{-5mm}
\caption{\color{blue} \textbf{Microbial dynamics}. 
(a) Our population dynamics model. 
(b) $\bar{x}$ vs.\ $\beta$ as obtained from Eq.\ (\ref{betaXEco}). The system features a bi-stable phase (grey shaded), then transitions to a single active fixed-point at $\beta \ge \beta_c$ .
(c) $R(x)$ vs.\ $x$ (purple) as obtained from (\ref{EcoRx}). The function is non-monotonic and hence $R^{-1}(x)$ (yellow) is ill-defined. The dual range in which $R^{-1}(x)$ admits ambiguous solutions (grey shaded) is directly related to the bi-stable phase of the system via Eq.\ (\ref{BetaRinv}). The relevant branch is only the \textit{lower} one, associated with the failed state of the system, on which we employ our reigniting.  
(d) We preserve only this lower branch by replacing the original $R(x)$ (purple dashed line) with a monotonous $\tilde{R}(x)$, in which we introduce a plateau instead of the non-monotonic range (purple solid line). The resulting function can now be unambiguously inverted (yellow).
(e) $F(x)$ (purple) from (\ref{EcoFx}) and $M_2(x)$ (yellow) from (\ref{EcoM2x}) for a structurally unrecoverable system ($\kappa = 5, \omega = 0.2$). 
(f) Increasing $\omega$ to $0.3$ the system becomes structurally recoverable, and can be successfully reignited for $\Delta \ge 10.5$ (right).
(g)-(i) Recoverability phase-diagrams in $\kappa,\omega$ plane, under different fixed values of $\Delta$. Here we used the parameters $F = 1/4,\ B = 1,\ C = 3,\ K = 10$ in (\ref{Ecology}). In this dynamics, due to the relatively long run-times, we set the network size at $N = 500$ and number of realizations in for each network to $5$ (this is instead of $N = 10,000$ and the $20$ realizations per data-point, used in the construction of the phase diagrams in Figs.\ 3 and 5 of the main text).
}
\label{FigEco}	
\end{figure}

The self dynamics describes migration at a rate $F$ coupled with logistic growth at rate $B$, with the system carrying capacity set to $C$, and the Allee effect \cite{allee1949principles} with strength $K$. The mutualistic interaction follows the Lotka-Volterra form $x_i x_j$. \cite{Holling1959} In our simulations on scale-free networks (Fig.\ \ref{FigEco}g-i) we set the parameters of (\ref{Ecology}) to $F = 1/4$, $B = 1$, $C = 3$ and $K = 10$; when simulating the microbiome we used $F = 5$, $B = 3$, $C = 3$ and $K = 10$. The reason for this difference is because for reigniting to be observed, we require a sufficiently broad bi-stable regime (Fig.\ \ref{FigEco}b, grey shaded), prompting us to adjust the parameters, accordingly, as we implement (\ref{Ecology}) on vastly distinct networks. Specifically, setting $F = 5,B = 3$ in the case of our scale-free networks led to the system recovering spontaneously, as the active state was broadly stable, while the bi-stable window was narrow. This prohibited us from observing, in practice, the partitioning of the bi-stable state into two separate phases - non-recoverable and recoverable. We therefore set these parameters lower to $F = 1/4$, $B = 1$, conditions under which the system has a sufficiently broad bi-stable state, for us to observe, with satisfactory resolution, its two sub-phases.  


\subsubsection{Free system}

Using Eq.\ (\ref{DynamicsMF5}) we write

\begin{equation}
\dod{\bar{\x}_\alpha}{t} = F + B \bar{\x}_\alpha 
\left( 1 - \dfrac{\bar{\x}_\alpha}{C} \right)(\bar{\x}_\alpha - K) 
+ \beta \bar{\x}_\alpha^2,
\label{EcologyMF}
\end{equation}

obtaining the fixed-points from

\begin{equation} 
\beta = 
\frac{-FC + B \bar{\x}_\alpha (\bar{\x}_\alpha - C)(\bar{\x}_\alpha - K)}
{C \bar{\x}_\alpha^2}.
\label{betaXEco}
\end{equation}

In Fig.\ \ref{FigEco}b we plot $\bar{\x}_\alpha$ vs.\ $\beta$ as obtained from Eq.\ (\ref{betaXEco}) finding two phases. For $\beta < \beta_c$ we observe two fixed-points, $\bar{\x}_0$ (red) and $\bar{\x}_1$ (green), a bi-stable phase, and for $\beta \ge \beta_c$ a single active state $\bar{\x}_1$.

\subsubsection{Reigniting} 
\label{SecEcoForced}	

Here the functions comprising $F(x)$ in (\ref{Fx}) are 

\begin{equation}
M_2(x) = x,
\label{EcoM2x}
\end{equation}

and 

\begin{equation}
R(x) = -\dfrac{F}{x} - B \left( 1 - \dfrac{x}{C} \right)(x - K).
\label{EcoRx}
\end{equation} 

The challenge is that in order to construct $F(x)$ we must invert $R(x)$, which as indicated in Fig.\ \ref{FigEco}c is non-monotonic and hence, in principle, non-invertible (purple solid line). The result is that $R^{-1}(x)$ is ill-defined for a certain range of $x$, matching potentially two values for the same $x$. This ambiguity is directly related to the bi-stability of $\bar{\x}_0$ and $\bar{\x}_1$ in the range $\beta < \beta_c$. Indeed, Eq.\ (\ref{betaXEco}) can be written in the form 

\begin{equation}
\beta \bar{\x}_\alpha = R(\bar{\x}_\alpha),
\end{equation} 

from which it follows that 

\begin{equation}
\bar{\x}_\alpha = R^{-1} (\beta \bar{\x}_\alpha),
\label{BetaRinv}
\end{equation}

in which the ambiguous value of $R^{-1}(\beta \bar{\x}_\alpha)$ for $\beta < \beta_c$ is precisely to root of the observed bi-stability. It, therefore, follows that of the two branches in $R^{-1}(x)$ the relevant branch is the one associated with $\bar{\x}_0$, as, indeed, the reigniting is applied on the failed state. Hence, to correct for this ambiguity we use a similar construction to the one shown in Fig.\ \ref{FigCobweb}d when treating the non-monotonic $F(x)$, namely, we eliminate the non-monotonic range of the function by replacing it with a constant plateau. This leads to the corrected $\tilde{R}(x)$ shown in Fig.\ \ref{FigEco}d, in which the original function (dashed-line) is replaced by a corrected monotonic $\tilde{R}(x)$ (purple solid line). Its inverse $\tilde{R}^{-1}(x)$ (yellow) is now well-defined, and, most importantly, suitable to predict the system's response to reigniting from a \textit{failed initial condition}. We can now use this corrected function to construct $F(x)$ in (\ref{Fx}) providing

\begin{equation}
F(x) = \dfrac{\tilde{R}(x)}{\omega} - 
\kappa \tilde{R}^{-1} \Big(  \omega x + \omega\kappa \bar{\x}_0 \Big).
\label{EcoFx}
\end{equation}	

This allows us to systematically use our cobweb plots to assess the system recoverability for different combinations of $\kappa$ and $\omega$ (Fig.\ \ref{FigEco}e,f). The resulting phase-diagrams in the $\kappa,\omega$ plane under different $\Delta$ are also shown (Fig.\ \ref{FigEco}g-i). Note that for sufficiently large $kappa,\omega$ the system exits the bi-stable regime and enters a state in which $\x_1$ is the only stable fixed-point (green phase at top right of phase-diagrams). In this state it spontaneously recovers even without reigniting.

\subsubsection{Recoverability of the gut microbiome}
\label{SecGutMicrobiome}

As a realistic application of Eq.\ (\ref{Ecology}) we considered the recoverability of a failed microbial network, using the interaction topology of the human gut microbiome. The construction of this network and its link weights is discussed in detail below. The resulting topology is highly diverse, both in terms of $P(k_{i,\rm in},k_{i,\rm out})$ and in terms of $P(w)$. Consequently, its shell statistics are non-uniform, with some shells characterized by higher degrees/weights than others. This represents a non-negligible discrepancy with respect to the configuration model ensemble, specifically violating characteristic 2 of Sec.\ \ref{ResultingApproximations}. Moreover, the network comprises a combination of cooperative $(\m Wij > 0)$ and adversarial links ($\m Wij < 0$), violating also characteristic 3. Hence, it offers an important testing ground for the broader applicability of our proposed framework.

\textbf{Constructing the microbiome interaction network}.\
To construct the gut microbiome we collected data on $766$ types of bacteria, $53$ archaea and $19$ eukaryotes, together comprising $N = 838$ microbial species. \cite{lim2020large} For each species the data include a description of all its consumed and produced nutrients among a set of $M = 283$ molecular compounds. This allows to construct two bipartite networks: the \textit{export} network 

\begin{equation}
\m Enm = \left\{ 
\begin{array}{ccl}
1 &  &  \textrm{In case species $n$ produces nutrient $m$}
\\[5pt]
0 &  &  \textrm{Otherwise}
\end{array}
\right.,
\label{Export}
\end{equation}

and the \textit{import} network

\begin{equation}
\m Inm = \left\{ 
\begin{array}{ccl}
1 &  &  \textrm{In case species $n$ consumes nutrient $m$}
\\[5pt]
0 &  &  \textrm{Otherwise}
\end{array}
\right.,
\label{Import}
\end{equation}

both networks of dimensions $N \times M$, namely $n = 1,\dots,N$ and $m = 1,\dots,M$. 

These two networks encode information on the inter-species interactions. For example, if $j$ produces $m$ ($\m Ejm = 1$) and $i$ consumes $m$ ($\m Iim = 1$) then $j$ positively contributes to $i$'s growth rate, capturing a directed cooperative link $j \rightarrow i$. In contrast if both $i,j$ consume the same nutrient $m$ ($\m Iim \m Ijm = 1$) then these two species compete over a shared resource, a bi-directional adversarial interaction $i \leftrightarrow j$. To quantify this we first normalize both matrices as 

\begin{equation}
\begin{array}{ccc}
\m {\E}jm = \dfrac{\m Ejm}{\displaystyle \sum_{n = 1}^N \m Enm}
& \,\,\,\,\,\,\,\, &
\m {\I}im = \dfrac{\m Iim}{\displaystyle \sum_{n = 1}^M \m Iin}
\end{array}.
\label{NormalEI}
\end{equation}

The normalized $\E_{jm}$ quantifies the relative contribution of $j$ to the availability of the nutrient $m$, as a fraction of all other $m$-exports in $E$, namely $\sum_{n = 1}^N \m Enm$. Similarly, $\I_{im}$ captures the reliance of $i$ on $m$, as one of $i$'s total of $\sum_{n = 1}^M \m Iin$ potential imports. Both terms become maximal in case $j$ is the sole exporter of $m$, or if $m$ represents a unique import of $i$. In contrast, if there are many alternative paths to produce $m$, or if $m$ is but one of $i$'s many consumed imports, the normalized weights in $\E_{jm}$ and $\I_{im}$ will be reduced, accordingly.  

We can now use (\ref{NormalEI}) to construct two $N \times N$ interaction networks. The first is the \textit{complementarity} network

\begin{equation}
\m Pij = \sum_{m = 1}^M \m {\I}im \m {\E}jm
\label{Complementarity}
\end{equation}

whose weights capture the positive contribution of species $j$ to species $i$, namely $\m Pij = P_{i \leftarrow j}$. Indeed, $j$ contributes to $\m {\E}jm$ percent of $m$'s availability, which, in turn contributes to $\m {\I}im$ percent of $i$'s growth rate. The second network captures the $i,j$ competition level via

\begin{equation}
\m Qij = \sum_{m = 1}^M \m {\I}im \m {\I}jm,
\label{Competition}
\end{equation}

quantifying $i$ and $j$'s mutual consumption of the shared resources. Note that, as opposed to $P$, $Q$ is symmetric, hence $\m Qij = \m Qji = Q_{i \leftrightarrow j}$. 

Together $P$ and $Q$ allow us to construct the complete interaction network $A$ and its weights $W$. For the topology we take $\m Aij = 1$ if \textit{either} $\m Pij$ \textit{or} $\m Qij$ are non-zero. Namely $A$ comprises all interactions, cooperative and adversarial. For the weights we assign

\begin{equation}
W = \omega_P P - \omega_Q Q,
\label{EcoWeights}
\end{equation}
 
a linear combination of the complementarity and competition networks. The rate constants $\omega_P$ and $\omega_Q$ represent a degree of freedom to set the average complementarity and competition \textit{rates}. Indeed, $P$ and $Q$ are only designed to capture the relative complementarity or competition between all node pairs, namely which pairs have a stronger and which have a weaker interaction. They are not, however, suited to provide the specific rates, which depend on units (time$^{-1}$), as well as on external conditions. For example, increasing the abundance of all nutrients is expected to weaken all competitive weights, as resources become affluent. Still it will preserve the relative strength of these weights, as expressed via (\ref{Competition}), since these relationships are embedded in the intrinsic structure of the export and import networks of (\ref{Export}) and (\ref{Import}) - indeed, a characteristic of the chemical nature of all species and nutrients. Therefore, $\omega_P$ and $\omega_Q$ offer a set of tunable parameters, by which to rescale all weights in $P$ and $Q$.     

As a guideline for setting $\omega_P,\omega_Q$, we consider the survival of the network's $N$ species. Indeed, as $\omega_Q$ is increased, competition becomes dominant and many species undergo extinction. We, therefore, simulated (\ref{Ecology}), taking $W$ from (\ref{EcoWeights}), under different values of $\omega_P$ and $\omega_Q$, seeking the window of relevant complementarity/competition strengths under which the majority of species survive. This represents conditions, \textit{e.g}., a level of resource availability, in which the system \textit{can} potentially function. Of course, even under these conditions the system may still fail, precisely the scenario for our recoverability framework. Still, the parameters in (\ref{EcoWeights}) must, at least in principle, enable a viable active state, which constrains us to a bounded range of $\omega_P$ vs.\ $\omega_Q$. We assume that these are the relevant conditions in which to examine the gut microbiome, as, indeed, under normal conditions, the majority of microbial species have non vanishing $x_i$. In the simulations of the microbiome recoverability we used these guidelines to set $\omega_P = 30$ and $\omega_Q = 1$. The remaining parameters in (\ref{Ecology}) were set to $F = 5, B = 3, C = 3$ and $K = 10$.   Under these conditions, the microbial network sustains a steady-state in which there are $568$ active species. The remaining $270$ species succumb to the adversarial links and undergo extinction.  

\textbf{Selective reigniting}.\
Within the configuration model framework of Sec.\ \ref{SecModelingFramework} most nodes are surrounded by statistically similar shells. As a consequence the specific source node $s$ is of little significance, and reigniting can be done with any selected node. In that sense, recoverability captures a characteristic of the \textit{system}, \textit{e.g}., the density/weights in the shells surrounding each node, and not of the specific choice of $s$ itself. This is rooted in characteristic 2 of Sec.\ \ref{ResultingApproximations}, stating that the $K_s(l)$ statistics are independent of $s$. Such uniformity is clearly observed in our simulations, for example in Fig.\ 3n-q of the main text, where recoverability rapidly \textit{jumps} from $\eta = 0$ to $\eta = 1$. This indicates that, typically, reigniting either succeeds for \textit{all} nodes, or fails for \textit{all} nodes, with only a narrow window in which there is a preferred fraction of nodes capable of reigniting. Hence, unless the system is only borderline recoverable, there is no significant preference for one node or the other.  

In the microbiome, however, $K_s(l)$ are diverse, and hence some nodes are better \textit{reigniters} than others. To rank the reigniting potential of a specific node $s$, we characterized the shells $K_s(l)$ surrounding it. The idea is that for $s$ to successfully activate the \textit{network} its forcing signal must penetrate not just its immediate neighborhood, but also reach the distant shells. Therefore $s$'s ranking is determined by all $K_s(l)$, rather than just by its degree. 

To evaluate this, consider the process of reigniting. The reactivating signal originates at $s$, and propagates along the shells $K_s(l)$. A typical node $i \in K_s(l)$ receives this reactivating signal through its immediate neighbor $j$ at $K_s(l - 1)$. Let us assume that $j$ has been successfully reactivated, \textit{i.e}., that $x_j$ is in a revived state $x_j = x_{\rm Rev} \in \Basin_1$, then ask what are the conditions for this reactivation to sweep also through $i$. Note that we do not assume that $x_{\rm Rev} = \bar{\x}_1$, since $\bar{\x}_1$ represents the average state in $K_s(l)$ after the entire network has been revived. In the current scenario, reigniting is in the midst of its propagation, and most of $j$'s neighbors, all in shells more distant than $K_s(l - 1)$, are still around $\Basin_0$. 

Taking $i$ to represent a \textit{typical} node, its residual degree can be extracted from (\ref{Kappa}), and hence in addition to $j$, which is closer to $s$, node $i$ is exposed to $\kappa$ other nodes that are all farther away at $K_s(l + 1), K_s(l + 2), \dots$. At this time these other nodes have not yet been affected by the reigniting signal, and hence they are all still at the $\x_0$ state. We can now approximate $i$'s response to $j$'s incoming signal via

\begin{equation}
\dod{x_i}{t} \approx M_0 \big( x_i(t) \big) + 
\omega \kappa M_1 \big( x_i(t) \big) M_2 \big( \bar{\x}_0 \big) + 
\m Wij M_1 \big( x_i(t) \big) M_2 \big( x_{\rm Rev} \big),
\label{jActivatingi}
\end{equation} 

where we split $i$'s $\kappa + 1$ interactions into two forms:\ $\kappa$ neighbors with an average weight $\omega$ at state $\x_0$, and a single neighbor $j$ at the state $x_{\rm Rev}$ and with weight $\m Wij$. The question is, therefore, what is the minimal weight $\m Wij$ that is sufficient to propagate the reigniting signal from $j$ to revive $i$. 

Taking $M_0,M_1,M_2$ from (\ref{Ecology}), Eq.\ (\ref{jActivatingi}) translates to

\begin{equation}
\dod{x_i}{t} = f\big( x_i(t) \big),
\label{EcojActivatingi}
\end{equation}

where

\begin{equation}
f(x) =  F + B x \left( 1 - \dfrac{x}{C} \right)(x - K) + 
\omega \kappa \bar{\x}_0 x + \tilde{W} x,
\end{equation}

and 

\begin{equation}
\tilde{W} = \m Wij x_{\rm Rev}
\label{Wtilde}
\end{equation} 

is the effective \textit{reviving force} acting on $x_i$. The resulting Eq.\ (\ref{EcojActivatingi}) tracks $x_i(t)$, as it interacts with a single activated neighbor $j$, and $\kappa$ suppressed ones.

\begin{figure}[h!]
\centering
\includegraphics[width=0.8\textwidth]{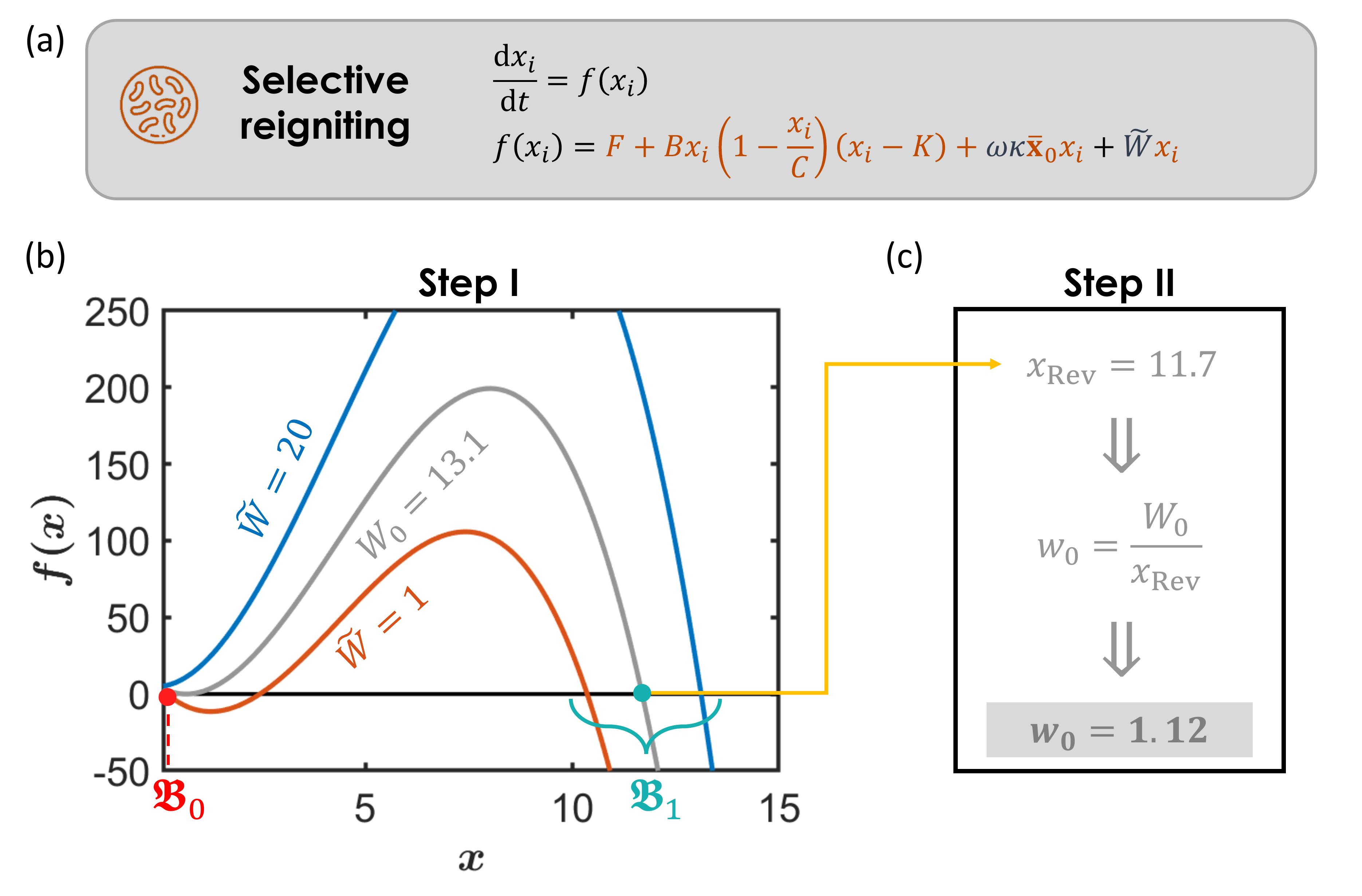}
\caption{\color{blue} \textbf{Extracting the critical weight $w_0$ and the revived state $x_{\rm Rev}$}. 
(a) We track the propagation of the reigniting signal from node $j \in K_s(l - 1)$ to $i \in K_s(l)$. This propagation is mediated by the reviving force $\tilde{W} = \m Wij x_{\rm Rev}$. 
(b) We seek a solution in which $x_i$ is revived, and hence must exit $\Basin_0$. This translates to two conditions:\ (i) $f(x)$ has a single intersection with the $x$-axis in the range of $\Basin_1$ (green); (ii) this intersection occurs at $x = x_{\rm Rev}$, as $i$ is activated towards the state of its revived neighbor $j$. Step I.\ First we seek $\tilde{W} = W_0$ that satisfies condition (i). This occurs at $\tilde{W} \ge 13.1$, predicting this to be the critical reviving force $W_0$ (grey). 
(c) Step II.\ Once we set $W_0$, we extract $x_{\rm Rev}$ from the intersection of $f(x)$ with the $x$-axis (green dot), fulfilling condition (ii). Here we find $x_{\rm Rev} = 11.7$, which using Eq.\ (\ref{CrtiticalWeight}) predicts the critical reviving weight at $w_0 = 1.12$. 
}
\label{FigReach}	
\end{figure}

Node $i$ will successfully reactivate if $j$'s reviving force $\tilde{W}$ is sufficient to transition $x_i$ towards $\x_1$, namely if Eq.\ (\ref{EcojActivatingi}) no longer has a stable state around $x_i = \bar{\x}_0$, but only around $x_i = \bar{\x}_1$. In Fig.\ \ref{FigReach}b we display $f(x)$ for different values of $\tilde{W}$. The fixed-points of the equation are captured by the intersections at $f(x) = 0$, and their stability by the condition $f^\prime(x) < 0$, \textit{i.e}.\ intersections where $f(x)$ has a negative slope. For small $\tilde{W}$ we observe two stable states (red), one around $\x_0$ and the other around $\x_1$. Under this condition, since $x_i$'s initial state is in $\Basin_0$, $i$ will not be revived, and revert to its pre-activation state. For large $\tilde{W}$ the only stable fixed-point is around the active state $\x_1$. This represents a successful activation, in which $j$ was able to transfer the reigniting signal to $i$, \textit{i.e}.\ the reviving force was sufficient (blue). In between, we observe a critical force $\tilde{W} = W_0$ (grey), in which $f(x)$ transitions from the bi-stable state to the single active fixed-point. This represents the minimal reviving force required for the reigniting at $s$ to spillover from one shell ($K_s(l - 1)$) to the next ($K_s(l)$). 

If, indeed $\tilde{W}$ is at the critical $W_0$, Eq.\ (\ref{EcojActivatingi}) predicts that $x_i$ will depart the inactive state, which is no longer stable, and transition to the revived state, $x_{\rm Rev}$. It will, in a sense, become the \textit{revived} node itself, now propagating the reigniting signal further on to its neighbors in $K_s(l + 1)$. This new state of $i$ is precisely captured by the intersection of $f(x)$ (grey solid line) with the $x$-axis in Fig.\ \ref{FigReach}b (green dot). Hence, this intersection point represents the revived state $x_{\rm Rev}$, which is an \textit{a priori} unknown variable of (\ref{jActivatingi}). We can now graphically extract $x_{\rm Rev}$ from the $f(x)$ intersection, and together with the fact that $\tilde{W} = W_0$, obtain the critical weight as

\begin{equation}
w_0 = \dfrac{W_0}{x_{\rm Rev}}.
\label{CrtiticalWeight}
\end{equation}

For $j$ to successfully revive $i$, their link weight must satisfy $\m Wij \ge w_0$.

This helps assess the reigniting capacity of a given node $s$:\ in case most links weights between $K_s(l - 1)$ and $K_s(l)$ are below $w_0$, reigniting may fail to propagate to $K_s(l)$. If, however, there are many links with $\m Wij \ge w_0$, then $K_s(l)$ will be revived by $s$, as it received the activation signal from $K_s(l - 1)$.


The analysis above indicates that the prime contributors to reigniting are the links whose weights are greater than $w_0$. If all shells are statistically identical, than they all include a similar density of such links, and hence they are all equally capable of propagating the reigniting signal (up to statistical discrepancies, which mainly occur at the bounds of recoverability, as seen, \textit{e.g}., in Fig.\ 3n-q of the main text). If, however, the shells differ significantly, as is the case for the microbiome, then we must rank them by the number of nodes they have that can be reach via links whose weight $\m Wij \ge w_0$. To achieve this we first generate the weight-selective network


\begin{table}[h!]
\centering
\includegraphics[width=0.9\textwidth]{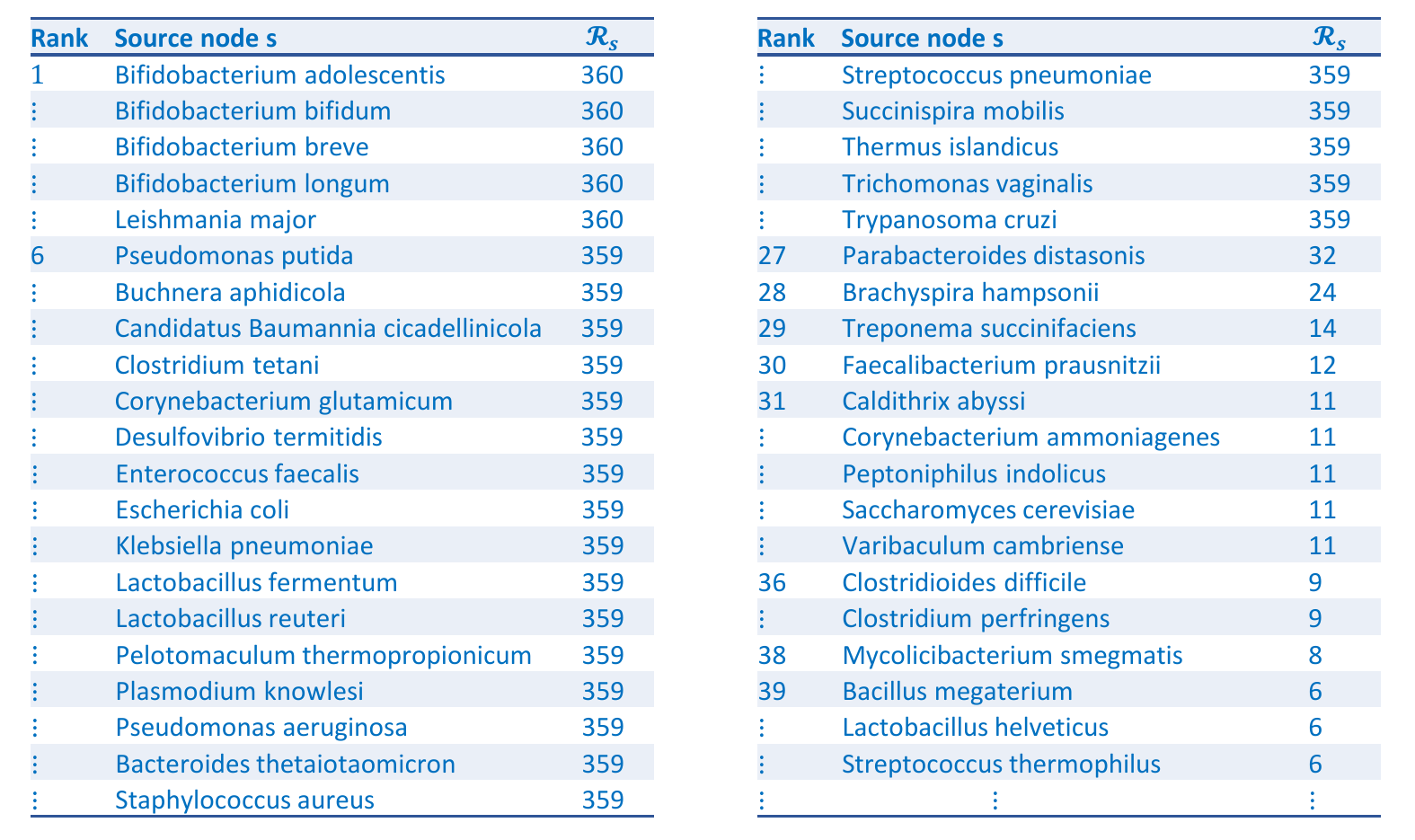}
\caption{\color{blue} \footnotesize \textbf{The reigniting capacity of all microbial species}.\
We used $\R_s$ in (\ref{Reach}) to rank all microbial species $n = 1,\dots,N$, according to their potential role as reigniting sources. We observe a very distinctive set of $26$ selective species, whose $\R_s$ is significantly higher than all others. Indeed, the majority of these $26$ species were capable of reigniting the failed microbiome. One outlier we consider below is \textit{P.\ putida}. Following restructuring this species too became a successful reigniter. In this table we only show the top $41$ reigniters; the complete table appears in Supplementary dataset 1.
}
\label{TableReigniters}	
\end{table}


\begin{equation}
\m Bij = \left\{
\begin{array}{ccc}
\m Aij & & \m Wij \ge w_0
\\[5pt]
0 & & \m Wij < w_0
\end{array}
\right.,
\end{equation} 
 
in which we only keep the links with sufficient (positive) weight, equal or above $w_0$. The selective network $B$ includes only the links that actively contribute to spreading the reigniting signal - the \textit{effectual links}. We can now use $B$ to obtain the \textit{effective shells} $K_s^B(l)$. These shells, unlike $K_s(l)$, contain only nodes that can be reached via paths of length $l$ form $s$, \textit{along the links of} $B$. Such paths are constructed from sequences of effectual links, whose weights are all above $w_0$, and hence nodes in $K_s^B(l)$ represent potential target nodes that are prone to reigniting by $s$. This allows us to define, for each source node $s$, the \textit{reach}


\begin{table}[h!]
\centering
\includegraphics[width=0.9\textwidth]{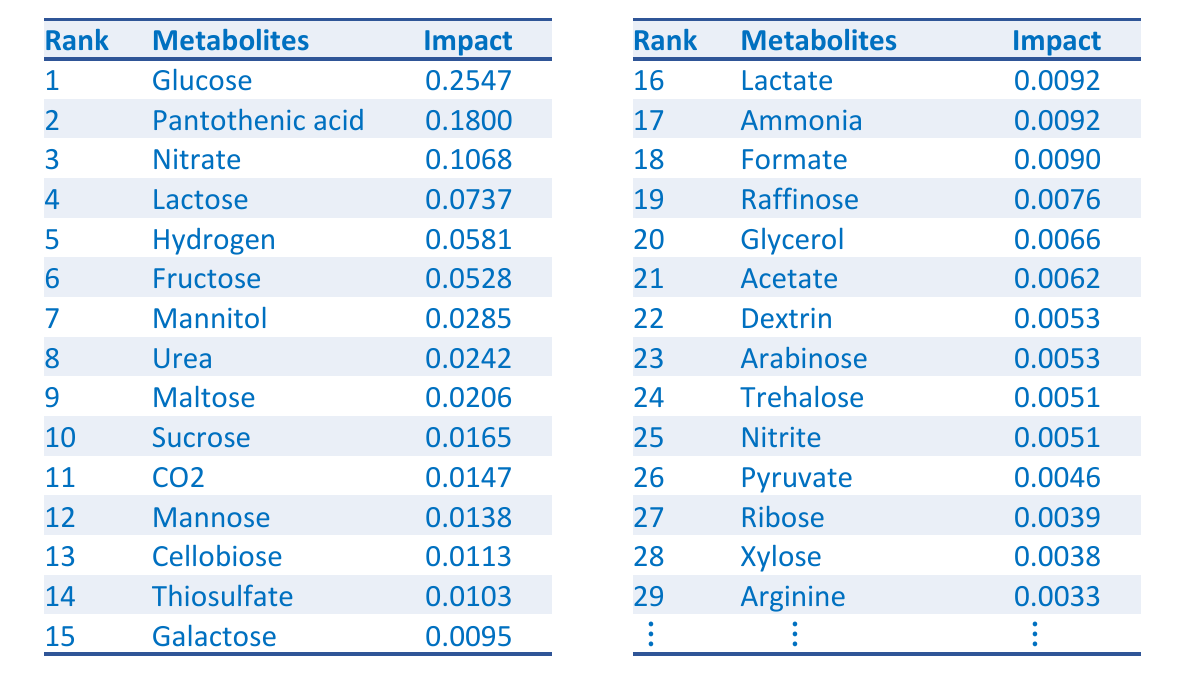}
\caption{\color{blue} \footnotesize \textbf{The adversarial impact of all nutrients}.\ 
Using $\Delta Q_m$ (\ref{DeltaQm}) we detect the nutrients that contribute most to the competition matrix $Q$. Nutritional interventions to ensure an abundance of the top ranked nutrients in the list can help restructure the network by reducing competition, thus enhancing the network's potential recoverability. Supplementing the top three nuterients, we find, was sufficient to allow recoverability via the \textit{a priori} failed reigniter node \textit{P.\ putida}. The complete table appears in Supplementary dataset 2.
}
\label{TableNutrients}	
\end{table}


\begin{equation}
\R_s = \sum_{l = 1}^{\infty} |K_s^B(l)|,
\label{Reach}
\end{equation}

which tracks the total number of nodes that $s$'s reigniting signal can, in fact, \textit{reach}. We predict that the ideal reigniters are nodes with the highest $\R_s$.  

Simulating the microbial dynamics with the inter-species interaction network constructed above, we find that the critical reviving force is $\tilde{W} = 13.1$ (Fig.\ \ref{FigReach}b, grey). The resulting $f(x)$ intersects the $x$-axis at $x_{\rm Rev} = 11.7$ (green dot), which, using (\ref{CrtiticalWeight}) predicts $w_0 = 1.12$ (Fig.\ \ref{FigReach}c). Constructing the selective network $B$, we calculated the reach $\R_s$ of all nodes and ranked them for their reigniting potential (Table \ref{TableReigniters}, Supplementary dataset 1). We find, indeed, a highly non random distribution, in which the top $26$ nodes have a significantly higher $\R_s$ than all remaining nodes - a direct consequence of the uneven distribution of weights in the microbial interaction network. Therefore, we designate these \textit{top-reigniters} to be our potential source nodes. 

\begin{figure}[t]
\centering
\includegraphics[width=0.75\textwidth]{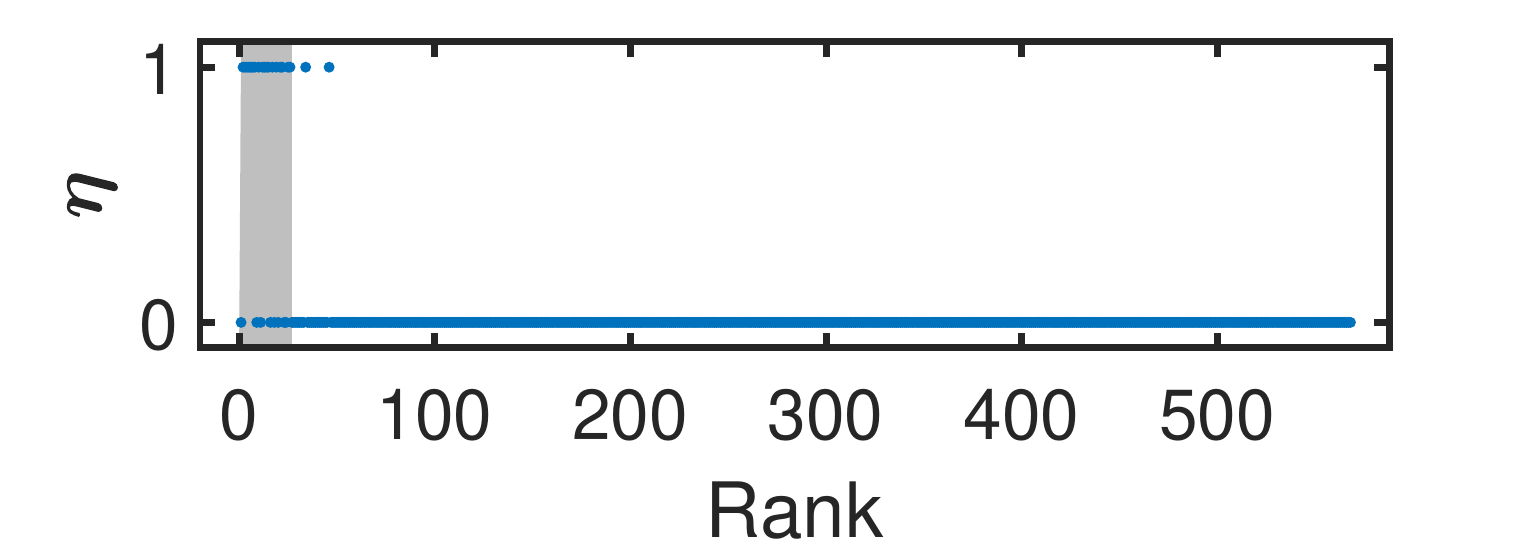}
\caption{\footnotesize \color{blue} \textbf{Selective reigniting in the microbiome}.\
The reigniting success rate $\eta$ vs.\ the $\R_s$-ranking as obtained from numerically simulating microbial dynamics. The top ranked reigniters (left) have a significantly higher probability to successfully revive the system. Highlighted (grey shade) are the top $26$ reigniters, of whom $18$ ($\sim 70\%$) successfully revived the network. Among the remaining $543$ species, only $2$ ($\sim 0.4\%$) were successful.
}
\label{FigSelectiveReigniting}	
\end{figure}

\textbf{Restructuring}.\
Our reigniting attempts with the top-ranked reigniters were mostly successful. We find that the top ranked $26$ species, all characterized by a large $\R_s$, of order $\sim 300$, (Table \ref{TableReigniters}) were, by and large, capable of steering the network back to $\Basin_1$, as predicted (success rate:\ $18$ out of $26$, \textit{i.e}.\ $\sim 70\%$). The remaining $543$ surviving species have $\R_s < 33$, namely more than an order of magnitude below the top $26$, and, therefore, quite expectantly, failed to reignite the microbiome (success rate:\ $2$ out of $543$, \textit{i.e}.\ $\sim 0.4\%$). In Fig.\ \ref{FigSelectiveReigniting} we present the reigniting success rate $\eta$ vs.\ a node's $\R_s$ ranking, showing, indeed, that all successful recovery instances are concentrated within the top ranking reigniters.

One of the failed nodes among the top ranked reigniters is the microbial species \textit{Pseudomonas putida} (\textit{P. putida} below), which, despite being ranked among the top $26$, failed to revive the system. We, therefore, used this species as grounds for testing our restructuring strategy.   

The observed reigniting failure of \textit{P.\ putida} is rooted in the abundance of adversarial interactions, in which $\m Wij < 0$, that occur within its surrounding shells. These negative interactions quench the impact of the forcing, and hinder recoverability. It is, therefore, the natural directive to restructure the network by weakening or removing the adversarial links. This can be achieved via two types of intervention:\ 

\begin{enumerate}
\item
\textbf{Node removal - via antibiotic treatment}.\ Using targeted antibiotics to reduce or remove altogether the population of selected nodes. Targeting the nodes which contribute most to the competition, namely that strengthen $Q$ in (\ref{Competition}), will help systematically quench $Q$ in (\ref{EcoWeights}), thus rendering the system \textit{more} recoverable.

\item
\textbf{Link deletion - via nutritional intervention}.\ Here we reduce the weight, or even delete, selected competitive links, by eliminating competition over specific nutrients. For example, if, using nutritional supplements, we generate an abundance of nutrient $n$, then the competition over this resource is quenched.    In technical terms, such intervention replaces $Q$ in (\ref{Competition}) by

\begin{equation}
\m Qij^{\{n\}} = \sum_{\substack{m = 1 \\ m \ne n}}^M \m {\I}im \m {\I}jm,
\label{Competition}
\end{equation}

where the contribution of $n$ to the weight of the competitive ($i,j$) link is eliminated from the sum.

\end{enumerate}

In our implementation we used option 2, which represents a more nuanced intervention, that we believe is more likely to be used in the face of a dysfunctional microbiome. Indeed, antibiotics, are, often the root cause of microbiome failure, \cite{willing2011shifting,blaser2016antibiotic,lichtman2016host} and removing additional species, from an already diluted microbiome may be deemed risky. Hence we ranked all nutrients $m = 1,\dots,M$ based on their overall contribution to the competition matrix, as evaluated by

\begin{equation}
\Delta Q_m = 1 - \dfrac{\displaystyle \sum_{i,j = 1}^N \m Qij^{\{m\}}}
{\displaystyle \sum_{i,j = 1}^N \m Qij},
\label{DeltaQm}
\end{equation}

capturing the percentage reduction in the overall competition that originates in nutrient $m$. The resulting ranking is shown in Table \ref{TableNutrients}, with the top three nutrients contributing to $Q$ identified as Glucose, Pantothenic acid (vitamin B5) and Nitrate. \cite{hord2009food} Ensuring an abundance of these three nutrients we reconstructed $Q$ and $W$ in (\ref{EcoWeights}), eliminating the competition over these selected modes. We then attempted again to reignite the failed microbiome by forcing \textit{P.\ putida}. We observe that following these, quite accessible, restructuring interventions, the microbiome, indeed, becomes recoverable through the \textit{P.\ putida}-reigniting. 


\subsection{Diffusive dynamics}
\label{SecFerromagnetism}


\begin{figure}[t]
\centering
\includegraphics[width=0.6\textwidth]{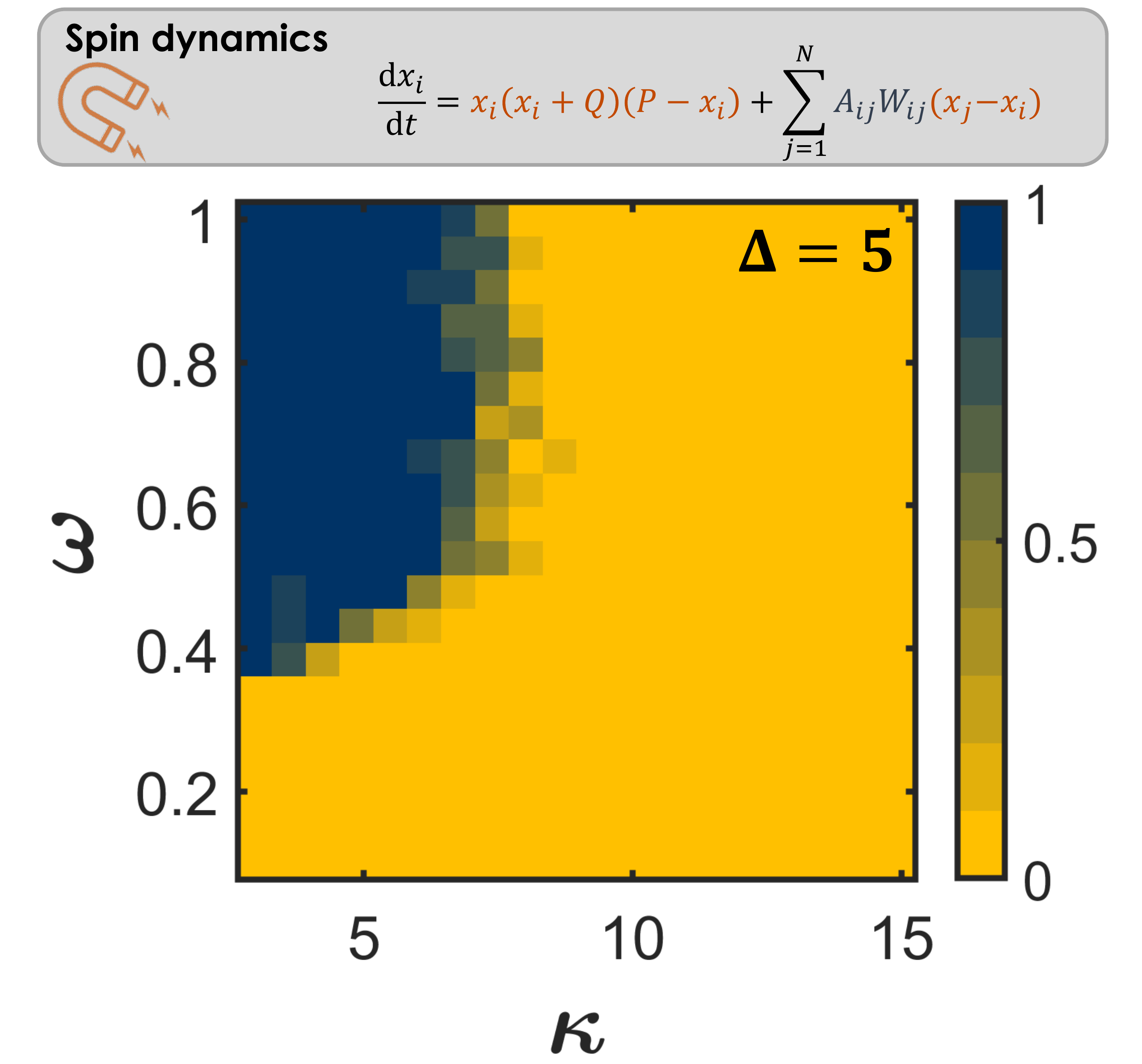}
\caption{\color{blue} \textbf{Reigniting in diffusive dynamics}.\
We show the phase diagram of recoverability in the $\kappa,\omega$ space under $\Delta = 5$. The recoverable phase ($\eta = 1$, blue) appears in the limit of small $\kappa$ and large $\omega$, \textit{i.e}.\ a sparse network with typically strong links. Here we set $P = 5$ and $Q = 1$ in Eq.\ (\ref{Ferro}).
}
\label{FigFerro}	
\end{figure}


We consider a system of coupled units (\textit{spins}) subject to an external force (\textit{magnetic field}), as modeled through the Ginzburg-Landau Equation. Here, the spin dynamics are captured via a continuum approximation, following \cite{Kessler1998pre} 
  
\begin{equation}
\dod{x_i}{t} = x_i(x_i + Q)(P - x_i) + \sum_{j = 1}^{N} \m Aij \m Wij (x_j - x_i).
\label{Ferro}	
\end{equation}

The first term on the r.h.s.\ represents the spin self-dynamics, alternating between the two spin-states $x_i = -Q$ and $x_i = P$. We focus on the case $P > Q$, capturing the breaking of symmetry due to the external field, which we arbitrarily take to be in the positive direction. The interaction term, diffusive in form, captures the magnetic pull between neighboring spins, seeking to align them towards $x_i = x_j$. The weighted topology $A \otimes W$ describes the adjacency patterns and strength of the spin-spin interactions. We consider initial conditions where all the spins are in the negative state $x_i = -Q$. Under a small external field, this state remains stable. We then select a random source spin $s$ and force it to sustain a positive state $x_s(t) = \Delta$, aiming to \textit{flip} the entire network towards $x_i = P$. This is analogous to single-node reigniting, transitioning the state of the entire network by pinning the activity of just one node.

Note that Eq.\ (\ref{Ferro}) does not fall under the framework of Eq.\ (\ref{Dynamics}), as, indeed, the interaction term cannot be factorized into the form $M_1(x_i)M_2(x_j)$. We, therefore, explore this dynamics numerically, aiming to examine the potential breadth of our proposed analysis, and its applicability beyond the limits of our analytical framework of Sec.\ \ref{SecModelingFramework}.

In Fig.\ \ref{FigFerro} we show the numerically obtained phase diagram in the $\kappa,\omega$ space, under a fixed $\Delta = 5$. Similarly to all other examined systems, also here we find a recoverable phase (blue) in which the $s$-forcing was, indeed, able to propagate to the entire network. Interestingly, in this system, recoverability favors lower $\kappa$, and hence the sparser the network $A$, the more recoverable it becomes. This is a unique feature of the diffusive dynamics in (\ref{Ferro}), which is \textit{attractive} ($x_j - x_i$) rather than \textit{cooperative} (\textit{e.g}., $x_i x_j$). In such dynamics $x_i$ tends to match with $x_j$, rather than simply grow with $j$'s activation. Therefore, having many neighbors in the $Q$-state makes flipping the network towards $P$ more difficult, as they all tend to oppose our reigniting force at $s$.


\clearpage

\section{Practical reigniting}
\label{SecPractical}

Our mathematical formulation focuses on a time-independent forcing of the form $\x_s(t) = \Delta$. This allows a tractable analysis, in which the forcing acts as an effective boundary condition, allowing us to analytically predict the long term response of the system via the recurrence (\ref{RecursionF}). In reality, it is often difficult to sustain such a constant reigniting signal, especially as the source node $s$ continues to interact with all other nodes via (\ref{Dynamics}). The crucial point is, however, that recoverability is, by and large, insensitive to the specific form of the forcing, as long as the reigniting node is kept at sufficiently high activity for a sufficiently long period. Indeed, an active $s$ reactivates its surrounding shells $K_s(l)$ even if its activity $x_s(t)$ fluctuates, as long as it maintains an \textit{average} high activity for enough time to drive all $K_s(l)$ into $\Basin_1$.

To examine this we consider a practical forcing, in which at given time intervals $T$ we externally administer \textit{jolts} to the activity of $s$ as

\begin{equation}
x_s(nT) \to x_s(nT) + \Delta,
\label{Boost}
\end{equation}
 
with $n = 0,1,\dots$. This can capture, for example, the periodic intake of medication or probiotics, which creates an instantaneous boost in the concentration of a specific protein, metabolite or microbial species. In neuronal or ferromagnetic networks, such forcing represents repetitive excitation, which can help gradually reactivate the entire system. Following each boost, $x_s(t)$ will, at first, be highly active, but then undergo relaxation back towards $\Basin_0$, driven by the natural system dynamics (\ref{Dynamics}). Let us consider $x_s(t)$'s natural relaxation time $\tau$, after which the effect of the $\Delta$-boost wares off. While this is not guaranteed, we can treat, for simplicity, $x_s(t)$'s relaxation following the $n$th reigniting boost as an exponential decay of the form

\begin{equation}
x_s(t) \approx \Delta e^{-\frac{t}{\tau}}.
\label{BoostRelaxation}
\end{equation} 

Other forms of decay are also possible, depending on the structure of Eq.\ (\ref{Dynamics}), but the analysis that follows remains equally relevant. For the periodic reigniting (\ref{Boost}) to succeed we require that $x_s(t)$ remains \textit{on average} above the critical forcing $\Delta_c$. Using (\ref{BoostRelaxation}) this translates to

\begin{equation} 
\dfrac{1}{T} \int_0^T \Delta e^{-\frac{t}{\tau}} \dif t \ge \Delta_c,
\label{BoostCondition}
\end{equation}

which in turn predicts

\begin{equation} 
\dfrac{\tau}{T} \left( 1 - e^{-\frac{T}{\tau}} \right) \ge \dfrac{\Delta_c}{\Delta}.
\label{BoostThreshold}
\end{equation}
 
Equation (\ref{BoostThreshold}) offers the bounds on the periodicity of the reigniting boosts. It takes $\tau$, a characteristic of the system's dynamics, $\Delta_c$, the critical forcing provided by our theory, and $\Delta$, our boost intensity, and provides the maximal waiting time between successive activation instances $T$. On the one hand, the larger the relaxation time $\tau$, the less frequent are the required activation boosts (large $T$). On the other hand, if $\Delta_c$ is large, \textit{i.e}.\ the system is difficult to reignite, we need more frequent instances of activation. This analysis can be readily generalized for different relaxation functions, other than the exponential decay of (\ref{BoostRelaxation}). 


\begin{figure}[t]
\centering
\includegraphics[width=0.95\textwidth]{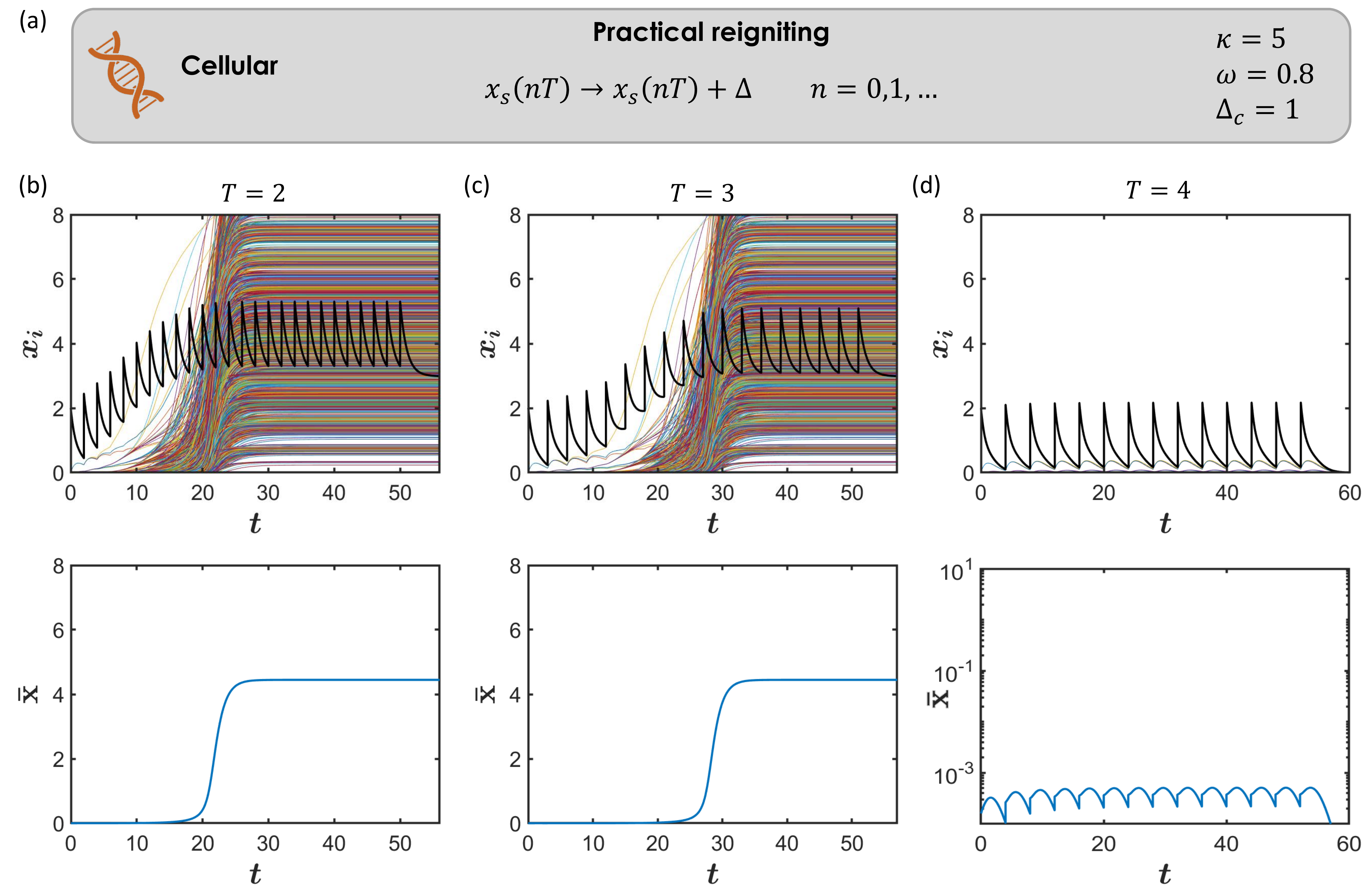}
\caption{\color{blue} \footnotesize \textbf{Practical reigniting}.\ 
(a) We applied reigniting through periodic boosts \textit{a l\`{a}} Eq.\ (\ref{Boost}) to our cellular dynamics. The network has $\kappa = 5$ and $\omega = 0.8$, for which our theory predicts a critical reigniting force of $\Delta_c = 1$.
(b) Using boosts of magnitude $\Delta = 2$ and periodicity of $T = 2$ we observe a successful reigniting. The source node (black) exhibits a chainsaw dynamics, with each boost followed by the natural system relaxation, hence it cannot sustain a constant forcing $\Delta$. Still, the boosts are sufficiently frequent to gradually reactivate all other node activities $x_i$ (colored solid lines), successfully steering the system into $\Basin_1$.
(c) Increasing the lag between successive boosts to $T = 4$, we now witness a failed reigniting, as predicted. 
}
\label{FigBoosts}	
\end{figure}


In Fig.\ \ref{FigBoosts} we examine this reigniting strategy on our Cellular dynamics of Sec.\ \ref{SecCellular}. We evaluate the relaxation time $\tau$ via the nodes' self-dynamics, writing

\begin{equation}
\dod{x_s}{t} \approx -Bx_s^a(t) + \dots.
\end{equation}

The idea is that the interaction term in (\ref{Dynamics}) contributes little to $s$'s relaxation as compares to its own internal dynamics. Setting $B = 1$ and $a = 1$, as we do in Sec.\ \ref{SecCellular} we have in Eq.\ (\ref{BoostRelaxation})

\begin{equation} 
x_s(t) \approx \Delta e^{-t},
\end{equation}

a relaxation time of $\tau = 1$. For this system, with $\kappa = 5$ and $\omega = 0.8$, we have $\Delta_c \approx 1$, which in (\ref{BoostThreshold}) provides 

\begin{equation} 
\dfrac{1}{T} \left( 1 - e^{-T} \right) \ge \dfrac{1}{\Delta}. 
\end{equation}

Setting $\Delta = 2$, \textit{i.e}.\ boosts that are roughly double in size of the critical forcing, this predicts $T \lesssim 2$. Indeed, we find that reigniting via (\ref{Boost}) with $\Delta = 2$ and intervals of $T = 2$ successfully revives the system (Fig.\ \ref{FigBoosts}b). However, similar conditions with $T = 4$ fail to reignite the system, as the boosts taper off before the system exhibits a sufficient response (Fig.\ \ref{FigBoosts}d). 

In practice, Eq.\ (\ref{BoostThreshold}) provides an underestimation of the maximal lag $T$. This is because, as the reigniting boosts accumulate, the surrounding nodes increase their activity, providing feedback that further reinforces the system's recovery. Therefore, while a lag significantly above $T = 2$ may fail, as shown in Fig.\ \ref{FigBoosts}d, if $T$ is just slightly above the predicted threshold, we may still witness a successful recovery. Indeed, in Fig.\ \ref{FigBoosts}c we observe a revived system with $T = 3$, slightly above the predicted threshold of $2$.


\vspace{20mm}

\section{Numerical analysis}

\subsection{Numerical integration}
\label{numerical_inte}

To numerically test our predictions we constructed Eq.\ (\ref{Dynamics}) for each of the systems in Sec.\ \ref{SecDynamicModels}, using the appropriate network $A$ (Scale-free, Erd\H{o}s-R\'{e}nyi, empirical, etc.) and weights (random $P(w)$ or empirically constructed). We then used a fourth-order Runge-Kutta stepper (Matlab's \texttt{ode45}) to numerically solve the resulting equations. Starting from a pre-selected initial condition $x_i(t = 0)$, $i = 1,\dots,N$ we allowed the system to reach steady-state by waiting for $\dot x_i \rightarrow 0$. To numerically realize this limit we implemented the termination condition 

\begin{equation}
\max_{i = 1}^N \left| \frac{x_i(t_n) - x_i(t_{n - 1})}{\Delta t_n}\right| < \varepsilon,
\label{TerminationCondition}
\end{equation}

where $t_n$ is the time stamp of the $n$th Runge-Kutta step and $\Delta t_n = t_n - t_{n - 1}$. As the system approaches a steady-state, the activities $x_i(t_n)$ become almost independent of time, and the numerical derivative $\dot x_i = (x_i(t_n) - x_i(t_{n - 1})) / \Delta t_n$ becomes small. The condition (\ref{TerminationCondition}) guarantees that the maximum of $\dot x_i$ over all activities $x_i(t_n)$ is smaller than the pre-defined termination variable $\varepsilon$. In our simulations, across the different dynamics we tested, we set $\varepsilon \le 10^{-2}$ to ensure that our system is sufficiently close to its {\it true} steady-state.

In case of bi-stability we examined the convergence of the system from multiple initial conditions. For example, setting $\x(t = 0)$ to a low value in $\Basin_0$ ensures convergence to $\x_0$, in case $\x_0$ is stable; setting it in $\Basin_1$ ensures convergence to $\x_1$, in case $\x_1$ is also stable. If only one of the states is stable - all initial conditions will converge to that single fixed-point. 

\subsection{Reigniting}
\label{Reigniting}

To simulate reigniting we set the initial condition of the system to $\x(t = 0) = \x_0$. We then select a random node $s$, decouple it from the remaining $N$ equations and set its state to $x_s(t) = \Delta$. Together with the remaining $N - 1$ equations of (\ref{Dynamics}) we arrive at Eq.\ (\ref{ForcedDynamics}), which takes the form

\begin{equation}
\left\{
\begin{array}{cclr}
x_s(t) &=& \Delta &
\\[10pt]
\dod{x_i}{t} &=& M_0(x_i) +  \displaystyle \sum_{j = 1}^N \m Aij \m Wij M_1(x_i) M_2(x_j) &
i \neq s
\end{array}
\right..
\label{SimForcedDynamics}
\end{equation}

Integrating this equation until reaching steady-state, \textit{i.e}.\ condition (\ref{TerminationCondition}), we find the final state $\x_{\rm Forced}$ of the \textit{forced} system. We then relax our forcing, re-couple $x_s(t)$ to the remaining $N - 1$ equations, and allow the system to relax to its final (unforced) state. This is achieved by setting the new initial condition to $\x(t = 0) = \x_{\rm Forced}$, and numerically solving Eq.\ (\ref{Dynamics}) until reaching steady-state. In case $\x_{\rm Forced} \in \Basin_1$, a successful reigniting, the system will reach $\x_1$. If, however, our reigniting failed, and $\x_{\rm Forced}$ remains in $\Basin_0$, the system, after forcing ceases, will revert to the undesired $\x_0$.

To simulate the forcing via periodic boosts, as considered in Sec.\ \ref{SecPractical}, we simply replace the equation $x_s(t) = \Delta$ in (\ref{SimForcedDynamics}) with the dynamic forcing of Eq.\ (\ref{Boost}).  Sustaining this dynamic activation for $\sim 20 - 30$ cycles, we once again cease our forcing and observe the system convergence to its natural state - $\x_1$, successful, or $\x_0$, unsuccessful.

\subsection{Constructing the phase diagrams in Figs.\ 3 and 5}
\label{SecPhaseDiagrams}

We used the configuration model framework, described in Sec.\ \ref{SecModelingFramework}, to construct undirected networks with a pre-assigned $\kappa$ and $omega$. The networks all included $N = 10^4$ nodes, had a scale-free degree distribution $P(k) \sim (k - k_0)^{-\gamma}$ and a bounded weight distribution $P(w)$. This allowed us to control $\kappa$ by varying $k_0$ within the range $[1,3]$ and $\gamma$ within $[2.5,4]$. Together we generated networks corresponding to $50$ distinct values of $\kappa$, ranging from $\sim 5$ to $\sim 50$ (Fig.\ 3) or $\sim 3$ to $\sim 30$ (Fig.\ 5). Next we matched each of these networks network with weights, representing $50$ distinct values of $\omega$, providing us, together, with $2,500$ independent combinations of $\kappa$ and $\omega$, upon which to examine our reigniting analysis. Each such $\kappa,\omega$ combination represents a single pixel in the $\kappa,\omega$ phase diagram. For each $\kappa,\omega$ combination we constructed $20$ independent network realizations, \textit{i.e}.\ $2,500 \times 20 = 5 \times 10^4$ weighted scale-free networks in each phase-diagram, comprising $20$ networks per pixel. 

For each network, we obtained the steady-state/s as explained in Sec.\ \ref{numerical_inte}, observing whether they are in the active, inactive or bi-stable regime. We then also tested whether they are recoverable via single-node reigniting, following Sec.\ \ref{Reigniting}. Counting the successful reigniting attempts $C$ among the $20$ networks with a given $\kappa,\omega$, we extracted $\eta = C/20$, the fraction of successful reigniting instances in each pixel. In a similar fashion we also selected $50$ values of the forcing $\Delta$, providing us, once again with $2,500$ combinations of $\kappa,\Delta$ and an additional $2,500$ of $\omega,\Delta$, upon which to test reigniting. As above, each pixel, represents $20$ realizations of reigniting, here selecting a random source node.


\begin{table}[t]
\centering
\includegraphics[width=0.99\textwidth]{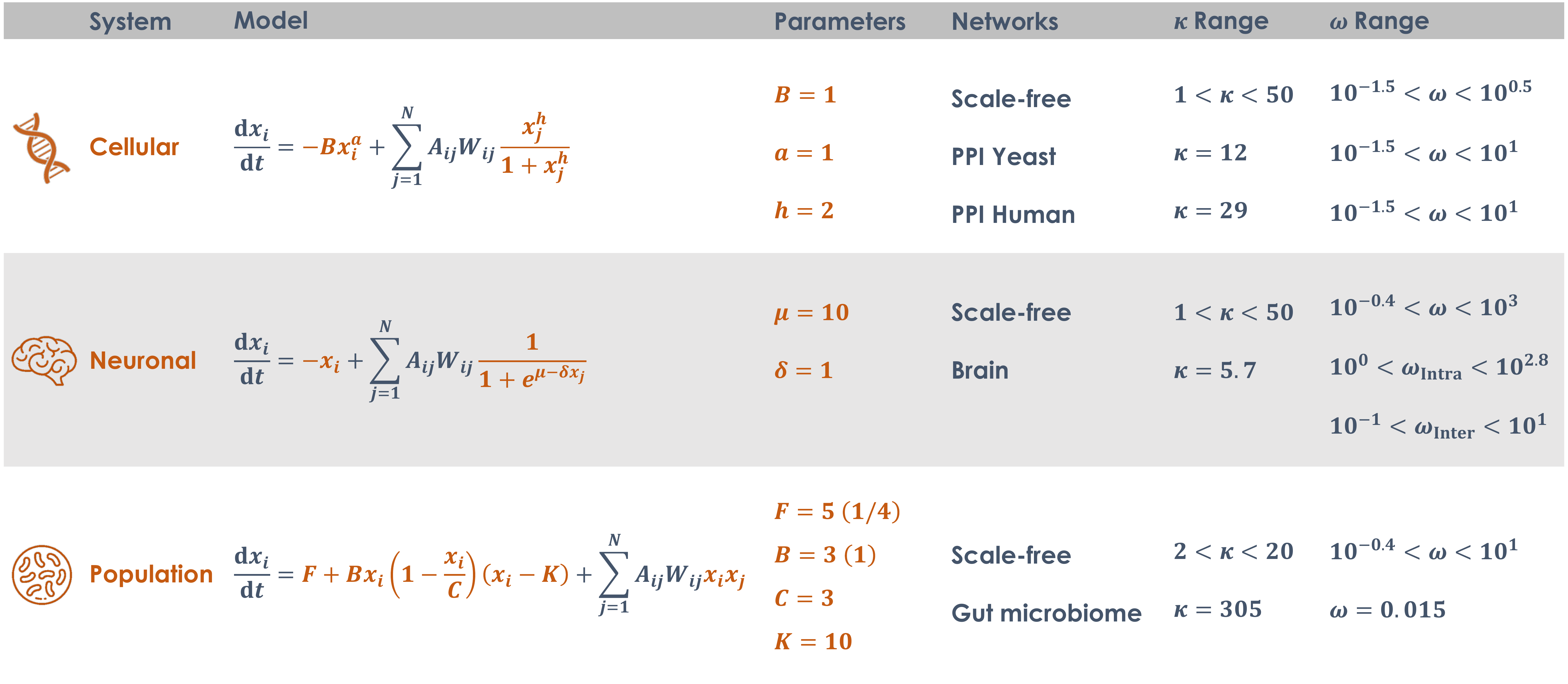}
\vspace{-3mm}
\caption{\footnotesize \color{blue} \textbf{Summary of models and parameters}.\
For each system we show the relevant dynamic equation (Model), and the parameters we used in our simulations. Each system was examined on model scale-free networks and relevant empirical networks. When constructing the phase-diagrams of Figs.\ 3 and 5 of the main text and Supplementary Fig.\ \ref{FigEco}g-i, we use scale-free networks with a range of $\kappa$ and $\omega$ values, as shown in the two rightmost columns. For the empirical networks $\kappa$ is fixed, yet we varied $\omega$ as listed in the relevant column:\ for PPI - see Fig.\ 3n-q of the main text; Brain - Fig.\ 5g of the main text. For the Gut-microbiome, $\omega$ represents the average weight of the empirically constructed network as described in Sec.\ \ref{SecGutMicrobiome}.
}
\label{ParameterTable}
\end{table}


\subsection{Model and empirical networks}
\label{Networks}

We used model and real networks, as summarized below:\

\textbf{Scale-free}.\
Scale-free networks, constructed via the configuration model, with $N = 10^4$ nodes, degree distribution $P(k) \sim (k - k_0)^{-\gamma}$. Varying the parameters $k_0$ and $\gamma$ we obtain an ensemble of scale-free networks, which we used to construct the phase-diagrams of Figs.\ 3 and 5 of the main text.

\noindent {\bf PPI Yeast} (Regulatory).\
The yeast protein-protein interaction network, an empirical scale-free network consisting of 
$1,647$ nodes (proteins) and $5,036$ undirected links, representing chemical interactions between proteins \cite{Yu2008}. 

\noindent {\bf PPI Human} (Regulatory).\ 
The human protein-protein interaction network, a scale-free network, consisting of $N = 2,035$ nodes (protein) and $L = 13,806$ protein-protein interaction links \cite{Rual2005}.

\textbf{Brain} (Neuronal).\ 
Mapping the physical fiber bundle connections between 998 brain regions, as measured using diffusion tensor imaging techniques \cite{bullmore2009complex}. 
The empirical network has a very broad weight distribution $P(w)$, ranging over several orders of magnitude, as $3 \times 10^{-5} < w < 0.23$. To extract the \textit{meaningful} connections, we set a threshold at $w_0 = 0.03$, and removed all links $(i,j)$ with $\m Wij < w_0$, resulting in a network constructed from the top $12.5\%$ strongest links. In this network the link weight $\m Wij$ does not represent the interaction rate between $i$ and $j$, but rather our confidence in the observed $(i,j)$ link. Therefore, once we selcted the top-scoring links, their weight is irrelevant in Eq.\ (\ref{Dynamics}), and hence we set it to be uniform.

This empirical network has a natural modular structure owing to the brain's two hemispheresthis allowed us to investigate the impact of molecularity on reigniting, by setting distinct weights for the commiseral vs.\ the intra-hemispheral links. The results are presented in Fig.\ 5 of the main text.

\textbf{Gut microbiome} (Microbial).\
This network construction is detailed in Sec.\ \ref{SecGutMicrobiome}.


\begin{table}[t]
\center
\includegraphics[width=0.6\textwidth]{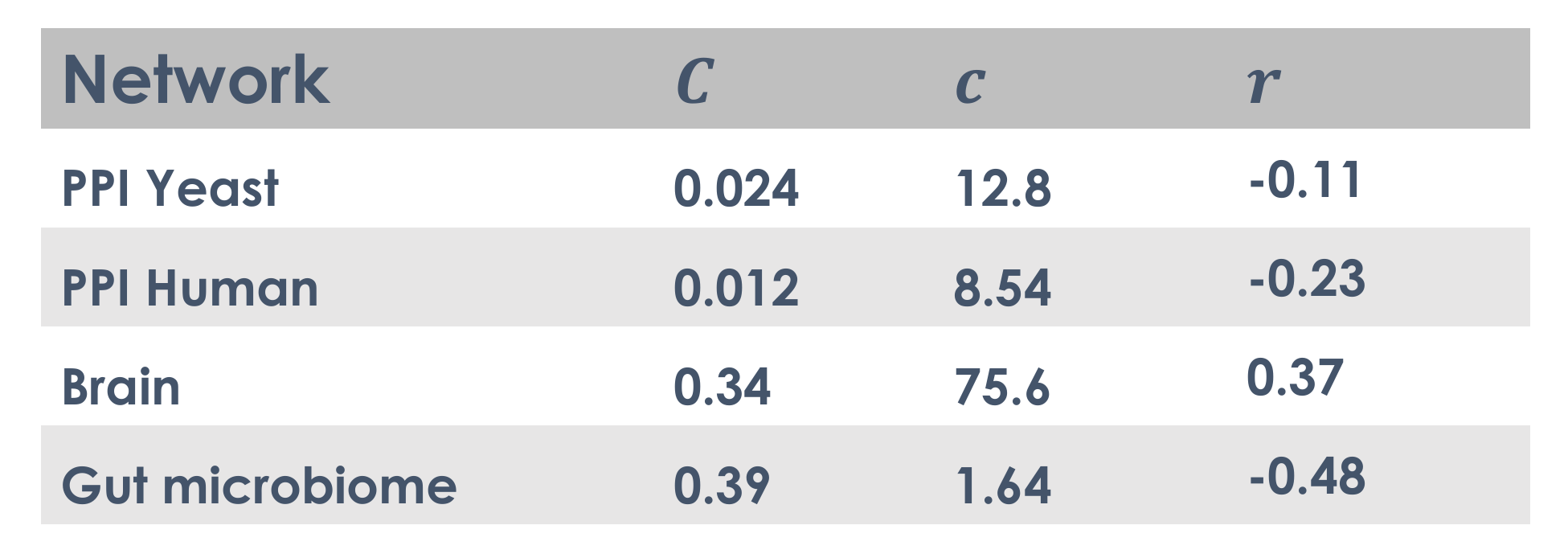}
\caption{\footnotesize \color{blue} \textbf{Robustness of our analytical framework}.\
For each of our empirical networks we present the clustering $C$, the normalized clustering $c$ and the degree-correlation coefficient $r$. Note that while the gut-microbiome is a directed network, here, for simplicity, we extracted $C$ and $r$ from its undirected equivalent $A^{\rm UD}_{ij} = \max(\m Aij,\m Aij^\top)$. This is because our only goal is to evaluate the level of discrepancy of our real-world $A$ from the configuration model, not to assess a specific characteristic of this or another networks.}
\label{TableDeviations}
\end{table}


\subsection{Deviations from the configuration model}
\label{SecDeviations}

While our analytical derivations are suited for the configuration model ensemble of Sec.\ \ref{SecModelingFramework}, in practice, we find that they offer rather accurate predictions and qualitative insights also on real-world networks, all of which exhibit deviations from the configuration model. Specifically, as explained in Sec.\ \ref{ResultingApproximations}, there are two features of the configuration model that are especially relevant in our analysis:\ (i) The tree-like structure, implying a scarcity of loops; (ii) the shell statistics, indicating the random nature of the connectivity patterns in $A$. 

We can quantify the extent to which a network violates (i), by measuring its clustering coefficient $C$ \cite{Newman2010}. This measure, indeed, tracks the abundance of short range loops in the network. For a randomly wired network we expect $C_0 = \av{k}/N$, which approaches zero in the limit $N \to \infty$. In Table \ref{TableDeviations} we show, for each network its actual clustering coefficient $C$, and also its normalized clustering $c = C/C_0$. The latter helps us quantify the level of deviation from our assumed tree-like network. We find that our empirical networks show significant discrepancies from the random wiring assumption, with $C$ often an order of magnitude above $C_0$, and yet our predictions continue to hold. This indicates that our theory is, by and large, robust against the presence of loops. In the case of Brain ($c = 75.6$), the deviation is truly consequential. And, as we have shown in the main text, it has a distinctive modular structure, which, indeed, requires specific treatment (Fig.\ 5 of main text). 

Characteristic (ii) can be examined via the network's degree-correlation coefficient $r$, designed to capture the distinct statistical properties of the shells surrounding different degree nodes. Once again, we expect, in the configuration model ensemble, to have $r \to 0$. In reality, our networks exhibit non-negligible $r$, ranging from $-0.48$ to $+0.37$, an array of non-negligible deviations from characteristic 2 of Sec.\ \ref{ResultingApproximations}. This, again, is a testament to our framework's robustness against such microscopic and meso-scopic discrepancies.   

\clearpage

\footnotesize
\bibliographystyle{unsrt}
\bibliography{bibliography}